%% file: jsikora_ApBp02_nov13.tex
\documentclass[useAMS,usenatbib]{mn2e}

\usepackage{amsmath}
\usepackage{amssymb}
\usepackage{graphicx}
\usepackage{graphics}
\usepackage{float}
\usepackage{epstopdf}
\usepackage{multicol}
\usepackage{breakurl}
\usepackage{textcomp}
\usepackage{enumitem}
\usepackage{subfigure}
\usepackage[bf,labelsep=period,font=small]{caption}
\usepackage{subfloat}
\usepackage{subfigure}
\usepackage{gensymb}
\usepackage{afterpage}
\usepackage{supertabular}
\usepackage{longtable}
\usepackage[flushleft]{threeparttable}
\title[mCP Stars Within 100pc II]{A Volume Limited Survey of mCP Stars Within 100pc \\II: Rotational and Magnetic Properties}
\author[J. Sikora et al.]
	{J.~Sikora,$^{1,2}$ G.~A.~Wade,$^{2}$ J.~Power,$^{1,2,4}$ C.~Neiner$^3$\\
$^{1}$Department of Physics, Engineering Physics \& Astronomy, Queen's University, Kingston, ON Canada, K7L 3N6\\
$^{2}$Department of Physics and Space Science, Royal Military College of Canada, PO Box 17000 Kingston, Ontario K7K 7B4, Canada\\
$^3$LESIA, Observatoire de Paris, PSL University, CNRS, Sorbonne Université, Univ. Paris Diderot, Sorbonne Paris Cité, \\\hspace{1cm}5 place Jules Janssen, 92195 Meudon, France\\
$^4$Large Binocular Telescope Observatory, 933 North Cherry Avenue, Tucson, AZ 85721, USA}

\begin{document}

\date{Accepted 2018 Oct. 23}

\pagerange{\pageref{firstpage}--\pageref{lastpage}} \pubyear{2018}

\maketitle

\label{firstpage}

\begin{abstract}
Various surveys focusing on the magnetic properties of intermediate-mass main sequence (MS) stars 
have been previously carried out. One particularly puzzling outcome of these surveys is the 
identification of a dichotomy between the strong ($\gtrsim100\,{\rm G}$), organized fields hosted by 
magnetic chemically peculiar (mCP) stars and the ultra-weak ($\lesssim1\,{\rm G}$) fields associated 
with a small number of non-mCP MS stars. Despite attempts to detect intermediate strength fields (i.e. 
those with strengths $\gtrsim10\,{\rm G}$ and $\lesssim100\,{\rm G}$), remarkably few examples have 
been found. Whether this so-called ``magnetic desert", separating the stars hosting ultra-weak fields 
from the mCP stars truly exists has not been definitively answered. In 2007, a volume-limited 
spectropolarimetric survey of mCP stars using the MuSiCoS spectropolarimeter was initiated to test the 
existence of the magnetic desert by attempting to reduce the biases inherent in previous surveys. Since 
then, we have obtained a large number of ESPaDOnS and NARVAL Stokes $V$ measurements allowing this 
survey to be completed. Here we present the results of our homogeneous analysis of the rotational 
periods (inferred from photometric and magnetic variability) and magnetic properties (dipole field 
strengths and obliquity angles) of the 52 confirmed mCP stars located within a heliocentric distance of 
$100\,{\rm pc}$. No mCP stars exhibiting field strengths $\lesssim300\,{\rm G}$ are found within the 
sample, which is consistent with the notion that the magnetic desert is a real property and not the 
result of an observational bias. Additionally, we find evidence of magnetic field decay, which confirms 
the results of previous studies.
\end{abstract}


\section{Introduction}\label{sect:intro}

The generation and broader characteristics of magnetic fields of cool stars are reasonably well 
understood within the framework of stellar dynamo theory \citep[e.g.][]{Charbonneau2010}. In contrast, 
the origin of the magnetic fields of main sequence (MS) stars more massive than about $1.5\,M_\odot$ 
remains a profound mystery. Over the past several decades, many clues related to this problem have 
been reported.

It is now reasonably well established that all magnetic, chemically peculiar stars (i.e. Ap/Bp stars, 
hereinafter referred to as mCP stars) host organized magnetic fields with strengths as large as 
$30\,{\rm kG}$ \citep[e.g.][]{Landstreet1982,Shorlin2002}. In general, the large-scale structures of 
these fields are relatively simple \citep[e.g.][]{Babcock1956,Kochukhov2015}, although a few obvious 
examples of more complex fields have been discovered \citep[e.g.][]{Kochukhov2011,Silvester2017}. 
Furthermore, both young and evolved MS mCP stars are known to exist 
\citep[e.g.][]{Wade1997,Kochukhov2006}, which suggests that these fields are stable over long time 
periods. Surface magnetic fields have been detected on some Herbig Ae/Be stars 
\citep[e.g.][]{Wade2007a,Alecian2013}, which are likely the progenitors of the MS mCP stars. 
All of these findings are consistent with the notion that the fields hosted by mCP stars are fossil 
remnants left over from an earlier stage in the star's evolution \citep[the \emph{fossil field} 
theory,][]{Cowling1945,Moss1984,Landstreet1987a}.

One property of stellar magnetism of upper MS stars that is not currently well explained by the fossil 
field theory is the fact that only $\sim10$~per~cent of all MS A- and B-type stars 
\citep[e.g.][]{Wolff1968,Smith1971} host strong, organized surface magnetic fields. \citet{Shorlin2002}, 
\citet{Bagnulo2006}, and \citet{Auriere2010} obtained a large number of magnetic measurements of 
non-mCP MS stars of spectral types A and B with median uncertainties of $20\,{\rm G}$, $95\,{\rm G}$, 
and $2\,{\rm G}$, respectively, however, no magnetic detections were reported. \citet{Makaganiuk2010} 
carried out a similar survey of HgMn stars -- obtaining typical longitudinal field uncertainties 
$\sim10\,{\rm G}$ and as low as $0.8\,{\rm G}$ -- but did not report any detections of circularly 
polarized Zeeman signatures. Recently, fields with strengths $\lesssim1\,{\rm G}$ (so-called 
ultra-weak, or Vega-type, fields) were detected on a small number of non-mCP stars 
\citep[e.g.][]{Lignieres2009,Petit2011,Blazere2016}. Based on these findings, \citet{Petit2011} 
speculate that a much higher fraction of MS A-type stars (i.e. $\gg10$~per~cent) may host ultra-weak 
surface fields. Regardless, the dichotomy between the strongly magnetic and the non-magnetic (or very 
weakly magnetic) MS A-type stars is unlikely to be entirely explained by the sensitivity of the current 
generation of spectropolarimeters. In the case of Vega, it is reported that its ultra-weak field 
exhibits a highly complex field structure \citep{Petit2010} that is atypical of the strongly magnetic 
mCP stars. It is therefore plausible that the ultra-weak fields are a distinct phenomenon, which may 
have an origin that differs from that of the strong, organized fields hosted by mCP stars 
\citep{Braithwaite2013b}.

In 2007, \citet{Auriere2007} explored the weak field regime of mCP stars by obtaining high-precision 
longitudinal field measurements of 28 such objects with reportedly weak or otherwise poorly constrained 
field strengths. All of the observed mCP stars were detected in their spectropolarimetric observations, 
and were inferred to exhibit dipolar field strengths of $B_{\rm d}\gtrsim100\,{\rm G}$ with the two 
weakest fields found to have $B_{\rm d}=100_{-100}^{+392}\,{\rm G}$ and 
$B_{\rm d}=229_{-76}^{+248}\,{\rm G}$. \citet{Auriere2007} hypothesized that there exists a critical 
field strength ($B_{\rm c}\approx300\,{\rm G}$), which corresponds to the minimum field strength that 
an mCP star must host in order to be invulnerable to a magnetohydrodynamic pinch-instability 
\citep{Tayler1973,Spruit2002}. In this scenario, every intermediate-mass MS star may be initially 
``assigned" a field strength (perhaps based on external factors, e.g. the local field properties at 
its location of formation, the presence of a companion, etc.) drawn from a probability distribution 
that increases towards lower field strengths; only those fields exceeding $B_{\rm c}$ are able to be 
maintained, which results naturally in the so-called ``magnetic desert" \citep[i.e. the dichotomy 
between the ultra-weak fields detected on a small number of non-mCP stars and the strong fields hosted 
by mCP stars,][]{Lignieres2014}.

While the detection of ultra-weak fields may not directly contradict the existence of a critical 
lower field strength limit, two stars have been found reportedly hosting fields with intermediate 
strengths \citep[i.e. $10\lesssim B_{\rm d}\lesssim100\,{\rm G}$, which is lower than the typical 
$B_{\rm c}\sim300\,{\rm G}$ proposed by][]{Auriere2007}. The massive early B-type star $\beta$~CMa 
reportedly hosts a field with $B_{\rm d}<230\,{\rm G}$ \citep{Fossati2015} while the primary component 
of the spectroscopic binary HD~5550 is reportedly an Ap star hosting a field having 
$B_{\rm d}<85\,{\rm G}$ \citep{Alecian2016}. We discuss these two examples in Sect. 
\ref{sect:discussion}; however, we note that the fact that nearly all mCP stars are found hosting 
fields $\gtrsim100\,{\rm G}$ despite the current detection limits that have been achieved remains 
conspicuous.

A potential problem with many of the reported empirical properties of mCP stars -- including the 
existence of the magnetic desert -- is the fact that they are generally inferred from intrinsically 
biased surveys: they are either biased towards brighter objects (magnitude limited surveys) or those 
hosting stronger, more easily detectable fields (field-strength limited surveys). In 2007, a volume-limited survey of mCP stars located within a heliocentric distance 
of $100\,{\rm pc}$ was initiated by \citet{Power2007_MSc} in order to reduce these observational biases. 
This work yielded the magnetic properties of a large number of mCP stars in the sample using 
measurements obtained with the now-decommissioned MuSiCoS spectropolarimeter at the Pic du Midi 
Observatory. However, at the completion of that investigation, nearly half of the sample remained 
either unobserved or had relatively poor constraints on their field strengths and geometries. We have 
recently completed this survey using measurements obtained by ESPaDOnS and NARVAL.

In Paper~I, we described in detail the sample of mCP and non-mCP stars included in the volume-limited 
sample. This sample was compiled using Hipparcos parallaxes \citep{ESA1997} to identify all MS stars 
with masses $\geq1.4\,M_\odot$ (i.e. all early-F, A-, and B-type MS stars) located within the adopted 
distance limit of $100\,{\rm pc}$. We then cross-referenced this list with the Catalogue of Ap, HgMn 
and Am stars \citep{Renson1991,Renson2009} as well as the Spectral Classifications compiled by 
\citet{Skiff2014} in order to identify confirmed and candidate mCP stars. Ultimately, 52 confirmed 
mCP stars were identified based on published, archived, and newly obtained photometric, spectroscopic, 
and spectropolarimetric (i.e. Stokes $V$) measurements. We derived fundamental parameters (effective 
temperatures, luminosities, masses, ages, etc.) of all of the intermediate-mass MS stars in the sample. 
Average surface chemical abundances of the mCP stars were also derived. The analysis presented in 
Paper~I serves as a starting point for the magnetic analysis presented here. The results included in 
this second paper (i.e. Paper~II) are organized as follows.

In Sect. \ref{sect:obs}, we discuss the newly obtained or previously unpublished MuSiCoS, ESPaDOnS, and 
NARVAL Stokes $V$ observations. In Sect. \ref{sect:Bz_obs}, we present our analysis of these 
measurements and how they are used to derive longitudinal magnetic field measurements; the measurements 
are then used to help identify each star's rotational period, as discussed in Sect. \ref{sect:Prot}. In 
Sect. \ref{sect:mag_param}, we derive the magnetic field strengths and geometries and in Sect. 
\ref{sect:evo}, we search for evolutionary changes of the field strengths. Finally, in Sect. 
\ref{sect:discussion} we discuss the results while presenting our conclusions drawn from the survey.

\section{New Observations}\label{sect:obs}

\subsection{MuSiCoS spectropolarimetry}\label{sect:MUS_obs}

The MuSiCoS \'echelle spectropolarimeter was installed on the $2\,{\rm m}$ T\'elescope Bernard Lyot (TBL) 
at the Pic du Midi Observatory in 1996 where it was operational until its decomissioning in 2006. It had 
a resolving power $\sim35\,000$ and was capable of obtaining circularly polarized (Stokes $V$) spectra 
from $3\,900$ to $8\,700\,{\rm \AA}$ \citep{Donati1999a}. For this study, we used a total of 151 Stokes 
$V$ observations of 23 stars that were obtained from Feb. 12, 1998 to June 8, 2006. These observations 
were reduced using the {\sc ESpRIT} software package \citep{Donati1997}.

We note that the raw MuSiCoS spectra used in this study are unavailable and we have relied on 
normalized and reduced spectra from a private archive. All of the available spectra span a wavelength 
range of $4\,500$ to $6\,600\,{\rm \AA}$ rather than the full range presumably associated with the raw 
spectra. Furthermore, an automatic normalization routine built into the {\sc ESpRIT} reduction package 
had been applied to the spectra.

\subsection{ESPaDOnS \& NARVAL spectropolarimetry}\label{sect:ESP_obs}

The ESPaDOnS and NARVAL \'echelle spectropolarimeters are twin instruments installed at the 
Canada-France-Hawaii Telescope (CFHT), and TBL, respectively. They have a resolving power $\sim65\,000$ 
and are optimized for a wavelength range of approximately $3\,600\,{\rm \AA}$ to 
$10\,000\,{\rm \AA}$.

We obtained 95 Stokes $V$ observations of 37 stars from Aug. 2, 2015 to Aug. 10, 2016 using ESPaDOnS. 
Twenty-three Stokes $V$ observations of 3 stars were obtained using NARVAL from Aug. 20, 2016 to 
Feb. 20, 2017. All of the observations obtained using ESPaDOnS and NARVAL were reduced with the 
{\sc Libre-ESpRIT} software package, which is an updated version of the {\sc ESpRIT} reduction package 
that was applied to the MuSiCoS data \citep{Donati1997}.

\input{tbl_det_obs.tex}

\input{tbl_non_det_obs.tex}

\section{Magnetic Measurements}\label{sect:Bz_obs}

Organized magnetic fields that are present in the photospheres of mCP stars may be detected by 
identifying Zeeman signatures in Stokes $V$ spectropolarimetric observations. While these signatures 
are typically weak in individual spectral lines, the SNRs can be significantly increased by calculating 
Least-Squares Deconvolution (LSD) profiles \citep{Donati1997,Kochukhov2010}. This cross-correlation 
technique involves essentially averaging a large number of spectral lines (typically $\gtrsim100$) 
having similarly-shaped profiles. It has been widely used in the study of mCP star magnetism 
\citep[e.g.][]{Wade2000,Shorlin2002}.

\subsection{Confirmed mCP Stars}\label{sect:mCP_Bz}

\begin{figure}
	\centering
	\includegraphics[width=1.0\columnwidth]{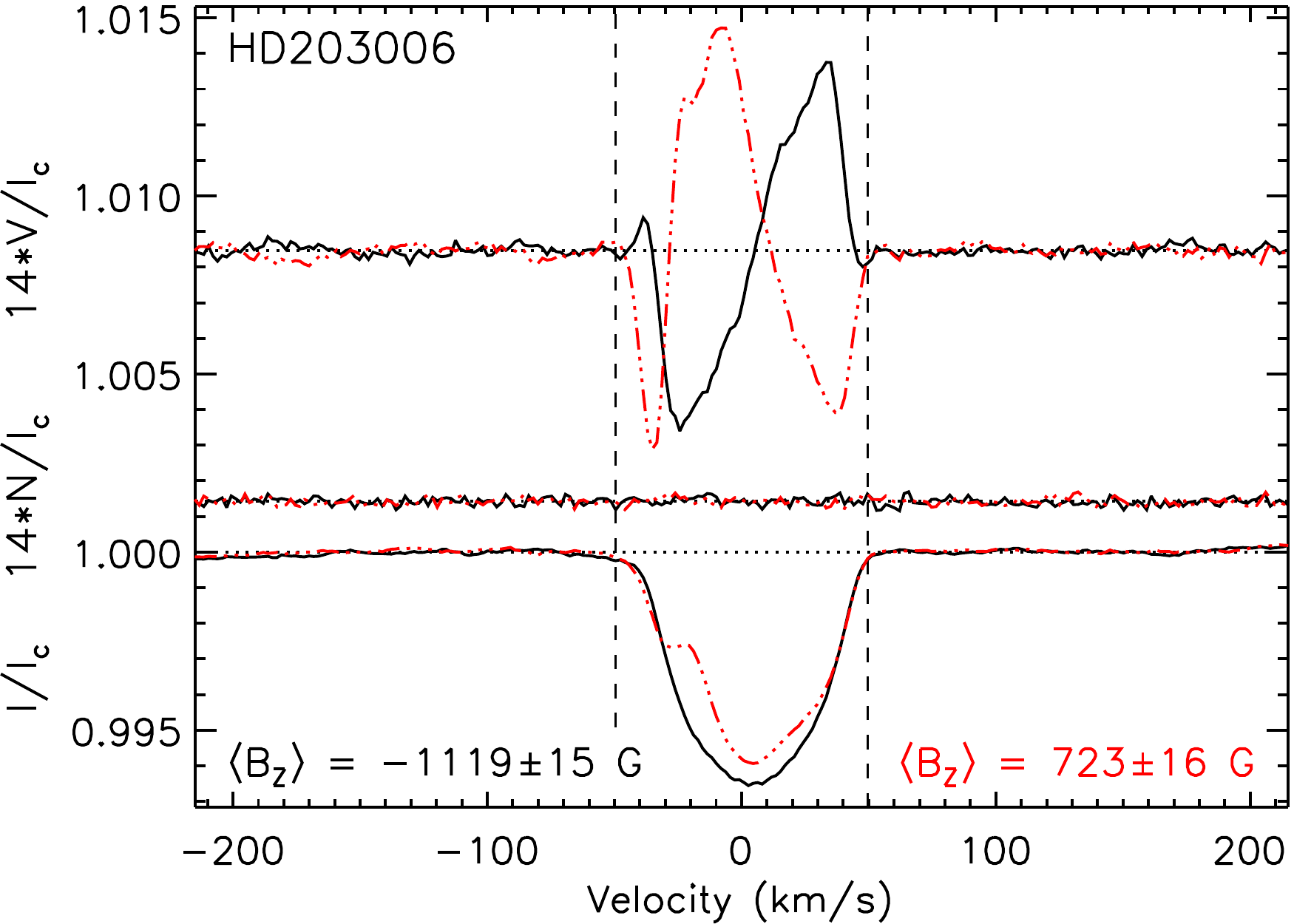}
	\caption{Two examples of the Stokes $V$ (top), diagnostic null (middle), and Stokes $I$ 
	(bottom) LSD profiles derived from the spectropolarimetric observations obtained using 
	ESPaDOnS. The vertical dashed lines indicate the adopted integration limits used to 
	derive the displayed $\langle B_{\rm z}\rangle$ values. Note that the Stokes 
	$V$ and diagnostic null profiles have been scaled by a factor of 14. Additional examples are 
	included in the electronic version of this paper.}
	\label{fig:lsd_ex}
\end{figure}

We generated LSD profiles for all of the available Stokes $V$ observations. This was carried out by 
first generating line lists containing wavelengths, depths, and Land\'e factors, from the Vienna 
Atomic Line Database (VALD) \citep{Ryabchikova2015a}. Custom lists specific to each star in the sample 
were obtained using Extract Stellar requests specifying the effective temperatures ($T_{\rm eff}$), 
surface gravities ($\log{g}$), and chemical abundances derived in Paper~I (solar abundances were 
adopted for those elements without estimated abundances); a microturbulence value ($v_{\rm mic}$) of 
$0\,{\rm km\,s^{-1}}$ was used along with a detection threshold of $0.05$ and a wavelength range of 
$4\,000$ to $7\,000\,{\rm \AA}$. Line masks were subsequently generated from each of the line lists and 
compared with the observed spectra: any lines in the line mask that were found to overlap with either 
telluric lines or broad Balmer lines were removed. The Stokes $V$, Stokes $I$, and diagnostic null 
\citep[i.e. the flux obtained by combining the subexposures such that the net polarization of the 
source is cancelled, Eqn. 3 of][]{Donati1997} measurements associated with each spectropolarimetric 
observation were normalized by fitting a multi-order polynomial to the continuum flux of each spectral 
order. An example of an LSD profile calculated using one of the observed spectra and its associated 
line mask is shown in Fig.~\ref{fig:lsd_ex}. Additional examples are shown in the electronic version of 
this paper.

The Stokes $I/I_{\rm c}$ and $V/I_{\rm c}$ LSD profiles were used to measure the disk-averaged 
longitudinal magnetic field ($\langle B_{\rm z}\rangle$) as given by equation 1 of \citet{Wade2000}. We 
used mean wavelengths ($\lambda_{\rm avg}$) and mean Land\'e factors ($z_{\rm avg}$) calculated from 
the customized line masks associated with each star. Prior to each $\langle B_{\rm z}\rangle$ 
measurement, the Stokes $I/I_{\rm c}$ and $V/I_{\rm c}$ LSD profiles were renormalized by fitting a 
$1^{\rm st}$ order polynomial (i.e. a linear function) to the regions where $I/I_{\rm c}\sim1$ and 
$V/I_{\rm c}\sim0$ (typically at $v\approx\pm100\,{\rm km\,s^{-1}}$). Any radial velocity shift that 
was apparent in the Stokes $I/I_{\rm c}$ LSD profile, as inferred from the calculation of the 
profile's ``center of gravity" (i.e. the integral of $vI/I_{\rm c}$ over that of $I/I_{\rm c}$), was 
removed. The $v$ integration limits were chosen to encompass the absorption profile as determined by 
eye. The derived values of $\langle B_{\rm z}\rangle$ associated with the confirmed mCP stars are 
listed in Table~\ref{tbl:det_obs_tbl}.

In addition to the previously unpublished $\langle B_{\rm z}\rangle$ measurements listed in 
Table~\ref{tbl:det_obs_tbl}, we also derived $\langle B_{\rm z}\rangle$ from archived ESPaDOnS and 
NARVAL Stokes $V$ observations. In these cases, we applied the same analysis that was used with the new 
observations reported in this study. This ensured that both the new and archived observations yielded 
consistent $\langle B_{\rm z}\rangle$ measurements such that any apparent variability cannot be 
attributed to the use of different line masks (i.e. all $\langle B_{\rm z}\rangle$ values are obtained 
using the same measurement system). In total, we used 400 measurements of 42 confirmed mCP stars 
derived using the line masks generated in this study -- corresponding to a median value of six 
observations per star. These $\langle B_{\rm z}\rangle$ measurements exhibit a median uncertainty of 
$\sigma_{\langle B_{\rm z}\rangle}=18\,{\rm G}$. Published $\langle B_{\rm z}\rangle$ measurements exist 
for the majority of the confirmed mCP stars. We compiled and included many of these measurements in our 
analysis when no corresponding archival Stokes $V$ observations were found. For ten out of the 
fifty-two confirmed mCP stars, only previously published measurements were available (i.e. no new or 
archived Stokes $V$ observations were available). Note that these published data are not derived using 
the same measurement system as used for the $\langle B_{\rm z}\rangle$ measurements that we derived 
from the Stokes $V$ observations and analyzed herein.

In summary, a total of 947 new, archived, and published $\langle B_{\rm z}\rangle$ measurements of the 
confirmed mCP stars were used in this study, corresponding to a median number of observations per 
star of 17. The measurements exhibit a median $\sigma_{\langle B_{\rm z}\rangle}$ of $49\,{\rm G}$ and a 
median minimum $\sigma_{\langle B_{\rm z}\rangle}$ per star of $15\,{\rm G}$. For four of the fifty-two 
stars, fewer than five observations are available. Two detections of HD~117025 are reported by 
\citet{Kochukhov2006} while, due to its relatively low declination of $-45\degree$, we were only able to 
obtain a single observation of HD~217522. For HD~29305, only four archived HARPSpol Stokes $V$ 
observations are available while for HD~56022, we obtained four new Stokes $V$ observations using 
ESPaDOnS.

\subsection{Null Results}\label{sect:null_Bz}

As discussed in Paper I, during the initial phase of this study, we identified a number of stars within 
the Catalogue of Ap, HgMn and Am stars \citep{Renson2009} reported as being potential mCP members. 
Additionally, several Am and HgMn stars were found to exhibit $\Delta a$, $\Delta(V_1-G)$, or 
$\Delta Z$ photometric indices consistent with those exhibited by mCP stars 
\citep[e.g.][]{Maitzen1998,Bayer2000,Paunzen2005}. We obtained Stokes $V$ observations for 19 of these 
stars using MuSiCoS, ESPaDOnS, and NARVAL in order to search for Zeeman signatures. The observations 
were analyzed using the same LSD technique that was applied to the confirmed mCP stars; however, the 
line masks generated from the VALD line lists used a surface gravity of $\log{g}=4.0\,{\rm (cgs)}$ and 
a solar metallicity (individual chemical abundances were not specified).

No Zeeman signatures were detected from the observations of the 19 stars. The minimum 
$\langle B_{\rm z}\rangle$ uncertainties obtained for each star ranged from $1.9\,{\rm G}$ to 
$69\,{\rm G}$ with a median value of $11.4\,{\rm G}$. \citet{Kochukhov2006} report a measured 
$\langle B_{\rm z}\rangle=-56\pm68\,{\rm G}$ for one of the 19 stars, HD~202627; we obtained a single 
observation of this star, which yielded a lower uncertainty and no detection 
($\langle B_{\rm z}\rangle=14\pm17\,{\rm G}$). The observations are summarized in 
Table~\ref{tbl:non_det_obs_tbl} where we list the measured longitudinal field values.

\section{Rotational Periods and Inclination Angles}\label{sect:Prot}

Magnetic CP stars are well known to be associated with the periodic variability of surface-averaged 
longitudinal magnetic field measurements \cite[e.g.][]{Pyper1969,Borra1982,Bohlender1993}. The Oblique 
Rotator Model (ORM) attributes these variations to a product of (1) the star's rotation and (2) the 
presence of a stable surface magnetic field that is non-axisymmetric with respect to the star's 
rotational axis \citep{Stibbs1950,Preston1967}. A similar explanation for the long-period 
($\gtrsim1\,{\rm d}$) photometric variability that is commonly associated with these stars is also 
widely accepted: the variations are understood to be produced by the presence of inhomogeneous 
structures (i.e. chemical abundance spots) located within the rotating star's atmosphere 
\citep[e.g.][]{Wolff1969,Adelman1992,Krticka2015}. Therefore, the characterization of both the 
rotationally modulated $\langle B_{\rm z}\rangle$ and photometric measurements may allow for an mCP 
star's rotational period ($P_{\rm rot}$) to be constrained.

Rotational periods of the majority of the confirmed mCP stars in our sample have been previously 
published \citep[e.g.][]{Catalano1998,Renson2001}. We performed a period search analysis (described 
below) on all of the $\langle B_{\rm z}\rangle$ data sets, which consist of published 
$\langle B_{\rm z}\rangle$ measurements along with those measurements derived from either new or 
re-analyzed archival Stokes $V$ spectra, as discussed in Sect. \ref{sect:mCP_Bz}. The analysis 
typically yielded a number of plausible rotational periods, which were then compared with those that 
have been previously reported in the literature. The same period search analysis was also carried out 
on Hipparcos Epoch Photometry ($H_p$) \citep{ESA1997}, which aided in the correct identification of 
$P_{\rm rot}$. For each of the 52 mCP stars, between 42 and 260 $H_p$ measurements are available 
spanning $3\,{\rm yrs}$. The minimum and average time intervals between each measurement are 
approximately $21\,{\rm min}$ and $11\,{\rm d}$, respectively. Each measurement has been assigned a 
quality flag, which indicates potential problems (e.g. high background flux or  inconsistent values 
obtained by the NDAC and FAST data reductions). Any measurements exhibiting quality flag numbers 
(referred to as `transit flags' in the Hipparcos catalogue) $>20$ were identified but not removed from 
the analysis. This decision to retain flagged measurements was based on the fact that, in certain 
cases, all of the star's measurements exhibited transit flags $>20$ despite the detection of 
variability that was consistent with that of the $\langle B_{\rm z}\rangle$ measurements. For most of 
the stars having flagged measurements, the number of flagged measurements was relatively insignificant 
and did not strongly influence the period search analysis.

Both the $\langle B_{\rm z}\rangle$ measurements and the Hipparcos Epoch Photometry were analyzed using 
two methods to identify the most probable rotational periods. First, normalized Lomb-Scargle periodograms 
were generated using an {\sc idl} routine based on the algorithm presented by \citet{Press2007}. This 
method yields the spectral power distribution, which is used to identify statistically significant 
frequencies (i.e. those having false alarm probabilities $<3$~per~cent) inherent to an unevenly sampled 
time series data set. A substantial benefit of this method is that it can be performed relatively 
quickly compared to the second period search analysis described below thereby allowing potentially 
relevant periods to be recognized efficiently. However, for the majority of the mCP stars, an 
insufficient number of $\langle B_{\rm z}\rangle$ measurements were available to yield statistically 
significant frequencies. This technique was found to be more useful when applied to the Hipparcos Epoch 
Photometry because of the larger number of data points available for each star. The 
$\langle B_{\rm z}\rangle$ measurements were then used to verify that the derived Hipparcos period 
provided an acceptable phasing of the magnetic data.

The periodogram calculation was followed by the application of a commonly used period search analysis 
described, for example, by \citet{Alecian2014}. The method involves fitting the time series data to 
a function consisting of the first two or three terms in a Fourier series using a range of fixed 
periods ($P$); plausible rotational periods are identified as those which yield the lowest $\chi^2$ 
values. We adopted a $2^{\rm nd}$ order sinusoidal fitting function given by 
\begin{multline}\label{eqn:sin_fcn}
	f(t)=C_0+C_1\sin{(2\pi[t-t_0]/P+\phi_1)}+\\
			C_2\sin{(4\pi[t-t_0]/P+\phi_2)}
\end{multline}
where $t_0$ is the epoch (set to zero during the period search analysis) and $C_0$, $C_1$, $C_2$, 
$\phi_1$, and $\phi_2$ are free parameters. We defined an initial grid of period values having a step 
size of $\Delta P=10^{-4}\,{\rm d}$ and spanning $0.1\leq P\leq25\,{\rm d}$. For each $P$ value in the 
grid, the best fit was derived and the associated $\chi^2$ values were recorded. This analysis was 
repeated with $C_2\equiv0$ (reducing Eqn. \ref{eqn:sin_fcn} to a $1^{\rm st}$ order sinusoidal fitting 
function), which was frequently found to decrease the number of statistically significant periods 
derived from the $\langle B_{\rm z}\rangle$ data sets. This is related to the fact that longitudinal 
field measurements of mCP stars are most sensitive to the dipole component 
\citep[e.g. Eqn. 68 of][]{Bagnulo1996}. Nevertheless, significant higher-degree contributions to 
$\langle B_{\rm z}\rangle$ curves are often detected in high-precision data 
\citep[e.g.][]{Kochukhov2004,Kochukhov2010,Silvester2015}.

Uncertainties in the adopted rotational periods ($\sigma_{P_{\rm rot}}$) were estimated by calculating 
the 3$\sigma$ confidence limits associated with the width of the $\chi^2$ trough; if 
$\sigma_{P_{\rm rot}}\leq\Delta P$, the grid's range ($P_{\rm max}-P_{\rm min}$) and $\Delta P$ was 
reduced, the grid was re-centered on the relevant period, and the grid of $\chi^2$ values was 
re-calculated. If the final $\sigma_{P_{\rm rot}}$ was found to be appreciably less than the published 
$\sigma_{P_{\rm rot}}$ -- or if no $\sigma_{P_{\rm rot}}$ was reported with the published 
$P_{\rm rot}$ -- the new $P_{\rm rot}$ and $\sigma_{P_{\rm rot}}$ was adopted.

After identifying $P_{\rm rot}$ and obtaining $\sigma_{P_{\rm rot}}$, either through a period search 
analysis or from the literature, final $1^{\rm st}$ and $2^{\rm nd}$ order sinusoidal fits to each 
star's $\langle B_{\rm z}\rangle(t)$ and $H_p(t)$ measurements were derived ($2^{\rm nd}$ order fits 
were only derived for those data sets consisting of more than 5 data points). The epoch of each star was 
defined such that $\langle B_{\rm z}\rangle(t_0)=|C_0+C_1|$ (i.e. the maximum, unsigned longitudinal 
field strength) while $\phi_1$ and $\phi_2$ were constrained such that $C_1>0$ and $C_2>0$. Note that 
the way in which the epoch is defined and the way in which $C_1$ and $C_2$ are constrained implies 
that, for the fits to $\langle B_{\rm z}\rangle$, $\phi_1=\pm\pi/2$ while $\phi_2$ is a free parameter; 
for the fits to $H_p$, both $\phi_1$ and $\phi_2$ are unrestricted free parameters.

Published periods for 18/52 of the mCP stars were found to be in agreement with those associated with 
the minimal $\chi^2$ value and/or maximal Lomb-Scargle spectral power yielded by our 
$\langle B_{\rm z}\rangle$ and Hipparcos Epoch Photometry period search analyses. In these cases, the 
stars' rotational period could be unambiguously identified. For 21/52 of the stars, the most probable 
periods inferred from the period search analysis were not consistent with the published periods. The 
rotational periods of these stars were determined by identifying those published periods, which are 
primarily inferred from photometric variability, that are consistent with local $\chi^2$ minima having 
values within 3$\sigma$ confidence limits of the global $\chi^2$ minima. We encountered complications 
regarding the identification of $P_{\rm rot}$ for the remaining 12/52 stars (discussed below in Sections 
\ref{sect:Prot_details01} to \ref{sect:Prot_details02}); however, we note that in most of these cases, 
final rotational periods were adopted.

In total, we adopted rotational periods for 48/52 of the mCP stars in the sample. The phased 
$\langle B_{\rm z}\rangle$ measurements and the associated best fitting sinusoidal functions are shown 
in Figures \ref{fig:phased_Bz01}, \ref{fig:phased_Bz02}, and \ref{fig:phased_Bz03}. The corresponding 
phased $H_p$ measurements are only included in the electronic version of this paper. The 
$\langle B_{\rm z}\rangle$ measurements as a function of HJD of the 4 stars with $>1$ measurement and 
for which we were unable to establish $P_{\rm rot}$ values are shown in Fig. \ref{fig:HJD_Bz01}.

In the following eight subsections (Sections \ref{sect:Prot_details01} to \ref{sect:Prot_details02}), 
we discuss those stars for which $P_{\rm rot}$ could not be unambiguously determined due to (1) an 
insufficient number of measurements, (2) no detection of photometric or $\langle B_{\rm z}\rangle$ 
variability, or (3) disagreement with published rotational periods.

\subsection{HD~27309 and HD~72968}\label{sect:Prot_details01}

The most precise published $P_{\rm rot}=1.5688840(47)\,{\rm d}$ \citep{North1995} for HD~27309 was found 
to be consistent with the most probable period inferred from the Hipparcos photometry; however, both 
this period and its second harmonic exhibit poor agreement with the variability of the 
$\langle B_{\rm z}\rangle$ measurements when fit to a $1^{\rm st}$ order sinusoidal function. A high 
quality $2^{\rm nd}$ order sinusoidal fit (Eqn. \ref{eqn:sin_fcn}) is obtained using the published 
$P_{\rm rot}$, which exhibits $C_1\sim C_2$ (i.e. comparable amplitudes of the $1^{\rm st}$ and 
$2^{\rm nd}$ order terms). HD~72968 is similar in that \citet{Maitzen1978} report a 
period of $11.305(2)\,{\rm d}$, however, this period is inconsistent with both the 
$\langle B_{\rm z}\rangle$ and Hipparcos measurements. Furthermore, the $v\sin{i}$ value and stellar 
radius derived in Paper I imply a maximum $P_{\rm rot}$ of approximately $8.2\,{\rm d}$. We find that 
halving the $11.305\,{\rm d}$ period ($P_{\rm rot}=5.6525\,{\rm d}$) yields acceptable $1^{\rm st}$ and 
$2^{\rm nd}$ order fits to the Hipparcos photometry and an acceptable $2^{\rm nd}$ order fit to 
$\langle B_{\rm z}\rangle$. We note that \citet{Auriere2007} adopt the same $5.6525\,{\rm d}$ period. 
Both the adopted magnetically-inferred rotational periods for HD~27309 and HD~72968 should be verified 
using additional measurements.

\subsection{HD~74067}

No published rotational period could be found for HD~74067. We were able to derive $P_{\rm rot}$ for 
HD~74067 based on the identification of a single statistically significant period in the 
$\langle B_{\rm z}\rangle$ $\chi^2$ distribution, which was found to be consistent with a local 
$\chi^2$ minima derived from the Hipparcos photometry.

\subsection{HD~128898}

As noted by \citet{Mathys1997}, the $\langle B_{\rm z}\rangle$ measurements of HD~128898 obtained by 
\citet{Mathys1991,Mathys1994} and \citet{Mathys1997} do not exhibit a trend that is consistent with the 
star's known rotational period \citep[$4.4790\,{\rm d}$,][]{Kurtz1994}. The authors attribute this to the 
low amplitude of $\langle B_{\rm z}\rangle$ variability. We did not obtain nor find any new 
$\langle B_{\rm z}\rangle$ measurements that could potentially better constrain the star's magnetic 
properties.

\subsection{HD~130559}

\begin{figure}
	\centering
	\includegraphics[width=1.01\columnwidth]{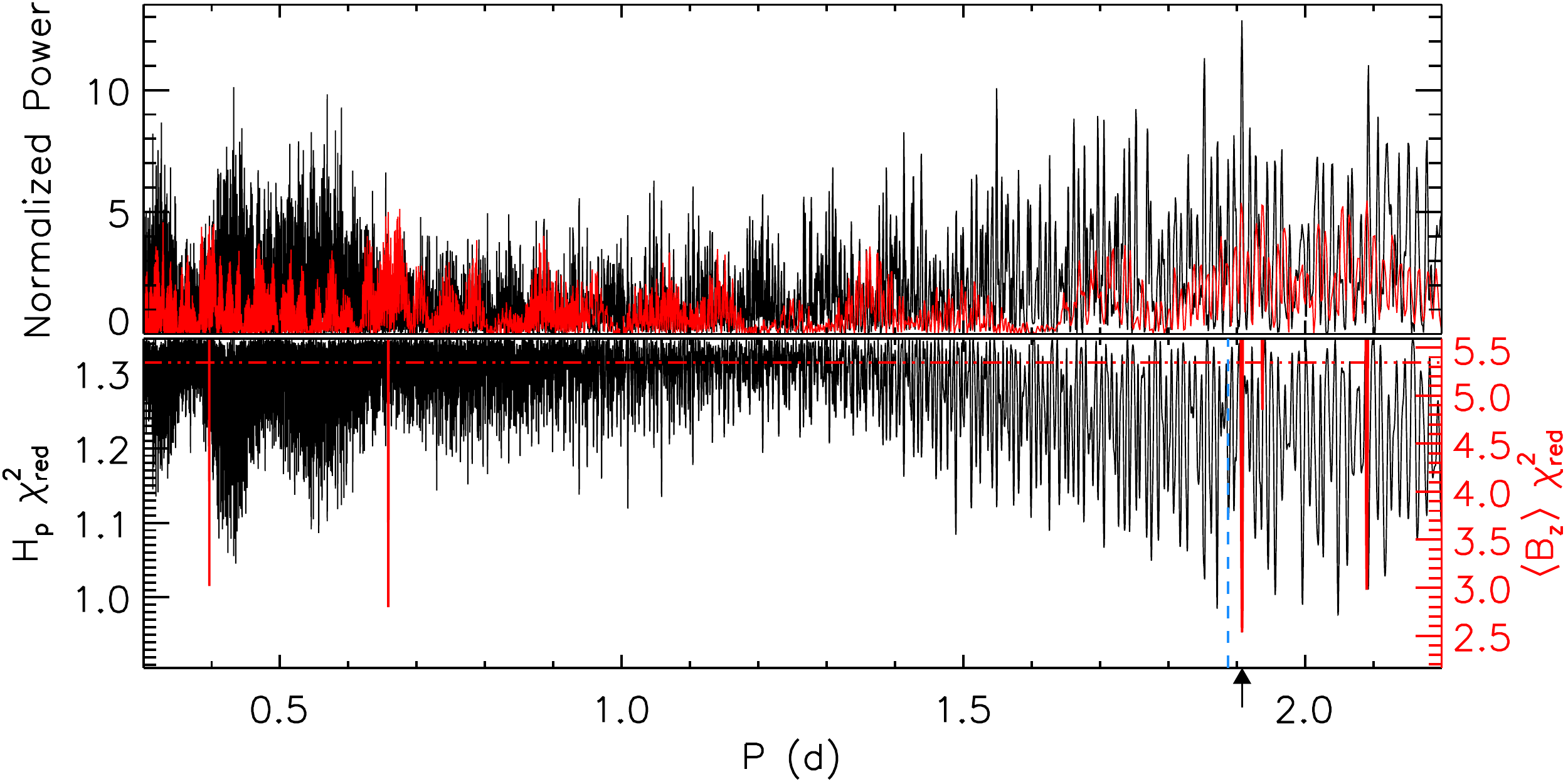}
	\caption{Normalized Lomb-Scargle periodograms (top) and $\chi^2$ distributions derived using the 
	$1^{\rm st}$ order sinusoidal function (bottom) associated with the Hipparcos (black) and 
	$\langle B_{\rm z}\rangle$ (red) measurements of HD~130559. The horizontal dot-dashed red line 
	corresponds to the $\langle B_{\rm z}\rangle$ 3$\sigma$ confidence limit calculated with respect to 
	$\chi^2_{\rm min}$; the $H_p$ periods shown in the $\chi^2$ distribution exhibit confidence 
	limits $<0.1\sigma$. The black arrow indicates the adopted $P_{\rm rot}=1.90798(71)\,{\rm d}$. The 
	vertical dashed blue line appearing in the $\chi^2$ plot corresponds to the $1.8871(8)\,{\rm d}$ 
	period identified by \citet{Wraight2012} based on STEREO photometry, which is not consistent 
	with the $\langle B_{\rm z}\rangle$ measurements.}
	\label{fig:HD130559_P}
\end{figure}

\begin{figure*}
	\centering
	\includegraphics[width=2.1\columnwidth]{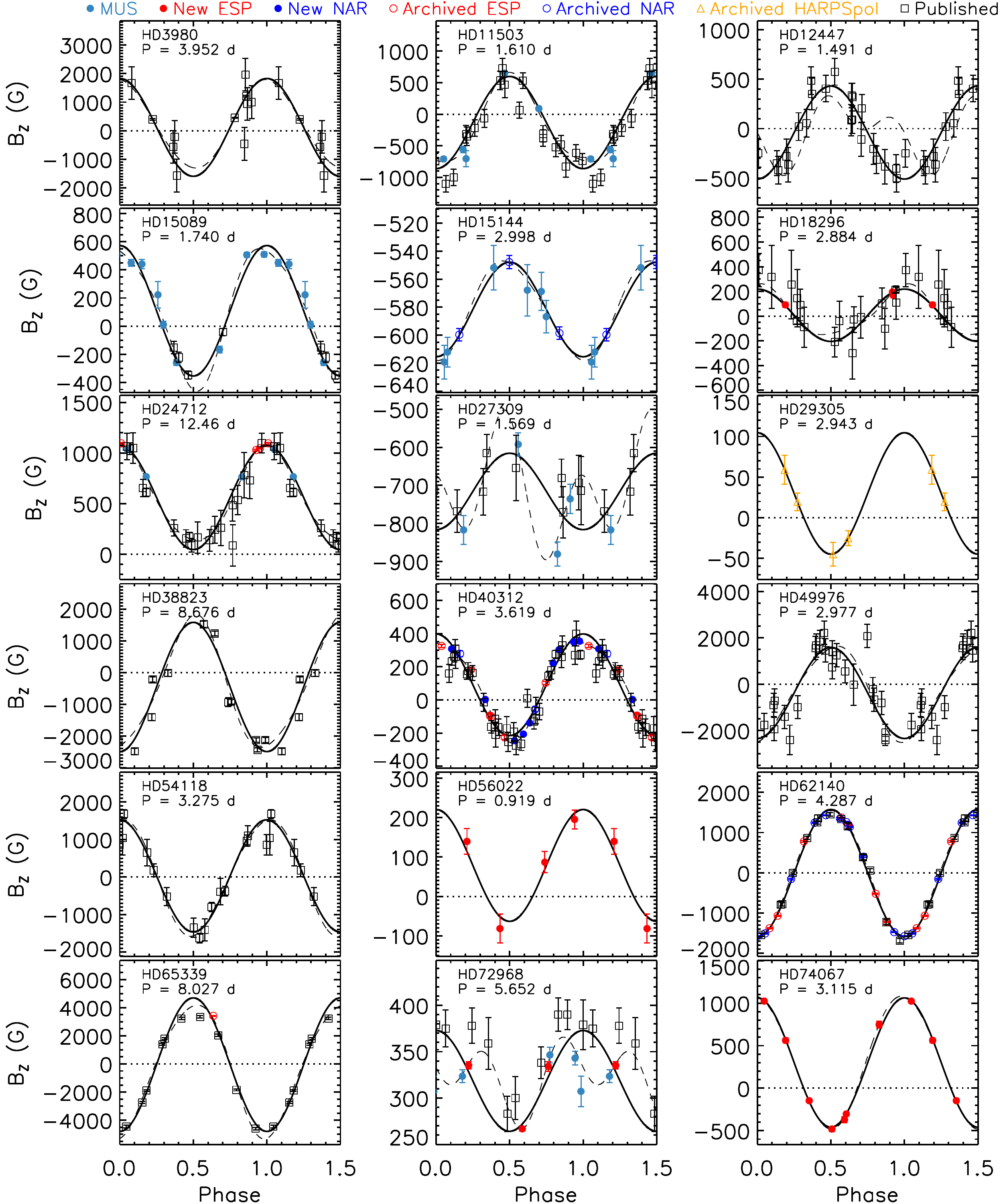}
	\caption{The $\langle B_{\rm z}\rangle$ measurements used in this analysis phased according to 
	each star's rotational period -- only those mCP stars with known $P_{\rm rot}$ values are shown. 
	The solid black curves and dashed black curves correspond to the best $1^{\rm st}$ and $2^{\rm nd}$ order 
	sinusoidal fits (defined by Eqn. \ref{eqn:sin_fcn}). Note that the periods listed in each figure 
	are rounded and do not correspond to the actual $P_{\rm rot}$ precision.}
	\label{fig:phased_Bz01}
\end{figure*}

\begin{figure*}
	\centering
	\includegraphics[width=2.1\columnwidth]{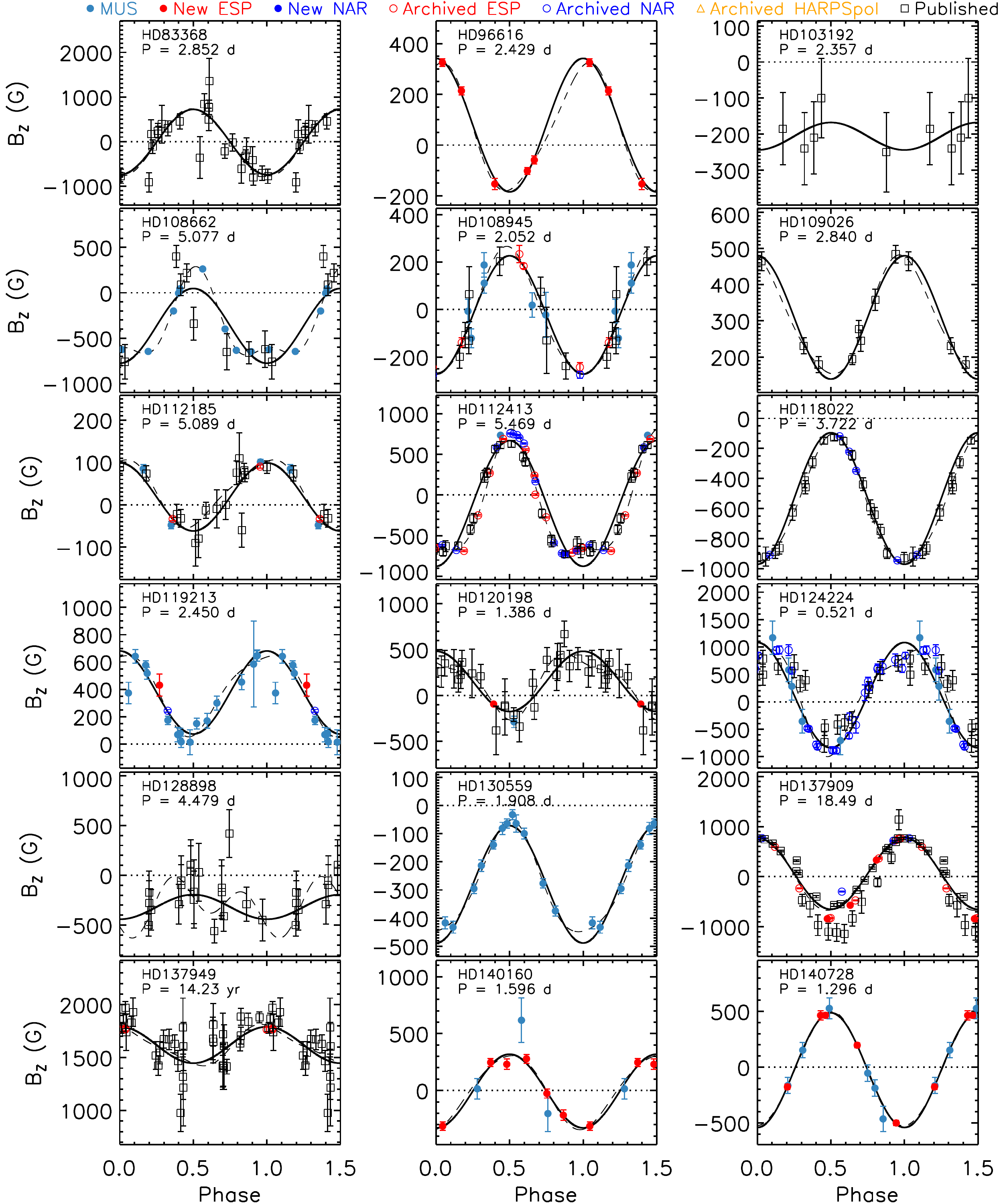}
	\caption{Continued from Fig. \ref{fig:phased_Bz01}.}
	\label{fig:phased_Bz02}
\end{figure*}

\begin{figure*}
	\centering
	\includegraphics[width=2.1\columnwidth]{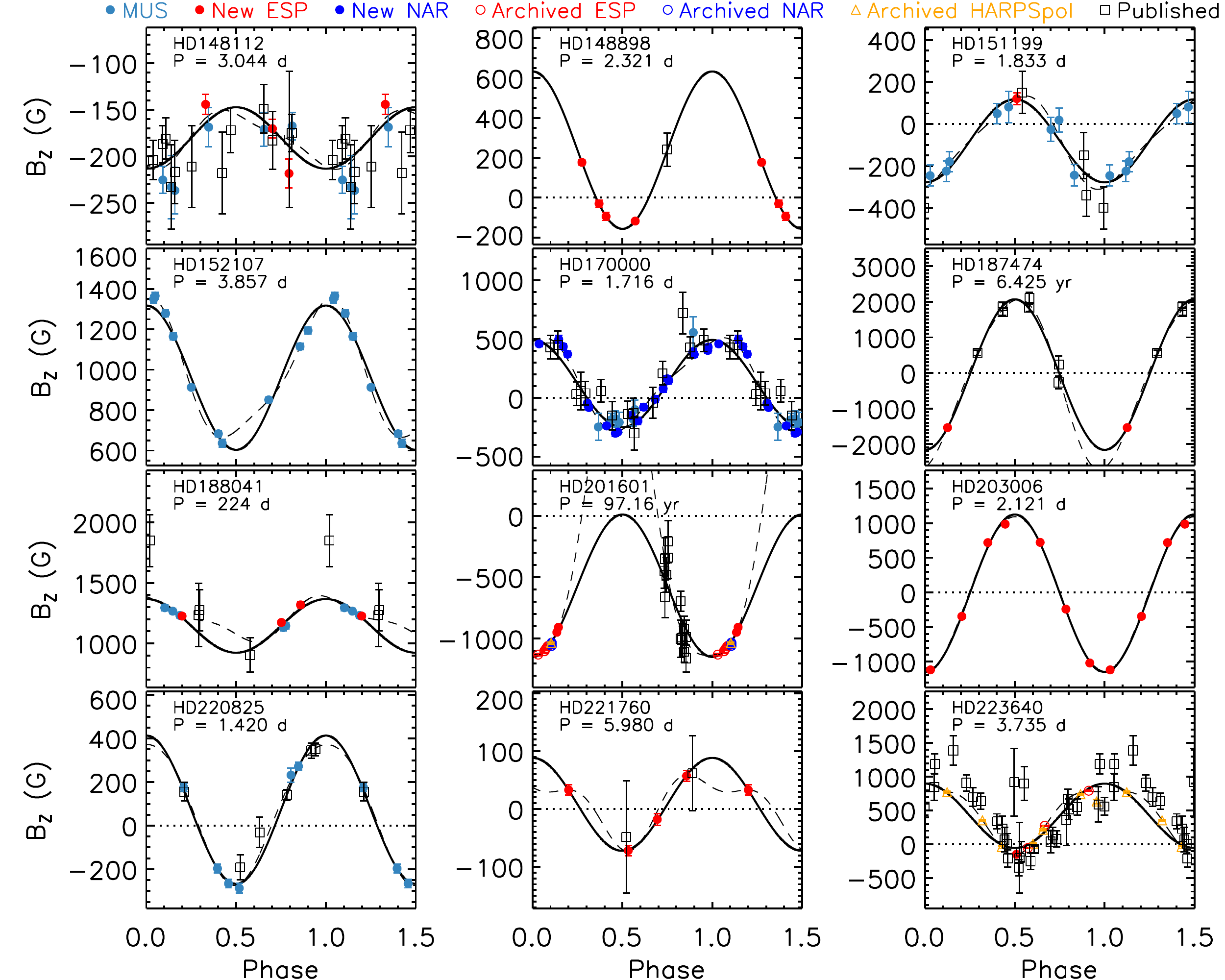}
	\caption{Continued from Fig. \ref{fig:phased_Bz02}.}
	\label{fig:phased_Bz03}
\end{figure*}

\begin{figure*}
	\centering
	\includegraphics[width=2.1\columnwidth]{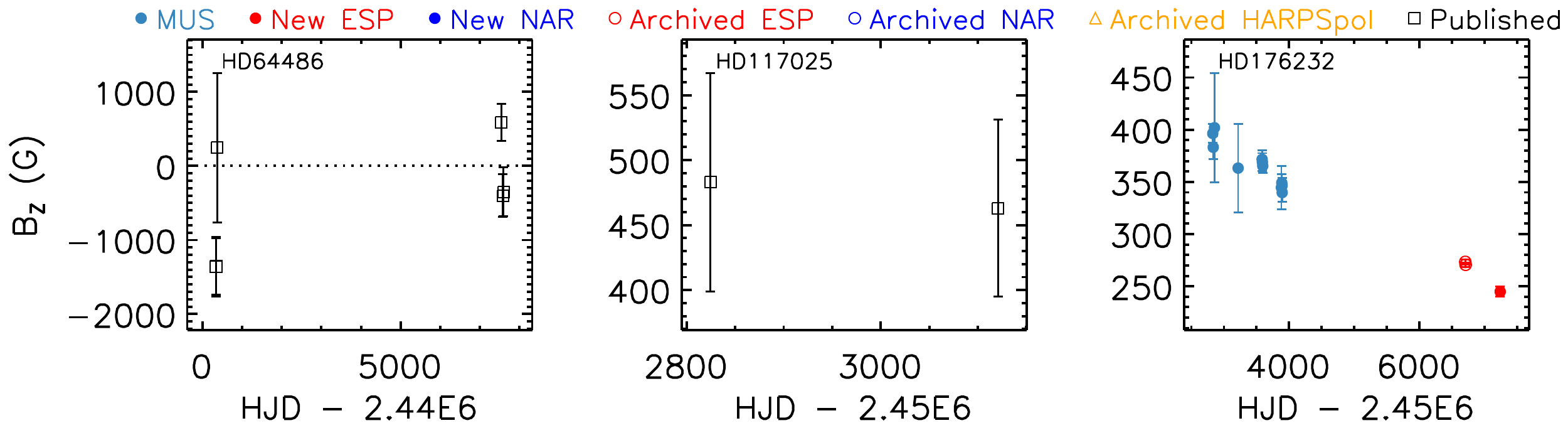}
	\caption{The $\langle B_{\rm z}\rangle$ measurements used in this analysis associated with those 
	stars with $>1$ measurement and without known $P_{\rm rot}$ values.}
	\label{fig:HJD_Bz01}
\end{figure*}

Two possible rotational periods ($1.8871(8)\,{\rm d}$ and $25.4(2)\,{\rm d}$) of HD~130559 are reported 
by \citet{Wraight2012} based on the detection of strong photometric variability using the STEREO 
spacecraft. No statistically significant variability was detected from the Hipparcos photometry. Our 
analysis includes 12 $\langle B_{\rm z}\rangle$ measurements obtained with MuSiCoS; the period search 
analysis of this data set yielded five plausible periods within $0.1<P<30\,{\rm d}$: 
$0.39661(5)\,{\rm d}$ ($\chi^2_{\rm red}=2.8$), $0.6585(2)\,{\rm d}$ ($\chi^2_{\rm red}=2.5$), 
$1.90798(71)\,{\rm d}$ ($\chi^2_{\rm red}=2.2$), $1.9377(13)\,{\rm d}$ ($\chi^2_{\rm red}=4.6$), and 
$2.0905(12)\,{\rm d}$ ($\chi^2_{\rm red}=2.7$). In Fig. \ref{fig:HD130559_P}, we show the Lomb-Scargle 
periodograms and $\chi^2$ distributions associated with the $\langle B_{\rm z}\rangle$ and Hipparcos 
measurements. It is evident that, although similar, the best-fit $1.90798(71)\,{\rm d}$ period, which 
yields a clear sine variation of $\langle B_{\rm z}\rangle$ versus phase, is not in agreement with the 
shorter rotational period identified by \citet{Wraight2012}: phasing the $\langle B_{\rm z}\rangle$ 
measurements with the $1.8871(8)\,{\rm d}$ period yields significant dispersion between points that are 
approximately coincident in phase (e.g. $\langle B_{\rm z}\rangle$ values of $-375\pm18\,{\rm G}$ and 
$-64\pm30\,{\rm G}$ appear separated in phase by $<0.03$). The authors note the possible influence of 
systematic effects on their inferred $P_{\rm rot}$ values, which could potentially explain the 
discrepancy; however, they suggest that the systematics are unlikely to strongly influence the reported 
periods.

We adopt $P_{\rm rot}=1.90798(71)\,{\rm d}$ as it (1) exhibits the closest agreement with one of the two 
reported photometric periods and (2) corresponds to the minimal $\chi^2$ sinusoidal fit to the 
$\langle B_{\rm z}\rangle$ measurements. Further observations are required to eliminate the alternative 
rotational periods identified here and to verify the adopted value.

\subsection{HD~148898}

\citet{Manfroid1985} report three plausible rotational periods for HD~148898: $1.79\pm0.02\,{\rm d}$, 
$2.33\pm0.02\,{\rm d}$, and $4.67\pm0.08\,{\rm d}$. Based on near infrared variability, 
\citet{Catalano1998} adopted the value of $P_{\rm rot}=0.7462(2)\,{\rm d}$ reported by 
\citet{Renson1978}. We obtained four new ESPaDOnS Stokes $V$ observations for this star, which we 
combined with the single measurement published by \citet{Kochukhov2006}. The five high-precision 
$\langle B_{\rm z}\rangle$ measurements could not be adequately phased using 
$P_{\rm rot}=1.79\pm0.02\,{\rm d}$ ($\chi^2_{\rm red}=38$); the $0.7462\,{\rm d}$, $2.33\,{\rm d}$, and 
$4.67\,{\rm d}$ periods yield high quality $1^{\rm st}$ order sinusoidal fits ($\chi^2_{\rm red}<0.01$) 
and are consistent with the derived radii and $v\sin{i}$ (i.e. 
$v_{\rm eq}>v\sin{i}$ for both periods). Here we adopt $P_{\rm rot}=2.3205(2)\,{\rm d}$ based on the 
marginally lower $\chi^2$ value associated with both the $\langle B_{\rm z}\rangle$ and Hipparcos 
measurements compared to the longer $4.682(1)\,{\rm d}$ period; however, we emphasize that additional 
observations are required to more confidently identify the correct $P_{\rm rot}$.

\subsection{HD~151199}

We only found one $P_{\rm rot}$ value of HD~151199 reported in the literature: \citet{Gokkaya1970} 
find that the star exhibits Ca~{\sc ii} K line variations having a period of $6.143\,{\rm d}$. The 
$\langle B_{\rm z}\rangle$ measurements exhibit a number of statistically significant periods with 
none appearing within $0.3\,{\rm d}$ of $6.143\,{\rm d}$. The $v\sin{i}$ value and stellar radius 
derived in Paper I imply a maximum $P_{\rm rot}$ of approximately $2.4\,{\rm d}$. We adopt the minimal 
$\chi^2$ period within $0.4-2.5\,{\rm d}$ ($P_{\rm rot}=1.83317(22)\,{\rm d}$ and $\chi^2_{\rm red}=1.0$); 
however additional magnetic, spectroscopic, or photometric measurements are required to verify this value 
due to the number of periods which yield reasonably high quality sinusoidal fits.

\subsection{HD~221760}

Four high-precision Stokes $V$ observations of HD~221760 were obtained with ESPaDOnS; the associated 
$\langle B_{\rm z}\rangle$ measurements were found to vary from $-72\pm9\,{\rm G}$ to $57\pm9\,{\rm G}$. 
The period search analysis performed using the $1^{\rm st}$ order sinusoidal function yielded a large 
number of statistically significant periods. None of the possible periods were found to be consistent with 
the $12-13\,{\rm d}$ rotational periods suggested by \citet{vanGenderen1971} and \citet{Catalano1998} 
based on their detections of photometric variability. Furthermore, the $v\sin{i}$ value of 
$22.4\pm0.7\,{\rm km\,s^{-1}}$ and stellar radius of $3.6\pm0.3\,R_\odot$ derived in Paper I imply a 
maximum rotational period of $\approx9.1\,{\rm d}$. We find that the $\langle B_{\rm z}\rangle$ 
measurements are coherently phased by a period that is one half that of one of the possible periods 
reported by \citet{Catalano1998} ($P=12.665\,{\rm d}/2=6.3325\,{\rm d}$).

While no additional archived Stokes $V$ observations or published $\langle B_{\rm z}\rangle$ 
measurements with sufficiently high precision could be found, 21 archived Stokes $I$ HARPS observations 
are available. We attempted to obtain additional constraints on HD~221760's $P_{\rm rot}$ by searching 
for spectral line variability using the combined HARPS and ESPaDOnS Stokes $I$ observations. No 
significant line profile variability could be detected (either visually or from equivalent width 
calculations) from various lines including those associated with Ti, Cr, and Fe.

We adopt $P_{\rm rot}=5.98\pm0.06\,{\rm d}$ based on the preceding discussion, however, we emphasize 
that further confirmation of this value is required.

\subsection{HD~64486, HD~117025, HD~176232, and HD~217522}\label{sect:Prot_details02}

No published rotational periods were found for these five mCP stars. We were unable to infer 
the $P_{\rm rot}$ values of HD~64486, HD~117025, or HD~217522 on account of (1) an insufficient number 
of available $\langle B_{\rm z}\rangle$ measurements and (2) the absence of any statistically 
significant variability in the associated Hipparcos photometric measurements. $P_{\rm rot}$ of 
HD~176232 could not be derived on account of insufficient phase coverage of its very long rotational 
period: the available $\langle B_{\rm z}\rangle$ measurements exhibit an approximately linear decrease 
from $400\,{\rm G}$ to $240\,{\rm G}$ over a $12\,{\rm yr}$ period. The $\langle B_{\rm z}\rangle$ 
measurements as a function of HJD are shown in Fig. \ref{fig:HJD_Bz01} (aside from HD~217522, 
for which only a single measurement was obtained).

\afterpage{
\onecolumn
\input{tbl_rot_param.tex}
\twocolumn
}

\subsection{Distribution of Rotational Periods}

\begin{figure}
	\centering
	\includegraphics[width=1.0\columnwidth]{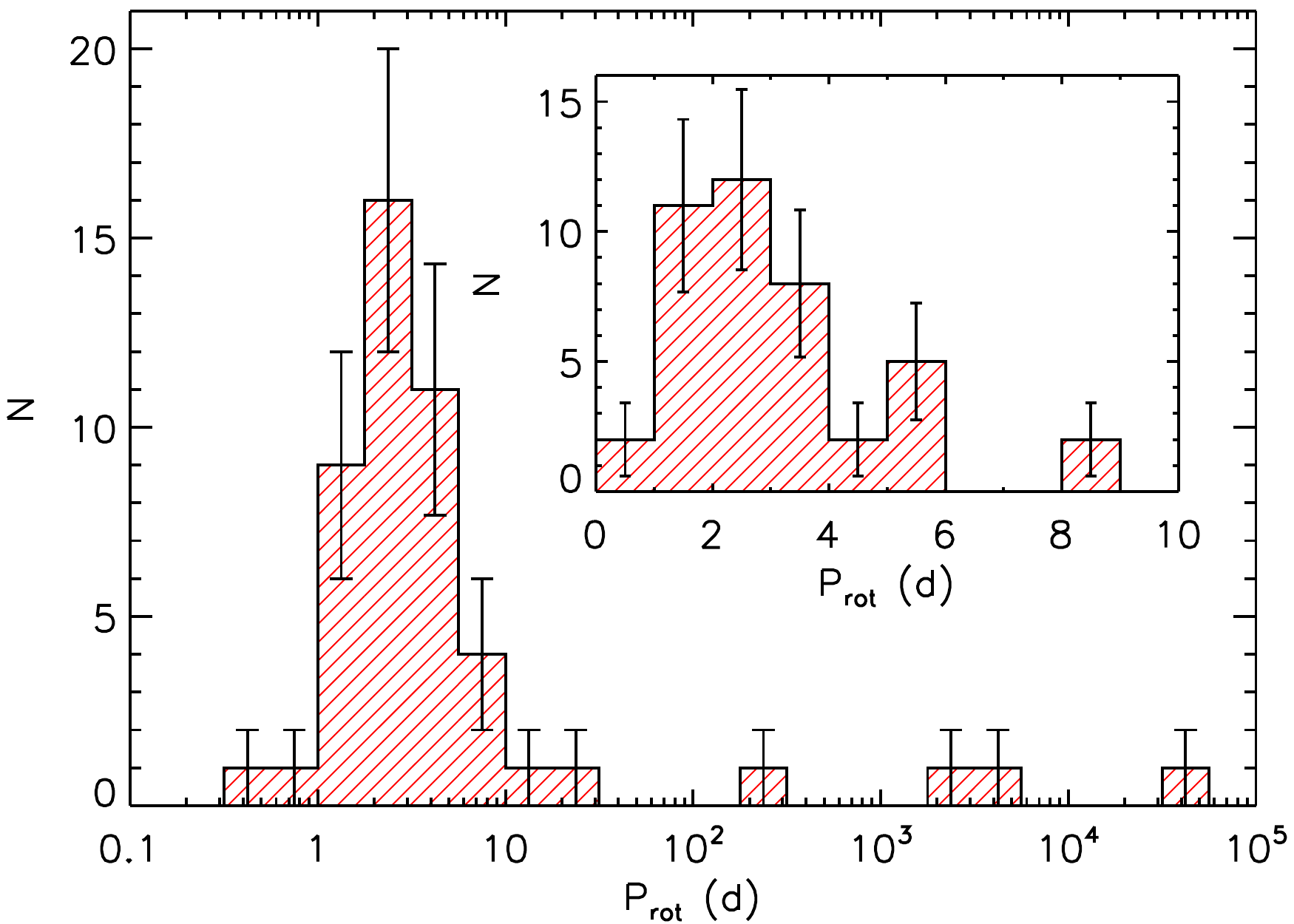}
	\caption{Distribution of rotational periods for 48/52 of the mCP stars. The inset plot shows the same 
	distribution for $P_{\rm rot}\leq10\,{\rm d}$.}
	\label{fig:Prot}
\end{figure}

In Fig. \ref{fig:Prot}, we show the distribution of rotational periods for those 48/52 stars with known 
values. It is evident that the sample consists of mCP stars exhibiting minimum and maximum periods that 
are comparable to the known fastest rotators 
\citep[$\sim0.5\,{\rm d}$, e.g.][]{Oksala2010,Grunhut2012b} and slowest rotators 
\citep[$\gtrsim100\,{\rm yrs}$, e.g.][]{Mathys2015}. We find that the distribution is consistent with a 
log-normal distribution (demonstrated by the derived Kolmogorov-Smirnov (KS) test statistic of 
$0.15\pm0.25$). Fitting a Gaussian function to the distribution yields a mean of $3.1\pm2.2$. We note 
that there exists a tail to very long periods, with the longest determined period in our sample being 
$97\,{\rm yrs}$.

We compared the distribution's peak $P_{\rm rot}$ with that yielded by larger previously published 
surveys. In Fig. 8 of \citet{Wolff1975}, the distribution of compiled rotational periods is 
concentrated below $10\,{\rm d}$ and exhibits a peak at $P_{\rm rot}<3.2\,{\rm d}$; applying the same 
binning ($\log{(\sigma_{P_{\rm rot}}/{\rm d})}=0.5$) to the periods associated with the volume-limited 
survey yields the same peak location and sharp decline in the number of stars with 
$P_{\rm rot}>10\,{\rm d}$. More recently, \citet{Netopil2017} reported rotational periods of more than 
$500$ confirmed or candidate mCP stars. We derived a mean of $2.4\,{\rm d}$ by fitting a Gaussian 
function to their distribution of reported periods; therefore, the two distributions' peak locations are 
in agreement within the estimated uncertainty. The preceeding discussion suggests that, in terms of 
the rotational periods, the survey presented here is representative of the larger population of known 
mCP stars.

\subsection{Inclination Angles}\label{sect:inc}

For each star in our sample with known $P_{\rm rot}$, we derived the inclination of the star's rotation 
axis assuming rigid rotation. The inclination angles were derived according to
\begin{equation}
	i=\arcsin\left[\frac{1}{50.6}\frac{v\sin{i}}{{\rm km\,s^{-1}}}\frac{P_{\rm rot}}{{\rm days}}\left(\frac{R}{R_\odot}\right)^{-1}\right]
\end{equation}
using the rotational periods in conjunction with the projected rotational velocities ($v\sin{i}$) and 
stellar radii ($R$) derived or adopted in Paper I. The $v\sin{i}$ values of those stars with long 
rotational periods ($P_{\rm rot}\gtrsim10\,{\rm d}$) could, in general, not be derived and have not 
been reported in the literature. This is related to the fact that, in these cases, the observed 
spectral line broadening is dominated by thermal broadening, Zeeman splitting, etc. thus preventing a 
determination of $v\sin{i}$ of useful precision. We were able to derive or adopt reported $v\sin{i}$ 
values for 43/47 of the stars with known rotational periods. Detailed analyses involving the derivation of 
$i$ associated with HD~24712 and HD~187474, which both exhibit $v\sin{i}<10\,{\rm km\,s^{-1}}$, have been 
previously published. For HD~24712, we adopt $i=43\pm2\degree$ derived by \citet{Bagnulo1995} using both 
circularly and linearly polarized spectra. For HD~187474, we adopt $i=86\degree$, which was derived by 
\citet{Landstreet2000} by modelling both $\langle B_{\rm z}\rangle$ measurements and mean field modulus 
measurements; no uncertainty is reported. In total, we were able to derive or compile inclination angles 
for 45/52 of the mCP stars.

\begin{figure}
	\centering
	\includegraphics[width=1.0\columnwidth]{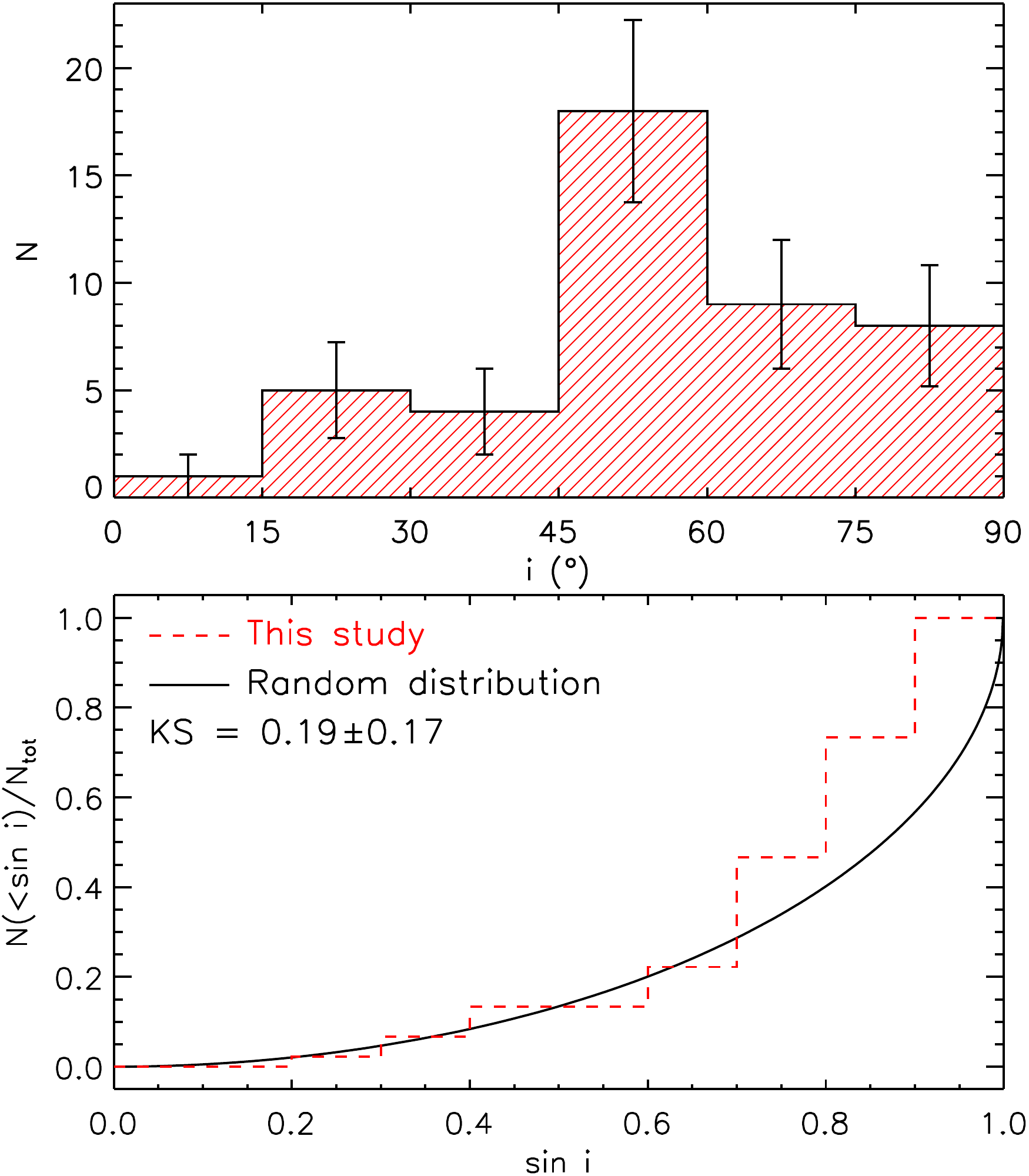}
	\caption{\emph{Top:} Distribution of inclination angles for the 44/52 mCP stars with known 
	$v\sin{i}$ and known $P_{\rm rot}$ values or with published $i$ values.
	\emph{Bottom:} Cumulative distribution function of $\sin{i}$ (dashed red) compared against that 
	associated with a distribution of randomly oriented rotational axes (solid black). The 
	Kolmogorov-Smirnov test statistic of ${\rm KS}=0.19\pm0.17$ suggests that the inclination angles 
	may not be distributed randomly.}
	\label{fig:inc_CDF_hist}
\end{figure}

In Fig. \ref{fig:inc_CDF_hist} (top), we show the distribution of the 45 known inclination angles. It is 
apparent that the distribution is strongly peaked at the $45\degree<i<60\degree$ bin. Such a feature is 
not associated with a distribution of inclination angles that are randomly oriented in space, which is 
characterized by a monotonic increase in frequency from $0\degree$ to $90\degree$. Furthermore, either 
an excess of moderate $i$ values ($30\degree<i<60\degree$) or a deficiency of high $i$ values 
($i>60\degree$) relative to a random distribution is apparent when comparing the cumulative 
distribution functions (CDFs) of $\sin{i}$ as shown in Fig. \ref{fig:inc_CDF_hist} (bottom). We 
computed a KS test statistic comparing the distribution of $i$ values with that associated with a random 
distribution ($0.19\pm0.17$), which suggests that the inclination angles of the mCP stars in this sample 
may not be drawn from a random distribution.

Previous studies have addressed the question of whether the inclination angles of mCP stars are in fact 
randomly oriented in space. \citet{Abt2001} and \citet{Netopil2017} compiled 102 and 180 inclination 
angles, respectively, and concluded that the resulting distributions are consistent with random 
distributions. The discrepancy between the observed and expected (random) $i$ distributions in the volume 
limited survey presented here may be caused by the fact that the observed distribution is incomplete: for 
7/52 stars, $i$ could not be derived or found in the literature. We attempted to estimate the statistical 
significance of the $0.19\pm0.17$ KS test statistic by carrying out a Monte Carlo (MC) simulation. This 
involved generating $10^5$ simulated distributions consisting of 45 $i$ values sampled from the 
theoretical random distribution. KS test statistics comparing each simulated distribution with the 
theoretical random distribution were then calculated. We found that 7~per~cent of the simulated 
distributions exhibited a test statistic $\geq0.19$; therefore, we conclude that the difference between 
the observed and random distributions is not statistically significant.

Although it is likely that the inclination angles presented here are randomly oriented, we note that the 
location of the maximum incidence of $i$ shown in Fig. \ref{fig:inc_CDF_hist} is similar to the location 
of the distribution's peak found in the results of the larger \citet{Abt2001} and \citet{Netopil2017} 
studies. In Fig. 1 of \citet{Abt2001}, the $\sin{i}$ distribution peaks at 
$\approx0.7$ ($i\approx45\degree$) while the inclination angles compiled by \citet{Netopil2017} exhibit 
a maximum frequency within $45\degree<i<60\degree$. This may not be entirely unexpected considering 
that a number of common mCP stars are included in all three studies: 20 and 21~per~cent of the 
\citet{Abt2001} and \citet{Netopil2017} samples of mCP stars with known $i$ are also included in our 
volume limited survey. The statistical significance of the position of this peak in our sample was 
estimated using the results of the MC simulation discussed above. For each of the $10^5$ simulated 
distributions consisting of 45 randomly oriented $i$ values, we determined the location of the maximum 
incidence when the distribution is binned using $\Delta i=15\degree$, as shown in Fig. 
\ref{fig:inc_CDF_hist}. Nineteen~per~cent of the simulated distributions exhibited a maximum incidence 
within $45\degree<i<60\degree$ suggesting that the location of the peak is not statistically significant. 
We also evaluated the statistical significance of the peak height relative to the neighbourhing bins (e.g. 
the peak shown in Fig. \ref{fig:inc_CDF_hist} exhibits a peak height of 9 relative to the two 
neighbourhing bins). We found that 7~per~cent of the simulated distributions exhibited peaks with 
relative heights $\geq9$. Therefore, while the significance of the peak height is higher than that 
associated with its location, we do not consider it to be statistically significant.

\section{Magnetic Parameters}\label{sect:mag_param}

The magnetic field strengths and geometries of the mCP stars can be estimated using the ORM 
\citep{Stibbs1950}. In particular, we use Equations 1 and 2 of \citet{Preston1967} to derive the 
strength of the field's dipole component ($B_{\rm d}$) along with the associated obliquity angle 
($\beta$, i.e. the angle between the dipole component's axis of symmetry and the star's rotational 
axis). This derivation depends on the star's inclination angle ($i$), linear limb-darkening coefficient 
($u$), and ratio of the minimum to maximum longitudinal field strengths 
($r\equiv\langle B_{\rm z}\rangle_{\rm min}/\langle B_{\rm z}\rangle_{\rm max}$).

Linear limb-darkening coefficients were derived using the grid calculated by \citet{Diaz-Cordoves1995}. 
This grid is calculated for a range of surface gravities ($0.0\leq\log{g}\leq5.0\,{\rm [cgs]}$), 
effective temperatures ($3\,500\leq T_{\rm eff}\leq50\,000\,{\rm K}$), and photometric filters (Johnson 
$UVB$ and Str{\"o}mgren $uvby$). We used the limb-darkening coefficients calculated for the Johnson $V$ 
filter because of the fact that this filter's transmission function approximately spans the wavelength 
range of the LSD line masks ($3\,000\leq\lambda\leq7\,000\,{\rm \AA}$) discussed in Sect. 
\ref{sect:mCP_Bz}. The grid of Johnson $V$ limb-darkening coefficients was interpolated for each star over 
$\log{g}$ and $T_{\rm eff}$ (using the $\log{g}$ and $T_{\rm eff}$ values derived in Paper I).

The ratios of minimum to maximum field strengths ($r$) for each star were derived from the fits to the 
phased $\langle B_{\rm z}\rangle$ measurements shown in Figures \ref{fig:phased_Bz01} to 
\ref{fig:phased_Bz03}. Both $\langle B_{\rm z}\rangle_{\rm min}$ and 
$\langle B_{\rm z}\rangle_{\rm max}$ were calculated using the mean ($B_0\equiv C_0$) and amplitudes 
($B_1\equiv C_1$) associated with the $1^{\rm st}$ order sinusoidal fits (i.e. Eqn. \ref{eqn:sin_fcn} 
with $C_2\equiv0$). The uncertainties in $B_0$ and $B_1$ (and thus, in $r$) were derived by applying the 
method of residual bootstrapping. The method involves calculating the residuals associated with the 
$\langle B_{\rm z}\rangle$ measurements and the ($1^{\rm st}$ order) sinusoidal fit. For each 
$\langle B_{\rm z}\rangle$ measurement, we add to it a randomly selected residual and the sinusoidal 
fit is recalculated. This process is repeated $10\,000$ times yielding approximately Gaussian 
fitting parameter distributions, which are used to estimate 3$\sigma$ uncertainties.

Finally, $B_{\rm d}$ and $\beta$ were derived using the calculated values of $i$, $u$, and $r$ according 
to Equations 1 and 2 of \citet{Preston1967}. Given the number of parameters involved in this derivation 
(e.g. $T_{\rm eff}$, $R$, $v\sin{i}$, etc.), it is difficult to evaluate how they are correlated. Without 
accounting for these correlations, the uncertainties in $B_{\rm d}$ and $\beta$ will likely be erroneously 
high. We estimated $\sigma_{B_{\rm d}}$ and $\sigma_\beta$ by extending the Monte Carlo (MC) uncertainty 
analysis carried out in Paper I. This involved calculating each star's $B_{\rm d}$ and $\beta$ for 
$\gtrsim1\,000$ data points each consisting of randomly selected effective temperatures and luminosities 
normally distributed according to their most probable values and their uncertainties (Shultz et al., 
in prep; a brief description is presented in Paper I). This analysis was extended by assigning $v\sin{i}$ and $r$ values -- randomly generated from 
normal distributions with widths defined by $\sigma_{v\sin{i}}$ and $\sigma_r$ -- to each of the previously 
generated MC data points. Ultimately, this method yields distributions of $B_{\rm d}$ and $\beta$ values, 
which can be used to infer $\sigma_{B_{\rm d}}$ and $\sigma_\beta$. In general, the resulting distributions 
are either positively or negatively skewed. Therefore, rather than defining $\sigma_{B_{\rm d}}$ and 
$\sigma_\beta$ using each distribution's standard deviation, we adopt minimum and maximum limits defined 
such that $99.7$~per~cent of the distribution is enclosed.

In six cases (HD~3980, HD~38823, HD~108662, HD~108945, HD~137909, and HD~223640) the most probable 
$v\sin{i}$ values derived in Paper I were found to exceed the equatorial velocities ($v_{\rm eq}$) 
calculated using $P_{\rm rot}$, and $R$; however, the $v\sin{i}$ and $v_{\rm eq}$ values of all six 
stars were found to be equal within the estimated uncertainties (i.e. they are consistent with 
$i\approx90\degree$). In these cases, we removed those MC data points for which $v\sin{i}>v_{\rm eq}$. 
The peak values of the resulting MC distributions were then used to define new, most probable $v\sin{i}$ 
values.

\subsection{Dipole Field Strengths}

\begin{figure}
	\centering
	\includegraphics[width=1.0\columnwidth]{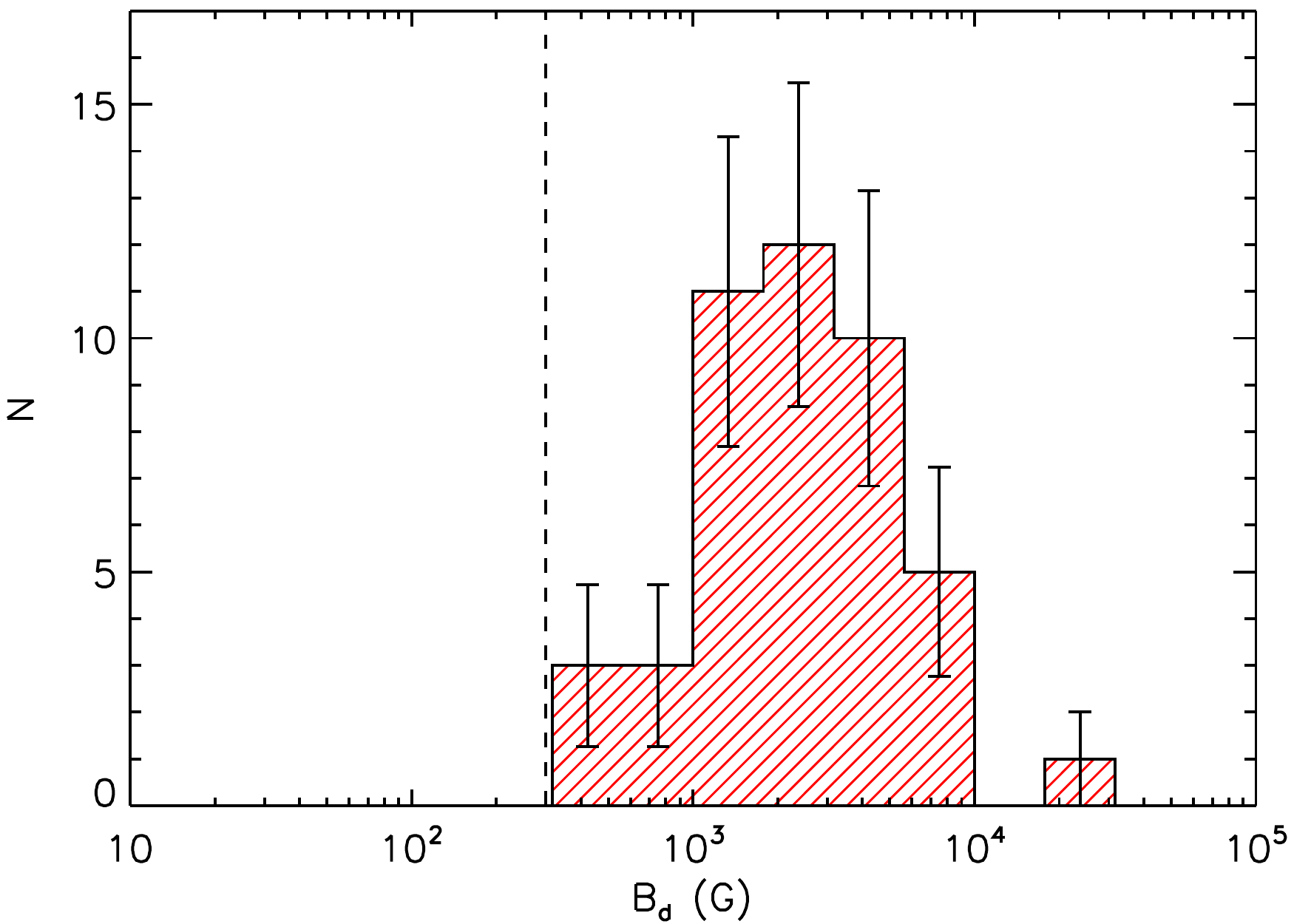}
	\caption{Distribution of dipole magnetic field strengths for 45/52 of the mCP stars. The 
	vertical dashed line corresponds to the critical field strength of a typical MS A-type star 
	($B_{\rm c}=300\,{\rm G}$) estimated by \citet{Auriere2007}.}
	\label{fig:Bd}
\end{figure}

In Fig. \ref{fig:Bd}, we show the derived dipole field strengths for 45/52 mCP stars in the sample (i.e. 
those with known rotational periods and inclination angles and for which multiple 
$\langle B_{\rm z}\rangle$ measurements are available). The $B_{\rm d}$ distribution is well 
characterized by a log-normal distribution as demonstrated by the derived KS test statistic of 
$0.10\pm0.19$. Fitting a Gaussian function to $\log{(B_{\rm d}/{\rm G})}$ yields a mean and 3$\sigma$ 
uncertainty of $3.4\pm0.2$ (corresponding to $2.6^{+1.9}_{-1.1}\,{\rm kG}$). The maximum derived 
$B_{\rm d}$ in the sample corresponds to $18.1_{-2.7}^{+3.4}\,{\rm kG}$ (HD~65339), which is in 
agreement with the value reported by \citet{Landstreet1988}. The minimum derived $B_{\rm d}$ 
corresponds to $330_{-60}^{+80}\,{\rm G}$ (HD~112185); however, the minimum dipole field strength 
derived when considering the upper $B_{\rm d}$ error limits corresponds to 
$B_{\rm d}^{\rm max}=390\,{\rm G}$ (HD~221760).

The survey carried out by \citet{Auriere2007} was specifically designed to search for mCP stars 
hosting weak dipole fields. They reported a minimum most probable field strength (i.e. minimum 
$B_{\rm d}$ without considering the estimated lower error limits) of $100\,{\rm G}$. The minimum 
$B_{\rm d}^{\rm max}$ found in their study is $477\,{\rm G}$, which is slightly higher but still 
comparable to that derived here. The fact that they did not find any dipole field strengths 
$\lesssim100\,{\rm G}$ led them to propose the existence of a critical dipole field strength 
($B_{\rm c}$), which defines a minimum field strength necessary for a field to maintain stability. 
They estimate that $B_{\rm c}\approx300\,{\rm G}$ for a typical A-type star and is indicated in Fig. 
\ref{fig:Bd}; it is clear that the majority of the 45 $B_{\rm d}$ values derived for the stars in our 
volume limited survey greatly exceed $300\,{\rm G}$.

\citet{Auriere2007} derived the following expression for the order of magnitude of 
$B_{\rm c}$ in terms of $P_{\rm rot}$, $R$, $T_{\rm eff}$, and the equipartion field strength 
of a typical main sequence (MS) A-type star ($B_{\rm eq}=170\,{\rm G}$):
\begin{equation}\label{eqn:Bc}
	B_{\rm c}\sim2B_{\rm eq}\left(\frac{P_{\rm rot}}{5\,{\rm d}}\right)^{-1}\left(\frac{R}{3\,R_\odot}\right)\left(\frac{T}{10^4\,{\rm K}}\right)^{-1/2}.
\end{equation}
We derived $B_{\rm d}/B_{\rm c}$ for each of the 45 stars having estimated dipole field strengths. All 45 
stars exhibit $B_{\rm d}/B_{\rm c}\gtrsim1$; four stars were found to have most probable 
$B_{\rm d}/B_{\rm c}\in[0.6,1)$. Three of these four stars (HD~29305, HD~56022, and HD~112185) have an 
estimated $B_{\rm d}/B_{\rm c}$ upper error limit $<1$ and therefore serve as the best candidates in 
our sample for either (1) potentially disproving the hypothesis that field strengths must exceed 
$B_{\rm c}$ or (2) refining the value of $B_{\rm c}$.

\begin{figure}
	\centering
	\includegraphics[width=1.0\columnwidth]{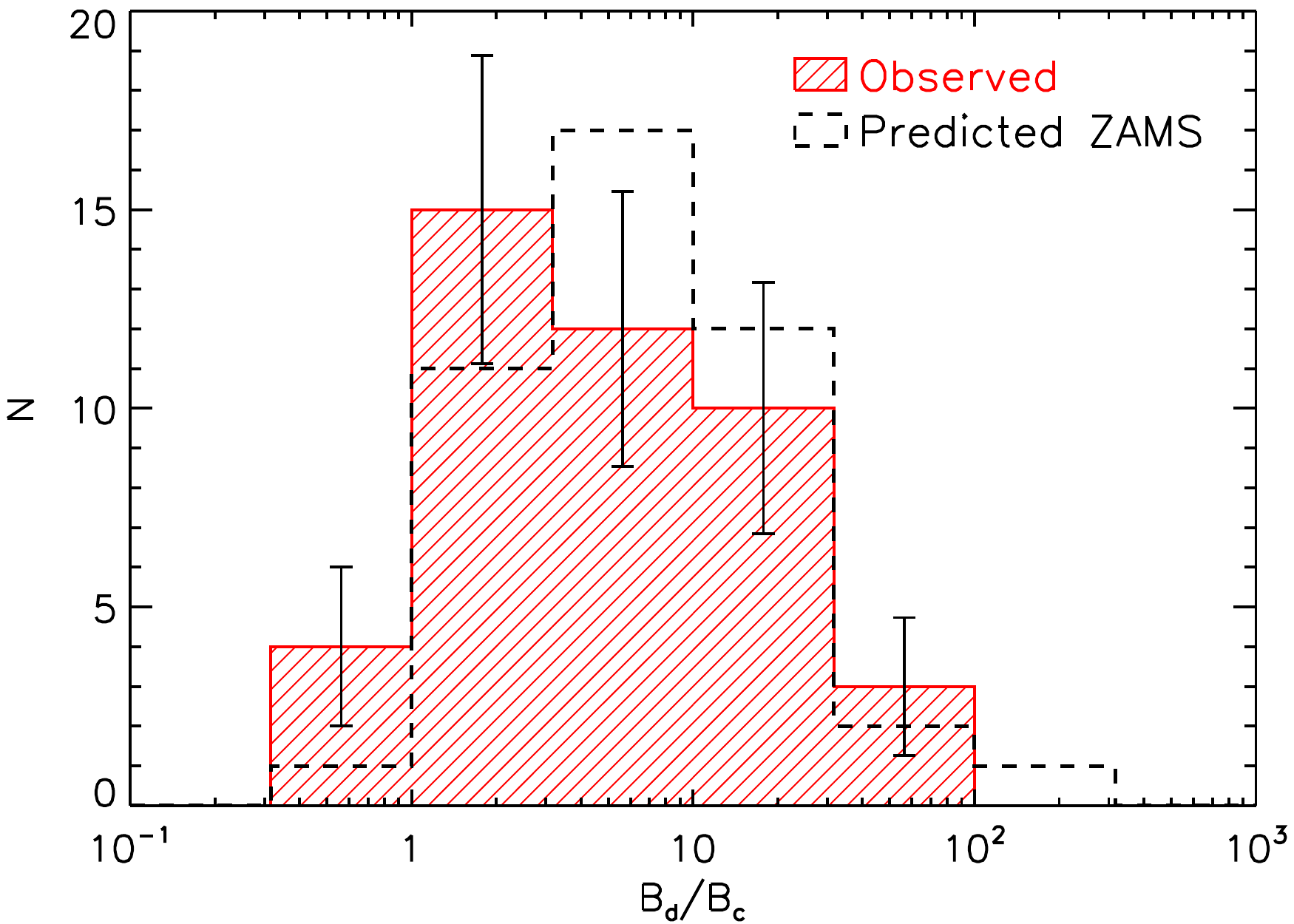}
	\caption{Distribution of the ratio of $B_{\rm d}$ to the critical field strength ($B_{\rm c}$) 
	derived by \citet{Auriere2007}; these authors hypothesize that no A-type stars should be found 
	exhibiting $B_{\rm d}/B_{\rm c}<1$ (indicated by the vertical dashed line).}
	\label{fig:Bd_Bc}
\end{figure}

The derivation of $B_{\rm c}$ by \citet{Auriere2007} applies to all A-type stars spanning the 
main sequence. Therefore, an additional test of the existence of $B_{\rm c}$ can be carried out by 
estimating each mCP stars' $B_{\rm d}/B_{\rm c}$ value as a function of fractional MS age ($\tau$) 
and determing if $B_{\rm d}/B_{\rm c}\ll 1$ at any point during its evolution across the MS. We 
estimated each stars' $R(\tau)$ and $T_{\rm eff}(\tau)$ by interpolating evolutionary tracks computed by 
\citet{Ekstrom2012} and \citet{Mowlavi2012}, which is discussed more thoroughly in Paper I. The change 
in $P_{\rm rot}$ occuring across the MS was estimated using two grids of rotating evolutionary tracks. 
For stars with masses $<1.7\,M_\odot$, we used the rotating solar metallicity ($Z=0.014$) evolutionary 
tracks computed by \citet{Ekstrom2012} for $v_{\rm eq}/v_{\rm c}$ equal to $0.0$ and $0.4$ where 
$v_{\rm eq}$ and $v_{\rm c}$ are the equatorial velocity and critical breakup rotational velocity at 
the zero age MS (ZAMS). For stars with masses $\geq1.7\,M_\odot$, we used the higher 
$v_{\rm eq}/v_{\rm c}$ density ($v_{\rm eq}/v_{\rm c}=0.0$, 0.1, 0.3, 0.5, 0.6, 0.7, 0.9, and 1.0) 
solar metallicity grids computed by \citet{Georgy2013a}. The change in the dipole field strength was 
estimated by assuming that magnetic flux is conserved. Under this assumption, $B_{\rm d}$ decreases 
with $R^{-2}$ as $R$ increases from the ZAMS to the terminal age MS (TAMS). We find that the 
predicted $B_{\rm d}/B_{\rm c}$ values decrease monotonically along the MS; as a result, only one of the 
45 stars with derived $B_{\rm d}$ values is predicted to have $B_{\rm d}/B_{\rm c}<1$ at earlier points 
during its MS lifetime. The distributions of the observed $B_{\rm d}/B_{\rm c}$ and the 
$B_{\rm d}/B_{\rm c}$ values predicted at the ZAMS are shown in Fig. \ref{fig:Bd_Bc}.

It is evident that both of the observed and predicted ZAMS $B_{\rm d}/B_{\rm c}$ distributions shown in 
Fig. \ref{fig:Bd_Bc} exhibit a sharp decrease in frequency at $B_{\rm d}/B_{\rm c}<1$. This is 
consistent with the notion that the current $B_{\rm d}/B_{\rm c}$ values are initially drawn from a 
wider distribution containing lower $B_{\rm d}/B_{\rm c}$ values: the initial distribution is truncated 
at $B_{\rm d}/B_{\rm c}=1$ resulting in a sharp decline towards lower values.

\subsection{Obliquity Angles}

\begin{figure}
	\centering
	\includegraphics[width=1.0\columnwidth]{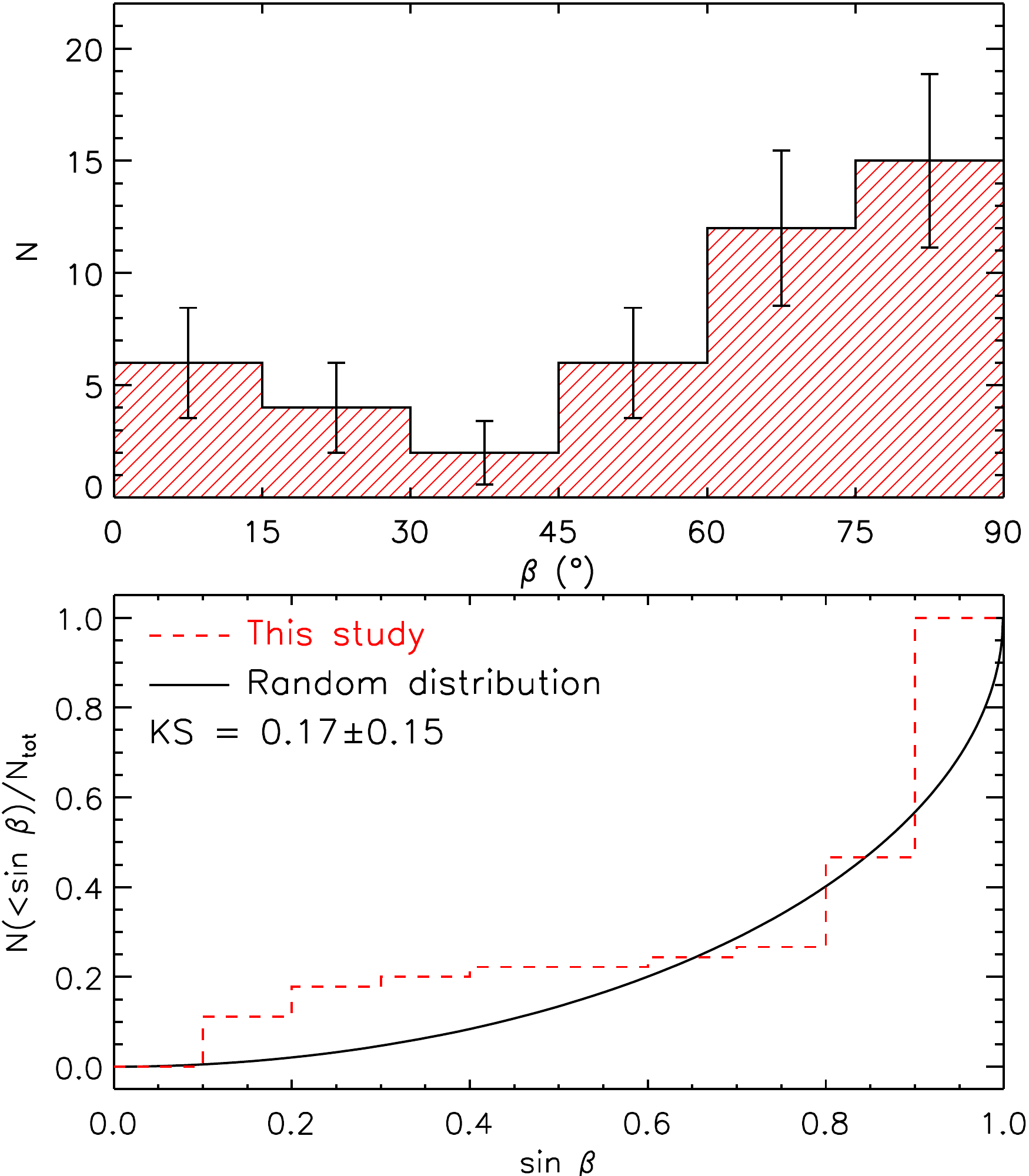}
	\caption{\emph{Top:} Distribution of obliquity angles for the 45/52 mCP stars with known 
	$P_{\rm rot}$ and $i$ values and for which multiple $\langle B_{\rm z}\rangle$ measurements are 
	\emph{Bottom:} Cumulative distribution function of $\sin{\beta}$ (dashed red) compared against that 
	associated with a distribution of randomly oriented magnetic dipole axes (solid black). The 
	Kolmogorov-Smirnov test statistic of ${\rm KS}=0.17\pm0.15$ suggests that the inclination angles 
	may not be distributed randomly.}
	\label{fig:beta_CDF_hist}
\end{figure}

In Fig. \ref{fig:beta_CDF_hist}, we show the distribution and CDF associated with the 45 obliquity 
angles ($\beta$) derived using Equation 3 of \citet{Preston1967}. The $\beta$ distribution exhibits a 
moderate increase from low to high $\beta$ values, which is qualitatively similar to that associated 
with a distribution of randomly oriented axes. A more quantitative comparison was carried out using the 
CDFs of the derived $\sin\beta$ and theoretical random distributions. We derived a KS test statistic of 
$0.17\pm0.15$ suggesting that the $\beta$ values may not be randomly oriented. The significance of this 
KS test statistic was evaluated using the same Monte Carlo simulation that was carried out in Sect. 
\ref{sect:inc} with the inclination angles. $10^5$ simulated distributions were generated, each 
consisting of 45 $\beta$ values drawn from the theoretical random distribution. A KS test statistic 
comparing each of the simulated random distributions with the theoretical random distribution were 
calculated. We found that 13~per~cent of the resulting KS values were $\geq0.17$; therefore, we 
conclude that the apparent difference between the derived $\beta$ values and the theoretical random 
distribution is statistically insignificant.

\input{tbl_mag_param.tex}

\section{Evolution of Magnetic Field Strength}\label{sect:evo}

In Paper I, we identified statistically significant trends in the average surface chemical abundances 
of certain elements (e.g. Si, Ti, Cr, and Fe) as functions of stellar age. Similar correlations between 
the atmospheric chemical abundances and ages of Bp stars have been previously reported by 
\citet{Bailey2014}. The authors also found that the same elements exhibiting coherent changes with age 
also exhibit changes with the measured magnetic field strengths; this is attributed to a decrease in 
field strength with age as previously reported by \citet{Landstreet2007,Landstreet2008} for both 
MS Ap and Bp stars ($8<T_{\rm eff}<20\,{\rm kK}$).

\begin{figure*}
	\centering
	\includegraphics[width=2.0\columnwidth]{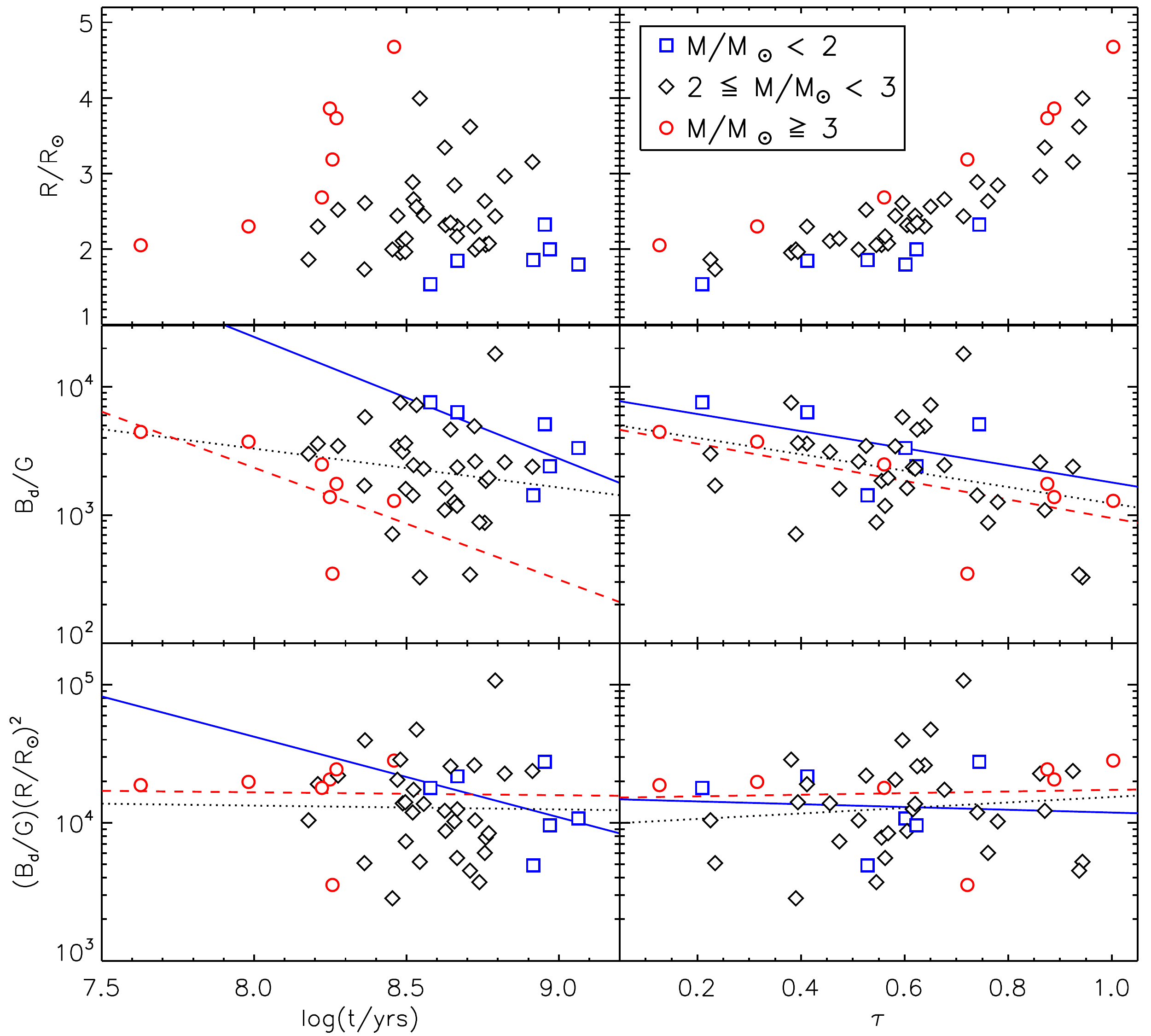}
	\caption{Radius ($R$, top row), dipole field strength ($B_{\rm d}$, middle row), and 
	$B_{\rm d}R^2$ (bottom row) for 45/52 of the mCP stars in the sample as functions of logarithmic 
	stellar age ($\log{t/{\rm yrs}}$, left column) and fractional MS age ($\tau$, right column). Three 
	mass ranges are identified: $M/M_\odot<2$, $2\leq M/M_\odot<3$, and $M/M_\odot\geq3$. The lines 
	correspond to the best fitting linear functions for the low-mass (solid blue), intermediate-mass 
	(dotted black), and high-mass (dashed red) stars. The derived slopes and their uncertainties are 
	listed in Table~\ref{tbl:age_B}.}
	\label{fig:age_B}
\end{figure*}

In Fig. \ref{fig:age_B}, we plot $R$, $B_{\rm d}$, and $B_{\rm d}R^2$ (i.e. the surface magnetic flux) 
as functions of age ($\log{t/{\rm yrs}}$) and fractional MS age ($\tau$). The 45 mCP stars represented 
in the figure are divided into low-mass ($M/M_\odot<2$), intermediate-mass ($2\leq M/M_\odot<3$), and 
high-mass ($M/M_\odot\geq3$) ranges. This is done for two reasons: (1) the increase in $R$ as each star 
evolves across the MS increases with mass; and (2) the width of the MS spanned by each of the three 
mass ranges decreases with decreasing mass. Therefore, under the assumption that magnetic flux is 
conserved, we expect to see larger changes in $B_{\rm d}$ with age in the high-mass range compared to 
the low-mass range. This can result in an increase in the dispersion of $B_{\rm d}$ and 
$B_{\rm d}R^2$ with increasing $\tau$ thereby decreasing our ability to detect such evolutionary 
changes.

The best fitting linear functions were derived for each of the $B_{\rm d}$ and $B_{\rm d}R^2$ values 
associated with the three mass intervals. We used an unweighted least-squares analysis because of the 
fact that the errors associated with $B_{\rm d}$, $\log{t/{\rm yrs}}$, and $\tau$ are typically large 
and asymmetric: $B_{\rm d}$ diverges as $|i-\beta|\to90\degree$ while $\log{t/{\rm yrs}}$ and $\tau$ are 
significantly more uncertain closer to the ZAMS than to the TAMS 
\citep[e.g. see Fig. 4 of][]{Kochukhov2006}. We found that the resulting fits yielded lower residuals 
compared to those obtained by considering both $x$ and $y$ uncertainties \citep[e.g. using the method 
described by][]{Williams2010}. We estimated 1$\sigma$ uncertainties in the fitting parameters by 
bootstrapping the residuals. The resulting linear fits are shown in Fig. \ref{fig:age_B} and the slopes 
are listed in Table \ref{tbl:age_B}.

We find that the dipole field strengths associated with all three of the mass intervals decrease over 
both $\log{t/{\rm yrs}}$ and $\tau$; we do not detect any changes in the magnetic flux ($B_{\rm d}R^2$) 
with stellar age. We note that the uncertainties in $R$ are relatively small ($\lesssim15$~per~cent) 
and that the estimated uncertainties in the slopes associated with $B_{\rm d}$ and $B_{\rm d}R^2$ are 
comparable (particularly for the slopes involving $\tau$). This suggests that the apparent differences 
in the rates of change of the field strength and the magnetic flux are not related to the uncertainty 
introduced by $R$. Therefore, we conclude that these results are statistically consistent with the 
notion that magnetic flux is conserved as an mCP star evolves across the MS.

\begin{table}
	\caption{Slopes and 1$\sigma$ uncertainties associated with the linear fits shown in Fig. 
	\ref{fig:age_B}.}
	\label{tbl:age_B}
	\begin{center}
	\begin{tabular}{@{\extracolsep{\fill}}l c r@{\extracolsep{\fill}}}
		\noalign{\vskip-0.2cm}
		\hline
		\hline
		\multicolumn{3}{c}{$\log{B_{\rm d}/{\rm G}}$ } \\
		\hline
		\noalign{\vskip0.5mm}
		Mass Interval & $\log{(t/{\rm yrs})}$ Slope & $\tau$ Slope \\
		\noalign{\vskip0.5mm}
		\hline	
		\noalign{\vskip0.5mm}
		$M/M_\odot<2$       & $-0.90\pm0.43$ & $-0.62\pm0.53$ \\
		$2\leq M/M_\odot<3$ & $-0.29\pm0.35$ & $-0.65\pm0.34$ \\
		$M/M_\odot\geq3$    & $-0.89\pm0.40$ & $-0.75\pm0.35$ \\
		\noalign{\vskip0.5cm}
		\multicolumn{3}{c}{$\log{[(B_{\rm d}/{\rm G})(R/R_\odot)^2]}$} \\
		\hline
		\noalign{\vskip0.5mm}
		Mass Interval & $\log{(t/{\rm yrs})}$ Slope & $\tau$ Slope \\
		\noalign{\vskip0.5mm}
		\hline	
		$M/M_\odot<2$       & $-0.53\pm0.54$ & $-0.06\pm0.59$ \\
		$2\leq M/M_\odot<3$ & $-0.02\pm0.36$ & $-0.19\pm0.31$ \\
		$M/M_\odot\geq3$    & $-0.04\pm0.42$ & $-0.04\pm0.36$ \\
		\noalign{\vskip0.5mm}
		\hline \\
	\end{tabular}
	\end{center}
\end{table}

The fact that $B_{\rm d}$ appears to decrease with increasing stellar age is qualitatively 
consistent with the findings of \citet{Landstreet2007}, whose survey only consisted of cluster members 
with well-constrained ages. Moreover, the rate of field strength decline that they derived for stars 
having $3\leq M/M_\odot\leq4$ ($-0.42\pm0.14$) is consistent with that derived here for our sample's 
high-mass stars ($-0.89\pm0.40$). This agreement provides evidence in support of the notion that the 
magnetic fields of MS mCP stars decay with age. The rate of change of magnetic flux for the same mass 
intervals are also in agreement within the uncertainties: \citet{Landstreet2007} derived a slope of 
$-0.22\pm0.14$ while we obtained $-0.04\pm0.42$. Despite the quantitative agreement, it is noteworthy 
that \citet{Landstreet2007} detect a decrease in magnetic flux over time whereas, for our sample, we do 
not. It is clear that our sample includes significantly fewer high-mass stars (7 compared to 25) and that 
the derived ages have a much higher uncertainty. On the other hand, the $B_{\rm d}$ values associated with 
the majority of the stars in our sample have been derived from reasonably well sampled 
$\langle B_{\rm z}\rangle$ curves; for most of the stars included in the \citet{Landstreet2007} study, 
only single $\langle B_{\rm z}\rangle$ measurements were obtained. The data set presented here can be 
used to assess the significance of this final point.

\begin{figure}
	\centering
	\includegraphics[width=1.0\columnwidth]{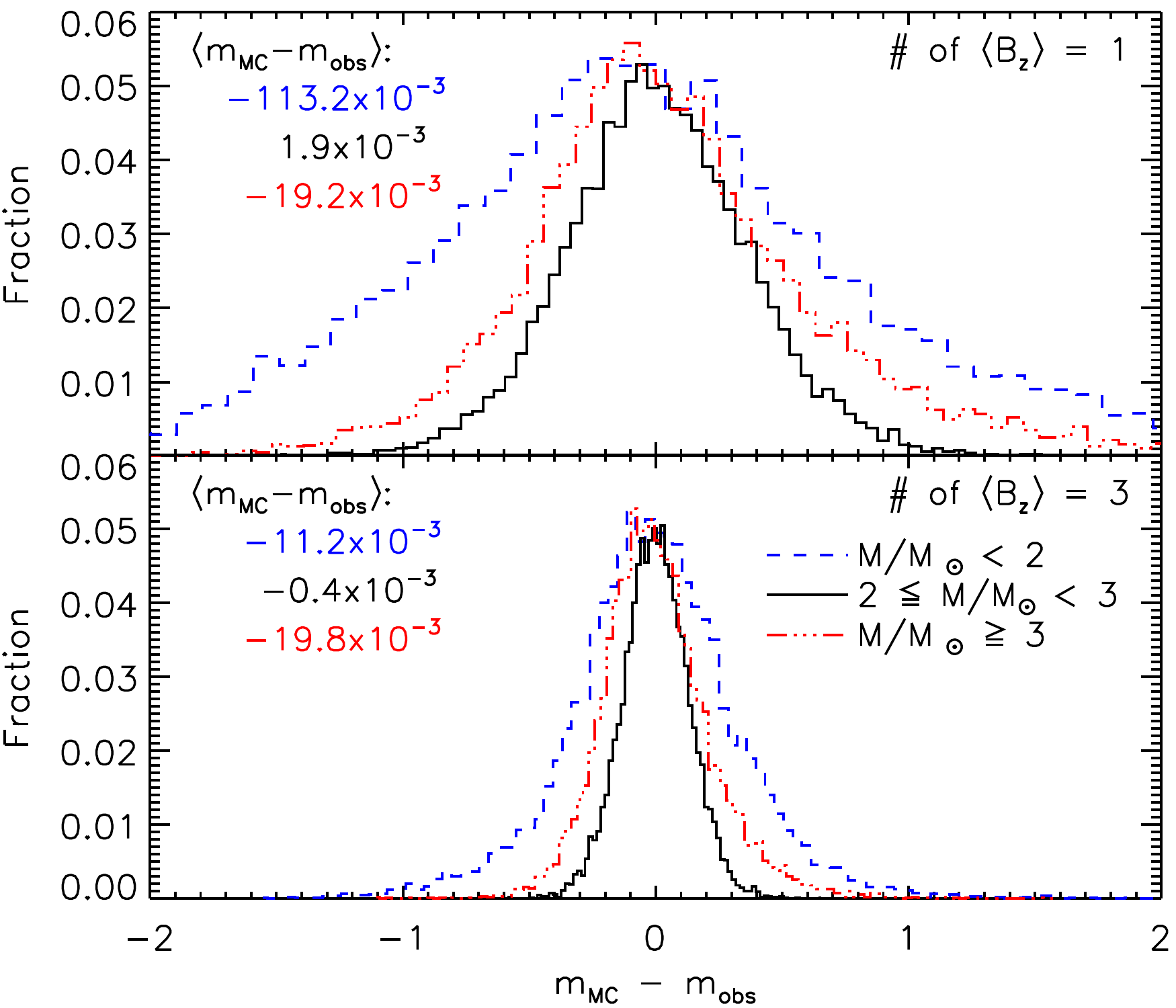}
	\caption{Distributions comparing the slope of $B_{\rm rms}R^2$ as a function of $\log{t/{\rm yrs}}$ 
	generated from the MC simulated data sets with that generated from the original data set. The top 
	distribution corresponds to the $B_{\rm rms}$ values calculated using 1 randomly generated 
	$\langle B_{\rm z}\rangle$ measurement while the bottom distribution uses 3 measurements. The 
	mean values of each distribution ($\langle m_{\rm MC}-m_{\rm obs}\rangle$) are listed in order 
	of increasing mass.}
	\label{fig:age_B_MC}
\end{figure}

We carried out an MC simulation in which the magnetic field strengths of the stars in our sample were 
estimated using only a small number of randomly sampled $\langle B_{\rm z}\rangle$ measurements. This 
involved generating $10^4$ simulated data sets consisting of either 1 or 3 
$\langle B_{\rm z}\rangle$ measurements for each of the 45 stars with known $B_0$ and $B_1$ 
(i.e. the mean and amplitude characterizing the $\langle B_{\rm z}\rangle$ curves). The 
$\langle B_{\rm z}\rangle$ measurements were generated using random phase values ($\theta\in[0,1]$) 
along with $B_0$ and $B_1$ such that $\langle B_{\rm z}\rangle(\theta)=B_0+B_1\sin{\theta}$. Each 
star's root-mean square field strength ($B_{\rm rms}$) was then calculated \citep[as done 
by][]{Landstreet2007}. Finally, the linear fitting analyses involving $\log{t/{\rm yrs}}$, 
$B_{\rm rms}$, and $B_{\rm rms}R^2$ were carried out and the derived slopes were compared with those 
generated using the original data set. In Fig. \ref{fig:age_B_MC}, we show the resulting distributions 
based on the $B_{\rm rms}R^2$ slopes.

The results of the MC simulation suggest that both the slopes of $B_{\rm rms}$ and $B_{\rm rms}R^2$ 
as functions of $\log{t/{\rm yrs}}$ and $\tau$ are biased towards lower values when $B_{\rm rms}$ 
is derived from a small number of $\langle B_{\rm z}\rangle$ measurements. However, we find that 
the bias is small and decreases with increasing sample size. Considering the large sample size of the 
survey carried out by \citet{Landstreet2007}, the bias is likely negligible as assumed by the authors. 
It is clear from the distributions shown in Fig. \ref{fig:age_B_MC} that, depending on the number of 
$\langle B_{\rm z}\rangle$ measurements used to derive $B_{\rm rms}$ of each star, the uncertainty in 
the slope may be significantly affected. We note that the MC simulation uses the stellar ages derived 
for the current sample in Paper I; therefore, the results of the simulation are certainly affected by 
the large age uncertainties to some extent.

It is plausible that the lack of detection of flux decay in our sample could result from (1) a small 
sample size for each of the mass bins (in particular for the high-mass stars which are expected to 
exhibit the largest decrease) and (2) large errors in the stellar ages. Additionally, it is noted that, 
in terms of both $\log{t/{\rm yrs}}$ and $\tau$, the mCP stars in our sample are generally older than 
those contained in the \citet{Landstreet2007} sample: the fraction of stars having $\tau<0.4$ is 
$>50$~per~cent in the \citet{Landstreet2007} sample compared with $15$~per~cent in our volume-limited 
sample. In Fig. 4 of \citet{Landstreet2008} it is apparent that the rate of flux decay associated with 
high-mass Ap stars is significantly higher for $\tau<0.2$ compared with $\tau>0.2$. This suggests that 
the apparent discrepancy in terms of the detection of flux decay between our sample and that of 
\citet{Landstreet2007} may be caused by a decay rate that is higher for younger MS stars coupled with 
the different age distribution of our sample.

\section{Discussion and Conclusions}\label{sect:discussion}

In Paper I, we presented an analysis of the fundamental properties and chemical abundances of 52 and 
45 confirmed mCP stars, respectively, located within a distance of $100\,{\rm pc}$. This study is the 
first of its kind in two specific ways. First, it is focused on a volume-limited sample and thus, is 
less affected by the biases inherent to previous studies of samples of mCP stars 
\citep[e.g.][]{Kochukhov2006,Hubrig2007}. Secondly, we have attempted to perform the analysis in a 
homogeneous manner such that any dispersion introduced by using varying techniques or theoretical 
models is minimized. The results presented here build on those of Paper I with the addition of an 
analysis of the confirmed mCP stars' rotational periods and magnetic properties. In the following, we 
discuss these results and present our conclusions.

Rotational periods for 48/52 of the confirmed mCP stars in the sample were adopted based on (1) 
the available $\langle B_{\rm z}\rangle$ measurements (i.e. newly obtained or unpublished measurements 
using ESPaDOnS, NARVAL, and MuSiCoS, newly analyzed measurements, and previously published 
measurements) and (2) previously published values typically derived from photometric variability. In 
general, we found that the rotational periods inferred from magnetic measurements are consistent with the 
published values. However, in several cases $P_{\rm rot}$ could not be identified ambiguously and we 
adopted $P_{\rm rot}$ values based on somewhat unreliable or tenuous evidence (e.g. the newly obtained 
$\langle B_{\rm z}\rangle$ measurements of HD~221760 were insufficient to derive a unique period and 
poor agreement was found with previously published values). Adopting unconfirmed rotational periods of 
certain stars in the sample may have contributed to the detection of an unusual and unexpected feature 
in the distribution of inclination angles. 

The feature in question is the large peak $i$ frequency occuring within $45$ to $60\degree$, which also 
corresponds to the distribution's global peak value. This is unexpected since it is not typically 
found in a distribution of $i$ values that are randomly oriented in space \citep[e.g.][]{Abt2001}. 
Although statistically insignificant based on an estimated p-value of $0.07$, it is perhaps noteworthy 
that a similar feature is found in the much larger data set of mCP $i$ values published by 
\citet{Netopil2017}. We note that the distribution shown in Fig. \ref{fig:inc_CDF_hist} is incomplete 
since the $i$ values of 7/52 sample stars could not be derived; however, their inclusion is unlikely to 
dramatically reduce the feature's statistical significance. No correlation between $\sin{i}$ and 
Galactic latitude is found \citep[e.g. Fig. 2 of][]{Abt2001}, suggesting that the origin of the 
unexpected feature found in the $i$ distribution is unlikely to be environmentally dependent.

\citet{Landstreet2000} derived $\beta$ values for a sample of 24 Ap stars and found that the slow 
rotators ($P_{\rm rot}>25\,{\rm d}$) tend to exhibit low $\beta$ values while only 2 faster rotators in 
their sample ($P_{\rm rot}<25\,{\rm d}$) were found with $\beta<60\degree$. Our sample consists of 5 
stars with $P_{\rm rot}>25\,{\rm d}$; only 1 of these stars was assigned a value of $\beta$ 
\citep[obtained by][]{Landstreet2000}. Obliquity angles were derived for all 44 of the stars with 
$P_{\rm rot}<25\,{\rm d}$. We did not identify any clear correlations between $\beta$ and $P_{\rm rot}$. 
We do confirm the findings of \citet{Landstreet2000} that $\beta$ tends to be large for these more 
rapidly rotating stars \citep[i.e. those with $P_{\rm rot}$ values that are more commonly found amongst 
mCP stars e.g.][]{Wolff1975,Bychkov2005,Netopil2017}. However, we find that the distribution of $\beta$ 
values is consistent with a theoretical distribution of randomly oriented axes. We also did 
not find any clear correlations between $\beta$ and absolute stellar age or fractional MS age or 
between $\beta$ and $B_{\rm d}$. Therefore, we find no evidence that the $\beta$ values of stars with 
$P_{\rm rot}<25\,{\rm d}$ are preferentially oriented as a result of some physical mechanism 
\citep[e.g.][]{Mestel1972,Moss1984}. We note that this result is consistent with the findings of 
\citet{Wade1997} who identified both young and evolved mCP stars that exhibit moderate $\beta$ values 
($\sim30\degree$).

We were able to constrain the dipole magnetic field strengths for 45/52 of the mCP stars in our 
sample. The minimum field strength found in our sample when considering the upper error limits of 
$B_{\rm d}$ corresponds to $390\,{\rm G}$. The fact that we did not find any stars with 
fields $\lesssim100\,{\rm G}$ is consistent with the notion that there exists a magnetic desert 
\citep[e.g.][]{Auriere2007,Lignieres2014}. We also derived the critical field strengths of each of the 
stars, which corresponds to the minimum field strength required for an mCP star's field to remain 
stable as hypothesized by \citet{Auriere2007}. Three stars (HD~29305, HD~56022, and HD~112185) were 
found exhibiting upper error limits of $B_{\rm d}/B_{\rm c}<1$; however, no stars were found having 
$B_{\rm d}/B_{\rm c}\ll1$. These stars may serve as useful targets for constraining $B_{\rm c}$, if 
this critical lower field strength limit does exist.

Although our volume-limited sample does not contain any examples of mCP stars with field strengths well 
below $B_{\rm c}$, several examples of intermediate- and high-mass stars have recently been reported. 
\citet{Fossati2015} derived a dipole field strength of $60<B_{\rm d}<230\,{\rm G}$ for the massive 
B1~II/III star, $\beta$~CMa, suggesting that the magnetic desert feature may be limited to the cooler, 
intermediate mass stars studied by \citet{Auriere2007}. On the other hand, \citet{Alecian2016} report a 
field strength of $B_{\rm d}=65\pm20\,{\rm G}$ for an Ap star with an effective temperature estimated 
to be $11.4\pm0.3\,{\rm kK}$. This star is the primary component of the spectroscopic binary, HD~5550, 
which exhibits an orbital period $\sim6.8\,{\rm d}$. They find that the magnetic component likely 
rotates with a period of $6.8\,{\rm d}$; no radius or luminosity is reported. Assuming that the star 
is positioned somewhere on the MS, we obtain a rough estimate of the critical field strength (Eqn. 
\ref{eqn:Bc}) of $130\lesssim B_{\rm c}\lesssim220\,{\rm G}$; therefore, it is likely that 
$B_{\rm d}/B_{\rm c}<1$. It is possible that the fact that this star is in a binary system with a 
relatively short period may somehow influence this result, however, considering the $>5\,{\rm d}$ 
orbital period, it is unlikely that any tidal interactions are taking place. It is also possible that 
the order of magnitude estimate of $B_{\rm c}$ estimated by \citet{Auriere2007} is simply too high or 
is in need of refinement.

A clear increase in the incidence rate of mCP stars with increasing mass was identified in 
Paper I (mCP stars account for $\approx3$~per~cent of MS stars with $M\approx1.5\,M_\odot$ and 
$\approx10$~per~cent of MS stars with $3.0<M/M_\odot<3.8$). The Monte Carlo simulation involving the 
\citet{Zorec2012} data did not reveal an increase in $B_{\rm c}$ with decreasing $M$, which might have 
otherwise explained the increased rarity of lower mass mCP stars. We conclude that, regardless of 
whether $B_{\rm c}$ exists, this particular property of mCP stars is likely a product of additional 
factors such as the increase in subsurface convection zone depth with decreasing mass.

In Paper I, we detected the decrease of average surface abundances of certain elements such as Si, Ti, 
and Fe over stellar age similar to the trends reported by \citet{Bailey2014} for Bp stars. Here we 
detect a decrease in $B_{\rm d}$ over both absolute age and fractional MS age. The rate of $B_{\rm d}$ 
decrease is strongest for the highest mass stars in our sample and is found to be in agreement with 
that reported by \citet{Landstreet2007} based on $B_{\rm rms}$ values. Contrary to the findings 
reported by \citet{Landstreet2007}, we do not detect any change in the surface magnetic flux over time; 
however, the reported decay rates are in agreement within the adopted uncertainties. We conclude that 
the lack of detection of surface flux decreases can plausibly be attributed to our smaller sample size 
and lower age precision.

\section*{Acknowledgments}

GAW acknowledges support in the form of a Discovery Grant from the Natural Science and 
Engineering Research Council (NSERC) of Canada.

\bibliography{ApBp_survey,ApBp_survey-other}
\bibliographystyle{mn2e}

\clearpage
\begin{figure*}
	\centering
	\includegraphics[width=2.1\columnwidth]{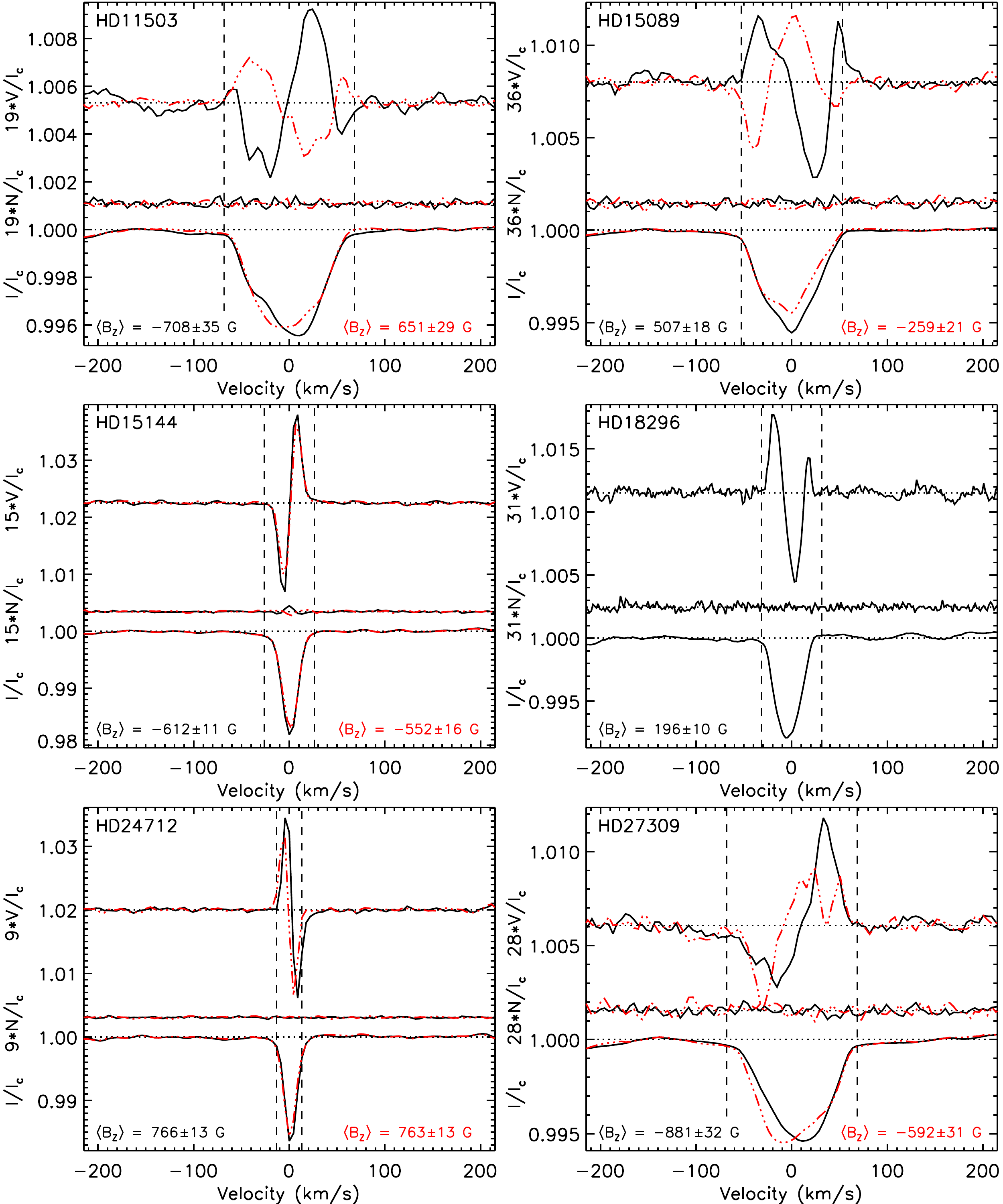}
	\caption{Examples of LSD profiles at one or two rotational phases.}
	\label{fig:lsd_ex_01}
\end{figure*}
\begin{figure*}
	\centering
	\includegraphics[width=2.1\columnwidth]{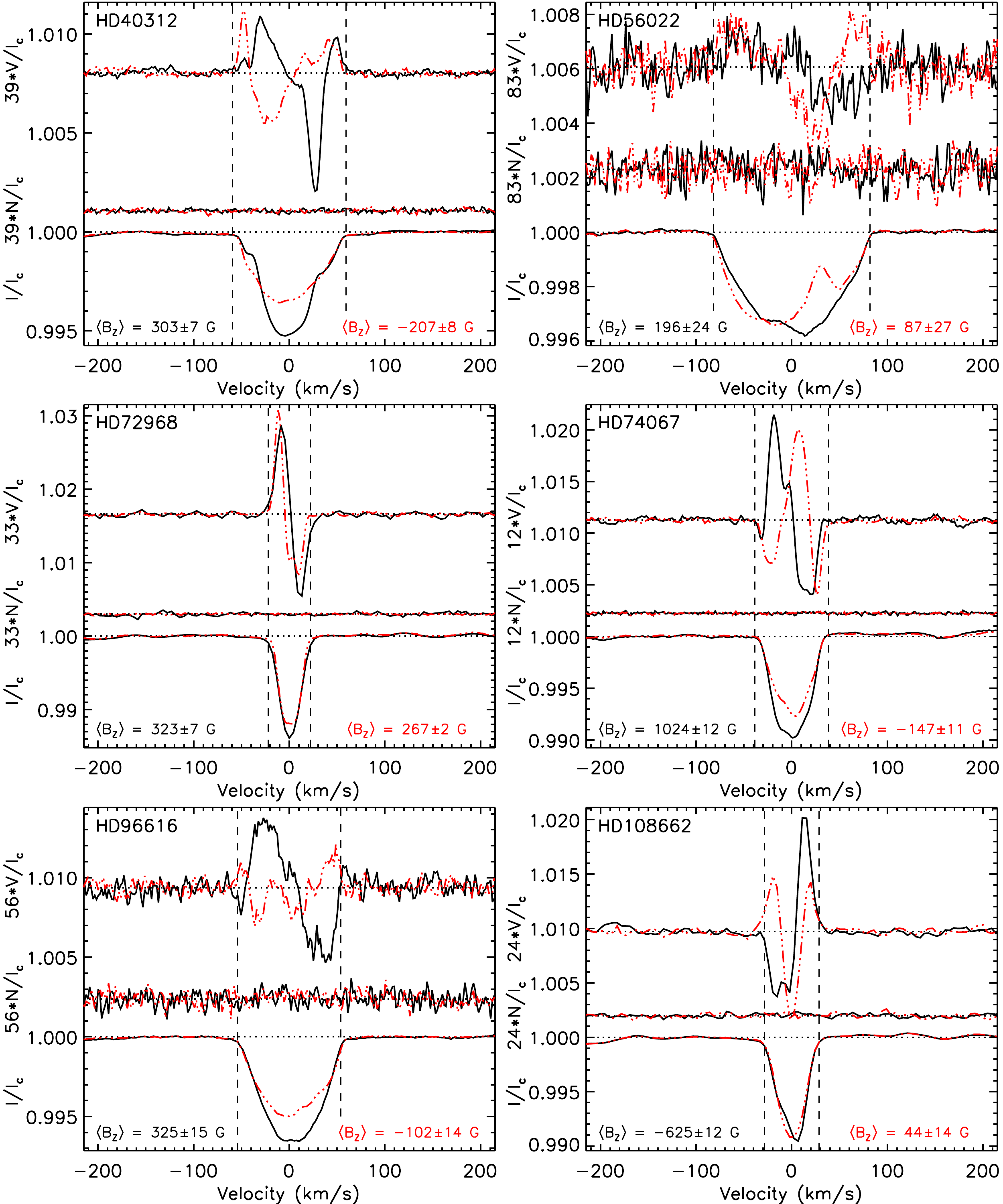}
	\caption{Examples of LSD profiles at one or two rotational phases.}
	\label{fig:lsd_ex_02}
\end{figure*}
\begin{figure*}
	\centering
	\includegraphics[width=2.1\columnwidth]{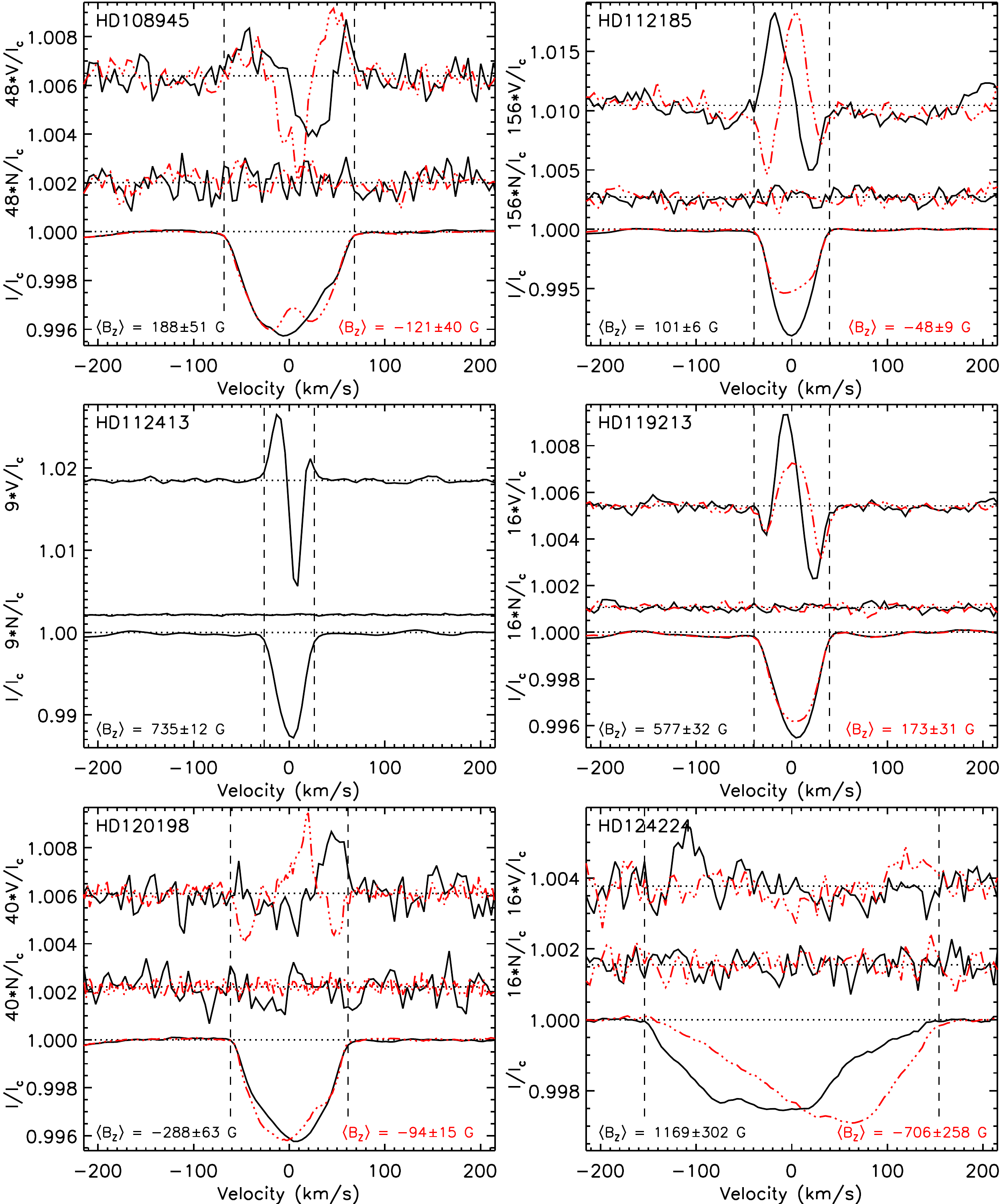}
	\caption{Examples of LSD profiles at one or two rotational phases.}
	\label{fig:lsd_ex_03}
\end{figure*}
\begin{figure*}
	\centering
	\includegraphics[width=2.1\columnwidth]{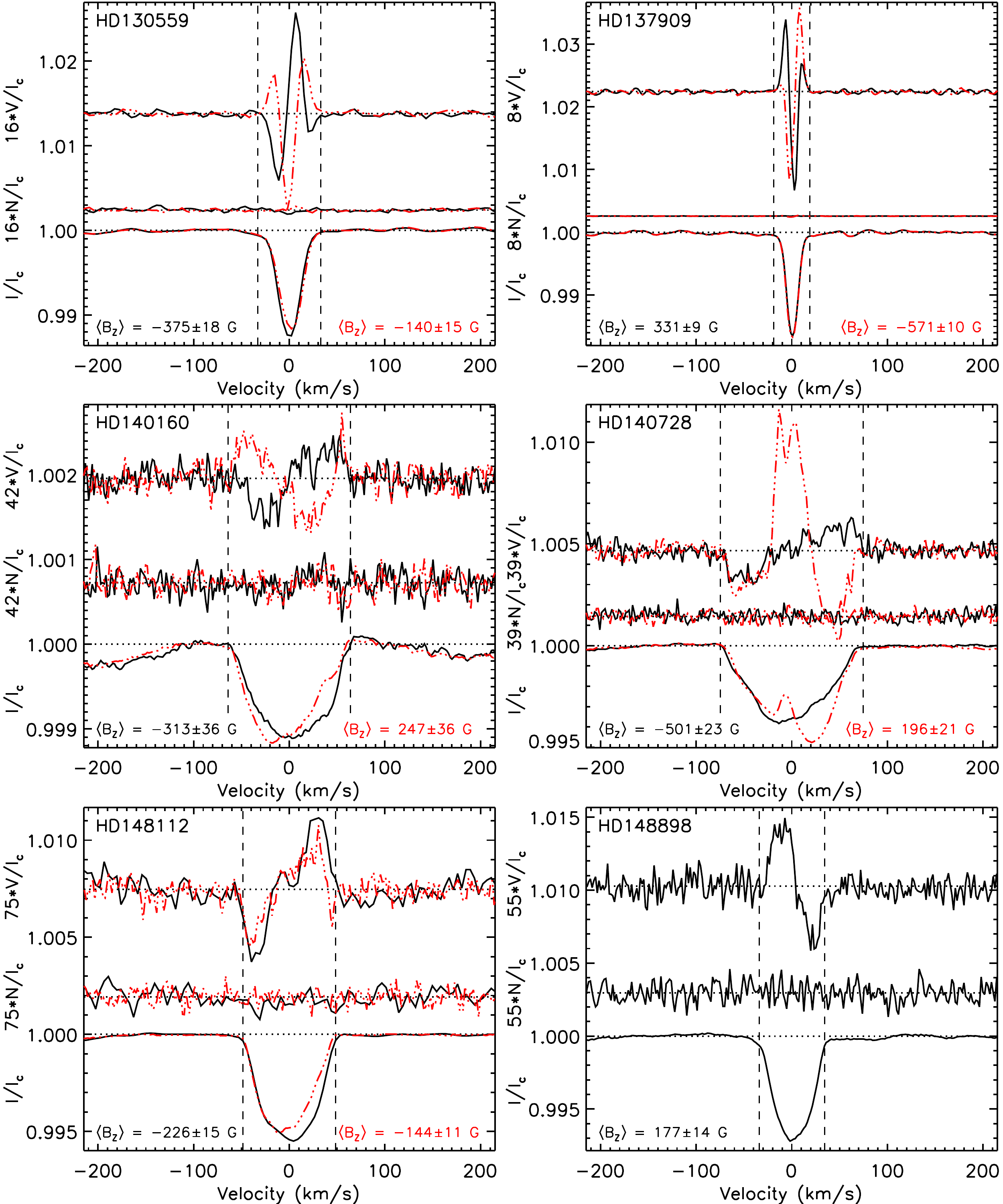}
	\caption{Examples of LSD profiles at one or two rotational phases.}
	\label{fig:lsd_ex_04}
\end{figure*}
\begin{figure*}
	\centering
	\includegraphics[width=2.1\columnwidth]{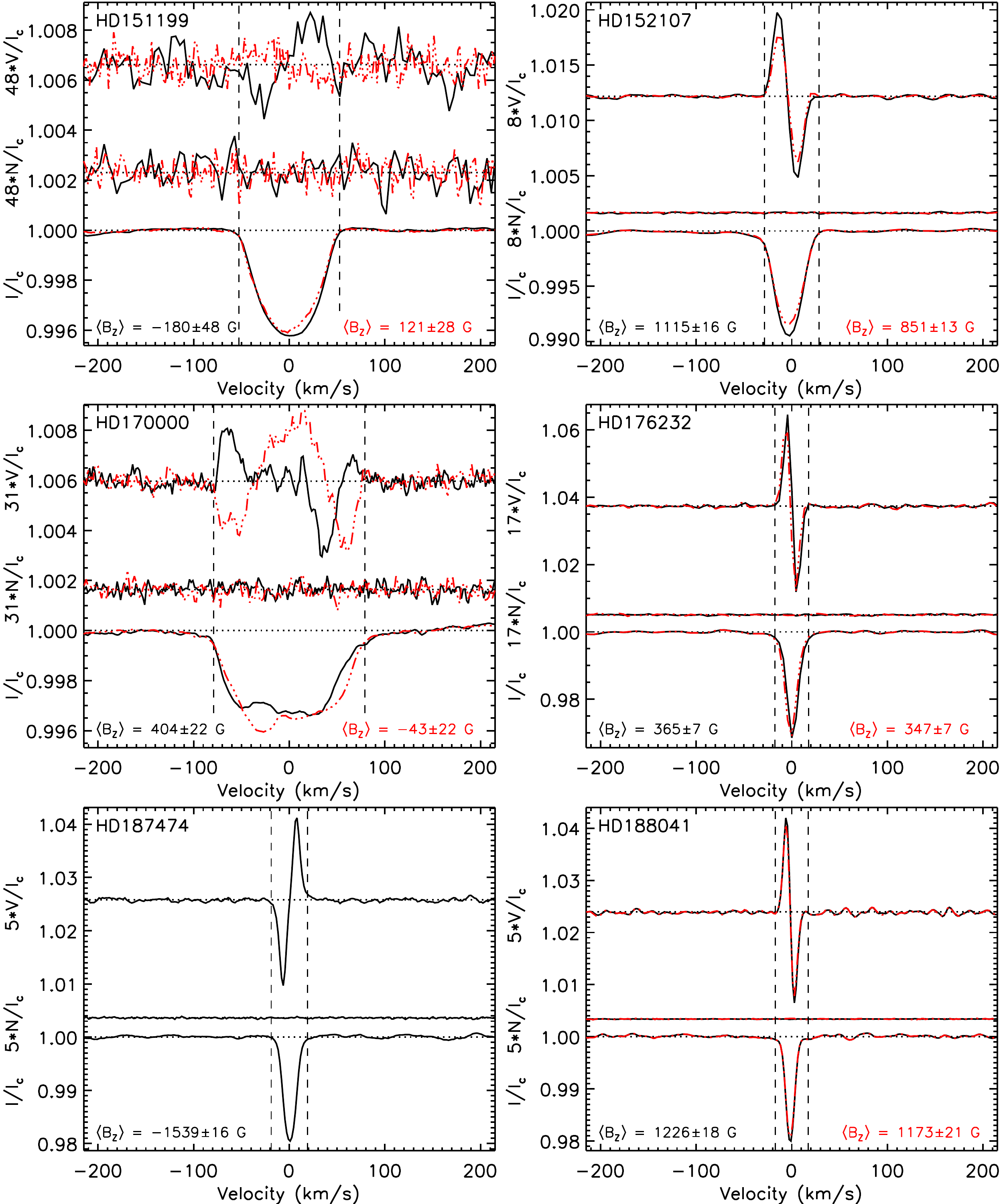}
	\caption{Examples of LSD profiles at one or two rotational phases.}
	\label{fig:lsd_ex_05}
\end{figure*}
\begin{figure*}
	\centering
	\includegraphics[width=2.1\columnwidth]{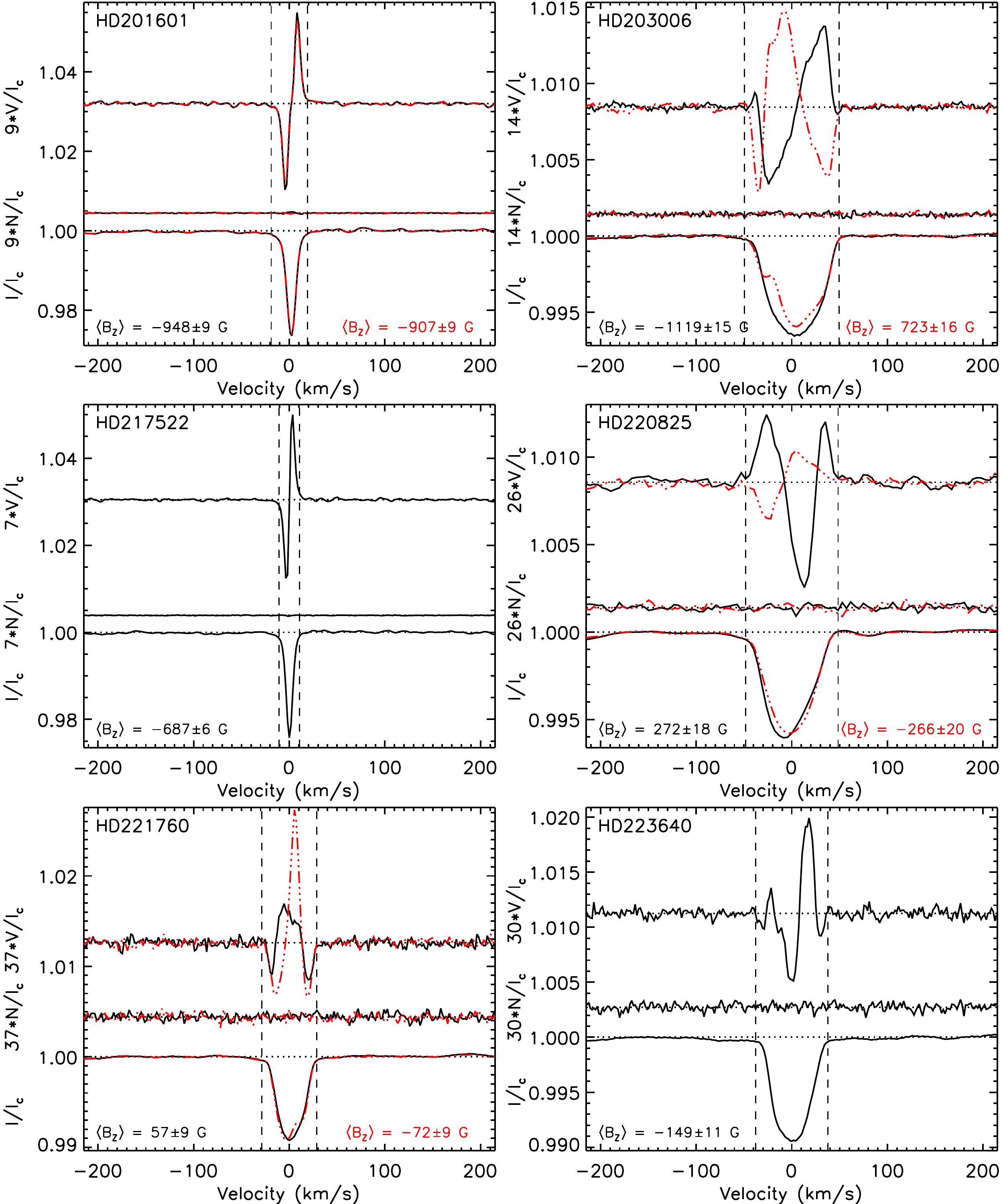}
	\caption{Examples of LSD profiles at one or two rotational phases.}
	\label{fig:lsd_ex_06}
\end{figure*}

\clearpage

\begin{figure*}
	\centering
	\includegraphics[width=2.1\columnwidth]{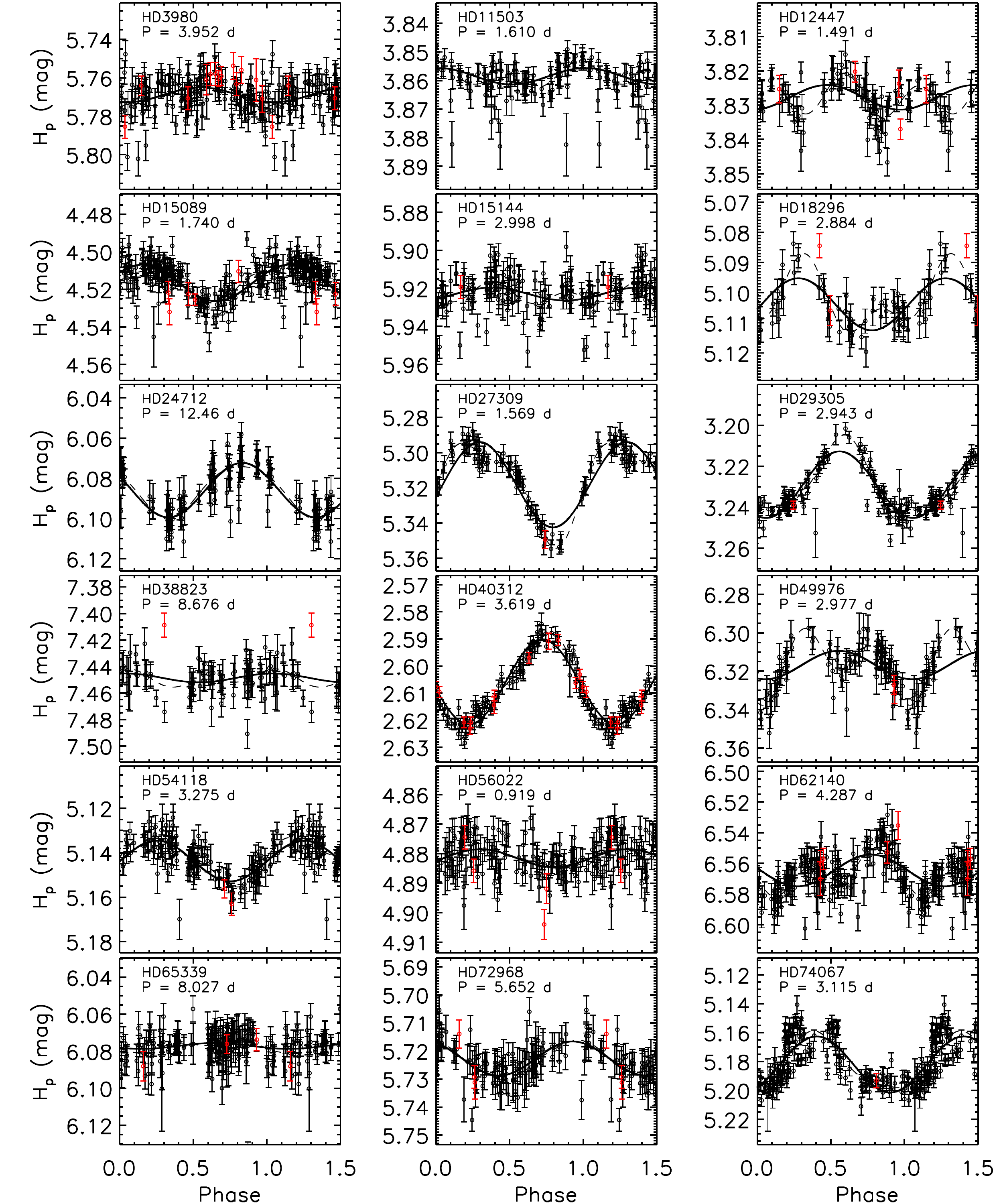}
	\caption{The Hipparcos Epoch Photometry associated with those mCP stars with known $P_{\rm rot}$.
	The solid black curves and dashed black curves correspond to the best $1^{\rm st}$ and $2^{\rm nd}$ 
	order sinusoidal fits (defined by Eqn. \ref{eqn:sin_fcn}). Note that the periods listed in each 
	figure are rounded and do not correspond to the actual $P_{\rm rot}$ precision.}
	\label{fig:phased_hipp01}
\end{figure*}

\begin{figure*}
	\centering
	\includegraphics[width=2.1\columnwidth]{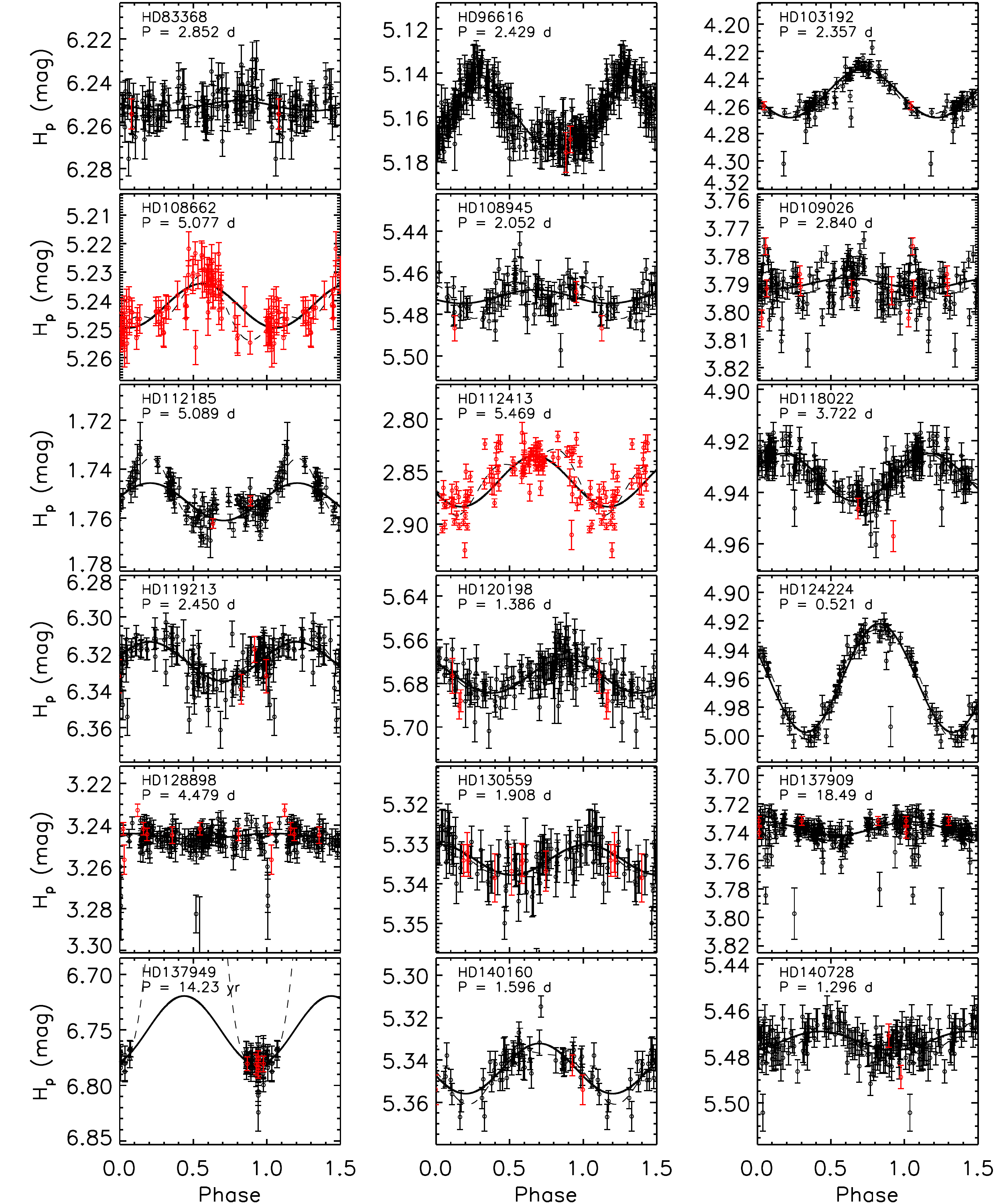}
	\caption{Continued from Fig. \ref{fig:phased_hipp01}.}
	\label{fig:phased_hipp02}
\end{figure*}

\begin{figure*}
	\centering
	\includegraphics[width=2.1\columnwidth]{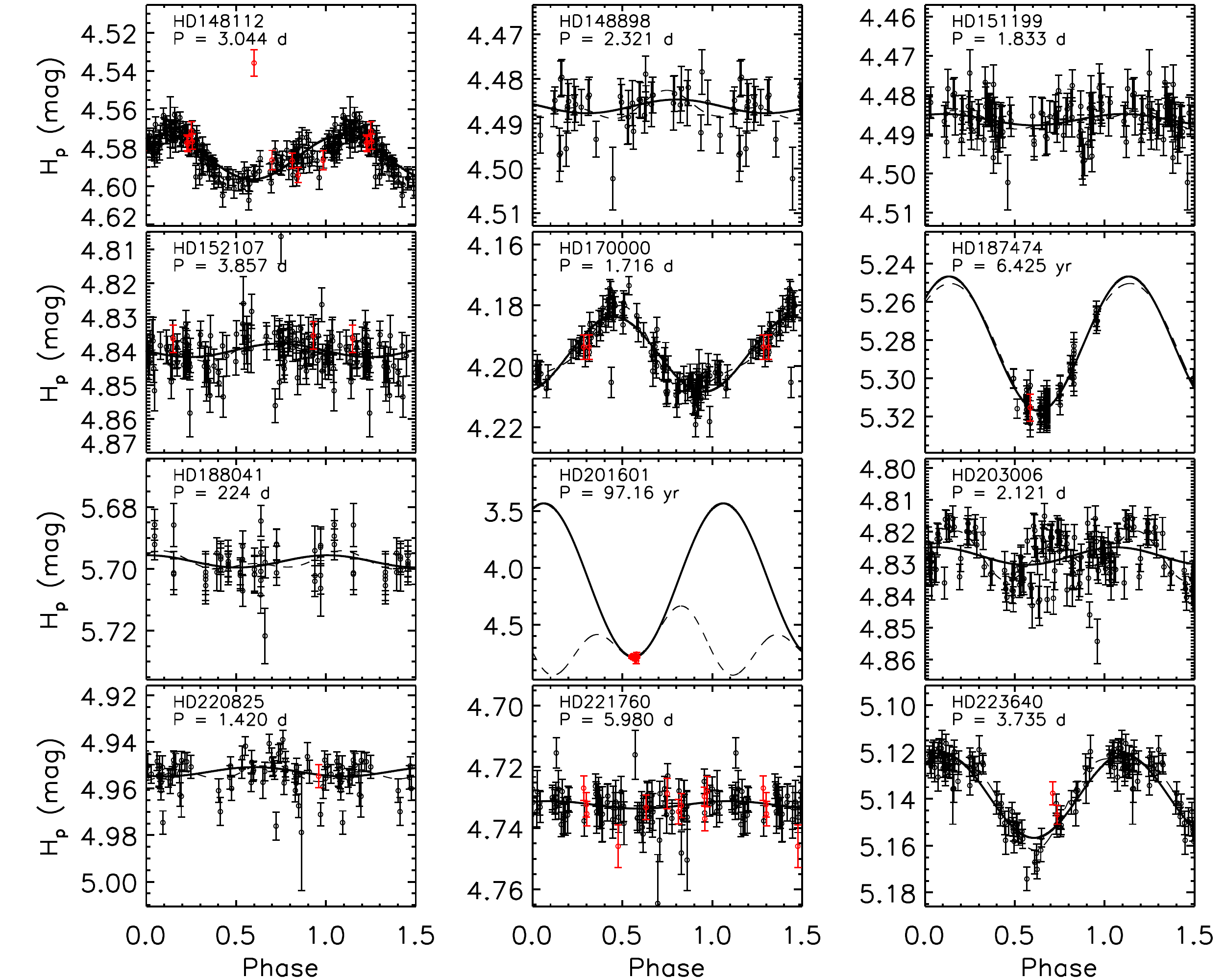}
	\caption{Continued from Fig. \ref{fig:phased_hipp02}.}
	\label{fig:phased_hipp03}
\end{figure*}

\begin{figure*}
	\centering
	\includegraphics[width=2.1\columnwidth]{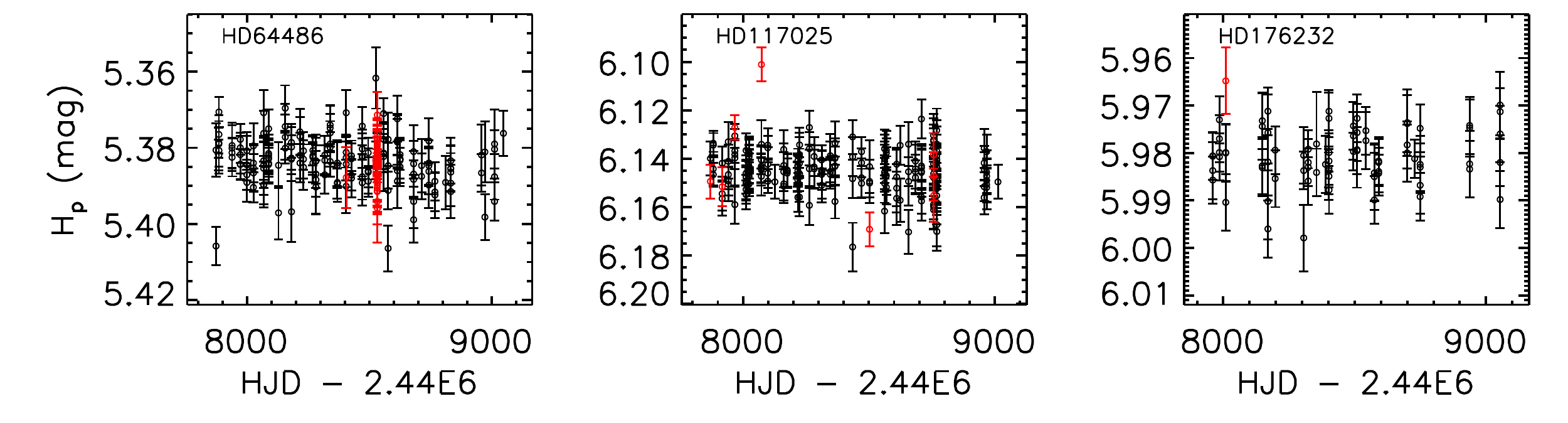}
	\caption{The Hipparcos Epoch Photometry associated with those stars without known $P_{\rm rot}$ 
	values.}
	\label{fig:HJD_hipp01}
\end{figure*}

\onecolumn
\input{tbl_det_obs_full.tex}
\twocolumn

\end{document}

%% file: tbl_det_obs.tex
\begin{table*}
	\caption{Observations of confirmed mCP stars -- those stars for which at least one definite 
	detection was obtained based on the criterion proposed by \citet{Donati1997}. Columns 1 to 5 contain 
	the HD number, instrument used to obtain the observation (ESP~=~ESPaDOnS, MUS~=~MuSiCoS, and 
	NAR~=~NARVAL), HJD, rotational phase, and the derived $\langle B_{\rm z}\rangle$ value and its 
	associated uncertainty. The full table will appear only in the electronic version of the paper.}
	\label{tbl:det_obs_tbl}
	\begin{center}
	\begin{tabular*}{2.1\columnwidth}{@{\extracolsep{\fill}}l c c c c @{\hskip 0.55cm} | @{\hskip 0.55cm} c c c c r@{\extracolsep{\fill}}}
		\noalign{\vskip-0.2cm}
		\hline
		\hline
		\noalign{\vskip0.5mm}
		HD & Inst. & HJD & Phase & $\langle B_z\rangle$ & HD & Inst. & HJD & Phase & $\langle B_z\rangle$ \\
		   &       &     &       & (G)                  &    &       &     &       & (G)                  \\
		\noalign{\vskip0.5mm}
		\hline	
		\noalign{\vskip0.5mm}
 15089 &      MUS &    3040.343 &      0.259 &           $223\pm93$ &        &      ESP &    7443.892 &      0.434 &           $-81\pm36$ \\
       &      MUS &    3586.543 &      0.077 &           $450\pm23$ &        &      ESP &    7447.848 &      0.739 &            $86\pm26$ \\
       &      MUS &    3589.652 &      0.864 &           $506\pm18$ &  72968 &      MUS &    3748.583 &      0.774 &        $346.4\pm8.4$ \\
       &      MUS &    3590.561 &      0.386 &          $-258\pm20$ &        &      MUS &    3749.549 &      0.945 &        $343.0\pm7.3$ \\
       &      MUS &    3591.597 &      0.981 &           $509\pm19$ &        &      MUS &    3755.429 &      0.985 &           $307\pm16$ \\
       &      MUS &    3594.551 &      0.678 &          $-166\pm24$ &        &      MUS &    3756.527 &      0.179 &        $323.5\pm7.0$ \\
       &      MUS &    3607.557 &      0.150 &           $441\pm32$ &        &      ESP &    7416.994 &      0.763 &        $334.1\pm5.2$ \\
       &      MUS &    3616.513 &      0.296 &            $11\pm24$ &        &      ESP &    7498.720 &      0.221 &        $335.2\pm3.8$ \\
 15144 &      MUS &    2253.385 &      0.716 &          $-568\pm13$ &        &      ESP &    7500.781 &      0.586 &        $266.8\pm2.5$ \\
       &      MUS &    2254.408 &      0.057 &          $-619\pm12$ &  74067 &      ESP &    7330.146 &      0.046 &          $1024\pm11$ \\
       &      MUS &    3410.324 &      0.620 &          $-567\pm18$ &        &      ESP &    7331.099 &      0.352 &          $-147\pm10$ \\
       &      MUS &    3613.503 &      0.392 &          $-551\pm15$ &        &      ESP &    7348.156 &      0.828 &           $748\pm38$ \\
       &      MUS &    3615.559 &      0.078 &          $-612\pm10$ &        &      ESP &    7415.993 &      0.605 &          $-303\pm12$ \\
       &      MUS &    3617.570 &      0.749 &          $-586\pm11$ &        &      ESP &    7440.875 &      0.592 &          $-370\pm34$ \\
 18296 &      ESP &    7556.127 &      0.190 &            $91\pm19$ &        &      ESP &    7445.857 &      0.192 &           $562\pm26$ \\
       &      ESP &    7561.124 &      0.923 &           $169\pm19$ &        &      ESP &    7446.838 &      0.506 &          $-480\pm28$ \\
       &      ESP &    7610.142 &      0.918 &        $195.7\pm9.7$ &  96616 &      ESP &    7358.169 &      0.668 &           $-58\pm15$ \\
 24712 &      MUS &     857.333 &      0.180 &           $765\pm13$ &        &      ESP &    7441.990 &      0.172 &           $213\pm16$ \\
       &      MUS &    1924.360 &      0.830 &           $763\pm12$ &        &      ESP &    7444.970 &      0.399 &          $-153\pm22$ \\
       &      MUS &    3247.675 &      0.052 &          $1033\pm17$ &        &      ESP &    7447.937 &      0.620 &          $-101\pm13$ \\
 56022 &      ESP &    7325.149 &      0.210 &           $139\pm32$ &        &      ESP &    7448.964 &      0.043 &           $325\pm14$ \\
       &      ESP &    7438.845 &      0.941 &           $195\pm24$ &        &          &             &            &                      \\
		\noalign{\vskip0.5mm}
		\hline \\
	\end{tabular*}
	\end{center}
\end{table*}

%% file: tbl_non_det_obs.tex
\begin{table*}
	\caption{Spectropolarimetric observations of those stars for which no Zeeman signatures were detected. 
	Columns 1 to 6 contain the HD number, instrument used to obtain the observation (ESP~=~ESPaDOnS, 
	MUS~=~MuSiCoS, NAR~=~NARVAL), HJD, exposure time, number of consecutive observations, and the derived 
	$\langle B_{\rm z}\rangle$ value and its associated uncertainty.}
	\label{tbl:non_det_obs_tbl}
	\begin{center}
	\begin{tabular*}{2.1\columnwidth}{@{\extracolsep{\fill}}l c c c c c @{\hskip 0.55cm} | @{\hskip 0.55cm} c c c c c r@{\extracolsep{\fill}}}
		\noalign{\vskip-0.2cm}
		\hline
		\hline
		\noalign{\vskip0.5mm}
		HD & Inst. &            HJD & $t_{\rm exp}$ (s) & \# & $\langle B_z\rangle$ (G) & HD & Inst. &            HJD & $t_{\rm exp}$ (s) & \# & $\langle B_z\rangle$ (G) \\
		   &       & $+2\,450\,000$ &                   &    &                          &    &       & $+2\,450\,000$ &                   &    &                          \\
		\noalign{\vskip0.5mm}
		\hline	
		\noalign{\vskip0.5mm}
   358 &      ESP &    7561.129 &    15 &   1 &            $31\pm19$ &        &      MUS &    3747.683 &  3200 &   1 &          $180\pm180$ \\
  4853 &      ESP &    7239.132 &   200 &   1 &            $24\pm19$ &        &      MUS &    3750.683 &  3200 &   1 &           $70\pm180$ \\
 27411 &      ESP &    7435.763 &     8 &   1 &             $0\pm11$ &        &      MUS &    3755.683 &  3200 &   1 &          $210\pm160$ \\
 27749 &      ESP &    7435.766 &     5 &   1 &          $5.4\pm9.5$ &        &      MUS &    3756.626 &  3200 &   1 &           $80\pm140$ \\
 67523 &      ESP &    7414.991 &     5 &   1 &          $0.5\pm1.9$ &        &      MUS &    3864.422 &  2400 &   1 &           $50\pm220$ \\
 78362 &      MUS &     858.604 &  1200 &   1 &         $-0.2\pm3.9$ &        &      MUS &    3874.440 &  2400 &   1 &          $-90\pm180$ \\
       &      MUS &    1202.554 &  1635 &   1 &            $-8\pm25$ &        &      MUS &    3885.401 &  2400 &   1 &          $120\pm240$ \\
       &      ESP &    7412.001 &     5 &   1 &         $-6.3\pm5.7$ &        &      MUS &    3892.366 &  2400 &   1 &          $-20\pm190$ \\
 90763 &      ESP &    7325.137 &     8 &   1 &            $34\pm54$ &        &      ESP &    7236.789 &   330 &   2 &            $39\pm26$ \\
       &      ESP &    7327.160 &     8 &   1 &            $79\pm47$ &        &      ESP &    7261.749 &   330 &   2 &           $-47\pm28$ \\
       &      ESP &    7328.157 &     8 &   1 &           $-18\pm50$ &        &      ESP &    7262.733 &   330 &   2 &            $10\pm25$ \\
       &      ESP &    7329.106 &     8 &   1 &            $60\pm53$ &        &      ESP &    7265.730 &   330 &   2 &             $3\pm43$ \\
       &      ESP &    7330.122 &    31 &   1 &             $6\pm27$ & 120025 &      ESP &    7414.075 &   123 &   1 &         $-4.6\pm8.2$ \\
       &      ESP &    7522.802 &    60 &   1 &            $12\pm23$ & 125335 &      ESP &    7408.155 &   200 &   1 &         $-4.0\pm3.2$ \\
102942 &      ESP &    7412.006 &    37 &   1 &            $-8\pm10$ & 136729 &      NAR &    7800.669 &  3188 &   1 &            $15\pm68$ \\
       &      ESP &    7497.913 &    37 &   1 &             $0\pm10$ & 139478 &      NAR &    7801.609 &  1176 &   1 &         $-2.9\pm7.6$ \\
       &      ESP &    7498.878 &   200 &   2 &         $-4.3\pm3.2$ & 149748 &      ESP &    7409.126 &   225 &   1 &            $20\pm13$ \\
       &      ESP &    7500.885 &   200 &   2 &         $-4.0\pm3.2$ & 156164 &      ESP &    7560.979 &    40 &   1 &          $190\pm290$ \\
105702 &      ESP &    7409.107 &     6 &   1 &            $22\pm17$ & 189849 &      ESP &    7476.130 &    19 &   1 &          $3.8\pm3.6$ \\
       &      ESP &    7495.949 &     6 &   2 &            $11\pm16$ & 202627 &      ESP &    7261.984 &   217 &   2 &            $14\pm16$ \\
       &      ESP &    7497.923 &     6 &   1 &           $-16\pm11$ & 206742 &      ESP &    7262.001 &    50 &   1 &            $-9\pm26$ \\
       &      ESP &    7498.955 &     6 &   1 &            $13\pm13$ &        &      ESP &    7262.001 &    50 &   1 &            $71\pm63$ \\
       &      ESP &    7500.902 &     6 &   1 &           $-15\pm11$ & 221675 &      ESP &    7554.124 &   100 &   1 &            $-3\pm13$ \\
115735 &      MUS &    1600.662 &  2595 &   1 &         $-100\pm190$ &        &          &             &       &     &                      \\
		\noalign{\vskip0.5mm}
		\hline \\
	\end{tabular*}
	\end{center}
\end{table*}

%% file: tbl_rot_param.tex
\renewcommand{\arraystretch}{1.2}
\begin{center}
\begin{longtable}{@{\extracolsep{\fill}}l c c c c c r@{\extracolsep{\fill}}}
\caption{Parameters associated with the $\langle B_{\rm z}\rangle$ curves shown in Figures 
	\ref{fig:phased_Bz01} and \ref{fig:phased_Bz02}. Columns 1 to 3 list each star's HD number, adopted 
	or derived rotational period, and the adopted epoch, respectively. References for those rotational 
	periods taken from the literature are listed in the table's footer; $P_{\rm rot}$ values without 
	references were derived in this study. Columns 4, 5, and 7 list the mean, amplitudes, and reduced 
	$\chi^2$ values associated with the first-order sinusoidal fits sinusoidal (i.e. $C_0$ and $C_1$ in 
	Eqn. \ref{eqn:sin_fcn} with $C_2\equiv0$). Column 6 lists the $r$ parameters 
	\citep[Eqn. 2 of][]{Preston1967}.} \label{tbl:rot_param_tbl} \\

\hline
\hline
\noalign{\vskip0.5mm}
HD  &  $P_{\rm rot}\,(d)$ & ${\rm HJD}_0-2.4\times10^6$ & $B_0\,({\rm G})$ & $B_1\,({\rm G})$ & $r$ & $\chi^2_{\rm red}$ \\
(1) &                 (2) &           (3) &              (4) &              (5) & (6) &                (7) \\
\noalign{\vskip0.5mm}
\hline	
\noalign{\vskip0.5mm}
\endfirsthead

\multicolumn{7}{l}{continued from previous page}\\
\hline
\hline
\noalign{\vskip0.5mm}
HD  &  $P_{\rm rot}\,(d)$ & ${\rm HJD}_0-2.4\times10^6$ & $B_0\,({\rm G})$ & $B_1\,({\rm G})$ & $r$ & $\chi^2_{\rm red}$ \\
(1) &                 (2) &           (3) &              (4) &              (5) & (6) &                (7) \\
\noalign{\vskip0.5mm}
\hline
\noalign{\vskip0.5mm}
\endhead

\noalign{\vskip0.5mm}\hline
\multicolumn{7}{l}{continued on next page}\\
\endfoot

\noalign{\vskip0.5mm}
\endlastfoot

3980   &         $3.9516(3)^{\,a}$ &         $40927.2031$ &         $120\pm1810$ &        $1710\pm4470$ &            $-0.87\pm0.44$ &        $1.5$ \\
11502  &              $1.60984(1)$ &        $43002.93904$ &         $-130\pm230$ &          $730\pm350$ &            $-0.69\pm0.21$ &        $3.8$ \\
12446  &        $1.4907(12)^{\,b}$ &         $43118.3498$ &           $-40\pm84$ &          $470\pm130$ &            $-0.84\pm0.07$ &        $0.8$ \\
15089  &       $1.74050(3)^{\,cd}$ &        $53039.89185$ &           $109\pm63$ &           $463\pm90$ &            $-0.62\pm0.10$ &        $6.6$ \\
15144  &              $2.99799(1)$ &        $52254.23776$ &       $-581.6\pm7.2$ &         $33.8\pm9.9$ &             $0.89\pm0.05$ &        $0.2$ \\
18296  &             $2.88416(15)$ &        $42999.22302$ &           $10\pm210$ &          $210\pm340$ &            $-0.94\pm0.16$ &        $0.7$ \\
24712  &       $12.4580(15)^{\,e}$ &         $47179.9838$ &          $560\pm160$ &          $510\pm250$ &             $0.04\pm0.39$ &        $3.6$ \\
27309  &     $1.5688840(47)^{\,f}$ &      $52247.1353483$ &          $-716\pm80$ &          $100\pm120$ &             $0.75\pm0.43$ &        $3.0$ \\
29305  &       $2.943176(3)^{\,g}$ &       $56967.257773$ &         $29.9\pm1.3$ &         $74.7\pm1.3$ &            $-0.43\pm0.01$ &       $<0.1$ \\
38823  &               $8.676(30)$ &          $51894.778$ &         $-460\pm430$ &         $2040\pm640$ &            $-0.63\pm0.16$ &       $24.7$ \\
40312  &              $3.61866(2)$ &        $42762.85334$ &            $91\pm44$ &           $307\pm62$ &            $-0.54\pm0.12$ &        $9.9$ \\
49976  &        $2.97666(8)^{\,h}$ &        $41401.97078$ &         $-380\pm610$ &         $1960\pm830$ &            $-0.68\pm0.20$ &        $3.2$ \\
54118  &       $3.27535(10)^{\,i}$ &        $42114.75746$ &           $30\pm250$ &         $1500\pm330$ &            $-0.96\pm0.01$ &        $1.5$ \\
56022  &        $0.91889(3)^{\,g}$ &        $57324.95641$ &            $79\pm39$ &           $142\pm75$ &            $-0.29\pm0.37$ &        $1.1$ \\
62140  &              $4.28677(3)$ &        $50505.89765$ &            $-5\pm57$ &          $1577\pm77$ &          $-0.993\pm0.001$ &       $13.7$ \\
65339  &        $8.02681(4)^{\,j}$ &        $50494.99521$ &          $-50\pm540$ &         $4740\pm840$ &          $-0.978\pm0.006$ &       $58.8$ \\
72968  &              $5.6525(10)$ &         $52251.9491$ &           $318\pm40$ &            $54\pm66$ &             $0.71\pm0.49$ &        $6.1$ \\
74067  &            $3.11511(226)$ &        $57326.88599$ &           $301\pm38$ &           $761\pm46$ &            $-0.43\pm0.04$ &        $1.6$ \\
83368  &       $2.851976(3)^{\,k}$ &       $45063.924739$ &          $-10\pm260$ &          $730\pm410$ &            $-0.97\pm0.03$ &        $1.7$ \\
96616  &              $2.42927(2)$ &        $57356.54706$ &            $79\pm18$ &           $263\pm25$ &            $-0.54\pm0.06$ &        $0.8$ \\
103192 &        $2.35666(2)^{\,i}$ &        $43736.07566$ &          $-206\pm68$ &            $38\pm99$ &               $0.7\pm1.2$ &        $0.3$ \\
108662 &             $5.07735(24)$ &        $42214.90968$ &         $-360\pm210$ &          $410\pm300$ &            $-0.06\pm0.61$ &       $53.5$ \\
108945 &             $2.05186(12)$ &        $51613.95547$ &           $-23\pm77$ &          $250\pm100$ &            $-0.83\pm0.11$ &        $1.2$ \\
109026 &          $2.84(22)^{\,l}$ &           $56336.96$ &           $309\pm19$ &           $170\pm29$ &             $0.29\pm0.13$ &        $0.7$ \\
112185 &       $5.0887(13)^{\,mn}$ &         $41794.5148$ &            $19\pm36$ &            $80\pm45$ &            $-0.62\pm0.31$ &        $1.9$ \\
112413 &              $5.46913(8)$ &        $50503.70120$ &          $-104\pm96$ &          $770\pm120$ &            $-0.76\pm0.06$ &      $184.8$ \\
118022 &       $3.722084(2)^{\,h}$ &       $50499.616970$ &          $-533\pm55$ &           $438\pm68$ &             $0.10\pm0.14$ &        $2.2$ \\
119213 &     $2.4499141(38)^{\,o}$ &      $53406.2587031$ &          $380\pm110$ &          $300\pm120$ &             $0.11\pm0.37$ &        $2.6$ \\
120198 &       $1.38576(80)^{\,p}$ &        $42769.49376$ &          $150\pm210$ &          $330\pm260$ &            $-0.36\pm0.61$ &        $1.1$ \\
124224 &  $0.52070308(120)^{\,qr}$ &     $42850.85176720$ &          $120\pm180$ &          $960\pm240$ &            $-0.78\pm0.09$ &        $7.0$ \\
128898 &         $4.4790(1)^{\,s}$ &         $42116.9439$ &         $-320\pm180$ &          $120\pm260$ &               $0.4\pm1.4$ &        $1.4$ \\
130559 &              $1.90798(1)$ &        $53407.61250$ &          $-280\pm25$ &           $208\pm32$ &             $0.15\pm0.13$ &        $2.0$ \\
137909 &       $18.4877(15)^{\,t}$ &         $46201.8254$ &           $60\pm150$ &          $710\pm190$ &            $-0.84\pm0.07$ &      $104.9$ \\
137949 &                    $5195$ &              $38166$ &         $1620\pm100$ &          $170\pm170$ &             $0.81\pm0.27$ &        $1.8$ \\
140160 &             $1.59587(11)$ &        $51607.01456$ &          $-10\pm150$ &          $320\pm180$ &            $-0.97\pm0.03$ &        $1.4$ \\
140728 &              $1.29559(2)$ &        $53864.86021$ &           $-27\pm35$ &           $514\pm42$ &            $-0.90\pm0.01$ &        $0.3$ \\
148112 &            $3.04416(112)$ &        $52094.28900$ &          $-180\pm23$ &            $33\pm35$ &             $0.69\pm0.46$ &        $1.2$ \\
148898 &               $2.3205(2)$ &         $52764.4371$ &           $238\pm83$ &          $390\pm110$ &            $-0.25\pm0.23$ &       $<0.1$ \\
151199 &             $1.83317(22)$ &        $53366.50581$ &           $-81\pm65$ &           $198\pm92$ &            $-0.42\pm0.33$ &        $0.8$ \\
152107 &      $3.857500(15)^{\,u}$ &       $53600.975034$ &           $961\pm49$ &           $357\pm64$ &             $0.46\pm0.13$ &        $9.8$ \\
170000 &        $1.71649(2)^{\,v}$ &        $42632.30626$ &           $123\pm60$ &           $370\pm82$ &            $-0.50\pm0.14$ &        $4.5$ \\
187474 &          $2345(15)^{\,w}$ &              $45534$ &          $-50\pm300$ &         $2120\pm420$ &            $-0.96\pm0.01$ &        $2.0$ \\
188041 &          $224.0(2)^{\,w}$ &            $46319.5$ &         $1140\pm210$ &          $220\pm430$ &             $0.68\pm0.79$ &        $1.6$ \\
201601 &        $35462.5(6)^{\,x}$ &            $52457.1$ &         $-570\pm560$ &          $580\pm680$ &              $-0.0\pm1.1$ &        $1.5$ \\
203006 &            $2.12073(135)$ &        $57238.62987$ &           $-11\pm48$ &          $1137\pm66$ &          $-0.981\pm0.002$ &        $4.8$ \\
220825 &             $1.42020(18)$ &        $52095.27809$ &            $73\pm46$ &           $340\pm60$ &            $-0.65\pm0.09$ &        $1.1$ \\
221760 &                 $5.98(6)$ &           $52790.80$ &             $8\pm15$ &            $80\pm15$ &            $-0.82\pm0.06$ &       $<0.1$ \\
223640 &      $3.735239(24)^{\,y}$ &       $42828.902150$ &          $420\pm350$ &          $480\pm420$ &            $-0.06\pm0.81$ &       $19.3$ \\
		\hline \\
\noalign{\vskip-0.3cm}
\multicolumn{7}{l}{$^{\rm a\,}$\citet{Maitzen1980}, $^{\rm b\,}$\citet{Borra1980}, $^{\rm c\,}$\citet{Musielok1980}} \\
\multicolumn{7}{l}{$^{\rm d\,}$\citet{Jasinski1981}, $^{\rm e\,}$\citet{Kurtz1982}, $^{\rm f\,}$\citet{North1995}} \\
\multicolumn{7}{l}{$^{\rm g\,}$\citet{Heck1987}, $^{\rm h\,}$\citet{Catalano1994}, $^{\rm i\,}$\citet{Manfroid1994}} \\
\multicolumn{7}{l}{$^{\rm j\,}$\citet{Hill1998}, $^{\rm k\,}$\citet{Kurtz1997}, $^{\rm l\,}$\citet{Alecian2014}} \\
\multicolumn{7}{l}{$^{\rm m\,}$\citet{Deutsch1947}, $^{\rm n\,}$\citet{Bohlender1990}, $^{\rm o\,}$\citet{Ziznovsky1995}} \\
\multicolumn{7}{l}{$^{\rm p\,}$\citet{Wade1998}, $^{\rm q\,}$\citet{Pyper1998}, $^{\rm r\,}$\citet{Sokolov2000}} \\
\multicolumn{7}{l}{$^{\rm s\,}$\citet{Kurtz1994}, $^{\rm t\,}$\citet{Bagnulo1999}, $^{\rm u\,}$\citet{Schoeneich1988}} \\
\multicolumn{7}{l}{$^{\rm v\,}$\citet{Musielok1986}, $^{\rm w\,}$\citet{Mathys1991}, $^{\rm x\,}$\citet{Bychkov2016}} \\
\multicolumn{7}{l}{$^{\rm y\,}$\citet{North1992}} \\

\end{longtable}
\end{center}
\renewcommand{\arraystretch}{1.0}

%% file: tbl_mag_param.tex
\renewcommand{\arraystretch}{1.2}
\begin{table*}
	\caption{Parameters associated with the magnetic field geometries and strengths. Columns 2 to 3 list 
	the inclination angles and obliquity angles. Columns 4 to 6 list the dipole field strengths 
	($B_{\rm d}$), critical field strengths ($B_{\rm c}$), and ratios of $B_{\rm d}$ to $B_{\rm c}$.}
	\label{tbl:mag_param_tbl}
	\begin{center}
	\begin{tabular*}{2.0\columnwidth}{@{\extracolsep{\fill}}l c c c c r@{\extracolsep{\fill}}}
		\noalign{\vskip-0.2cm}
		\hline
		\hline
		\noalign{\vskip0.5mm}
		HD  &  $i\,(\degree)$ & $\beta\,(\degree)$ &  $B_{\rm d}\,({\rm G})$ & $B_{\rm c}\,({\rm G})$ & $B_{\rm d}/B_{\rm c}$ \\
		(1) &             (2) &                (3) &                     (4) &                    (5) &                   (6) \\
		\noalign{\vskip0.5mm}
		\hline	
		\noalign{\vskip0.5mm}
3980   &           $84_{-32}^{+4}$ &           $84_{-82}^{+3}$ &   $6360_{-5570}^{+57570}$ &         $285_{-33}^{+41}$ &         $22_{-19}^{+206}$ \\
11502  &          $76_{-24}^{+13}$ &          $54_{-50}^{+27}$ &    $3000_{-750}^{+29130}$ &        $652_{-99}^{+139}$ &      $4.6_{-1.4}^{+47.9}$ \\
12446  &           $38_{-9}^{+13}$ &            $86_{-4}^{+3}$ &      $2450_{-520}^{+710}$ &      $1010_{-190}^{+220}$ &       $2.4_{-0.4}^{+0.5}$ \\
15089  &          $56_{-14}^{+23}$ &           $71_{-30}^{+8}$ &      $1850_{-160}^{+490}$ &         $725_{-61}^{+69}$ &       $2.5_{-0.3}^{+0.7}$ \\
15144  &                  $20\pm5$ &             $9_{-2}^{+3}$ &        $1951_{-45}^{+73}$ &         $425_{-38}^{+42}$ &       $4.6_{-0.5}^{+0.6}$ \\
18296  &           $29_{-9}^{+12}$ &     $89.0_{-13.7}^{+0.4}$ &     $1430_{-830}^{+1090}$ &        $567_{-99}^{+119}$ &       $2.5_{-1.5}^{+1.8}$ \\
24712  &          $43_{-10}^{+11}$ &          $45_{-12}^{+11}$ &      $3340_{-280}^{+380}$ &            $96_{-6}^{+7}$ &      $34.5_{-4.0}^{+4.6}$ \\
27309  &          $49_{-10}^{+16}$ &                   $7\pm4$ &     $3600_{-580}^{+1980}$ &       $780_{-120}^{+140}$ &       $4.6_{-1.3}^{+3.4}$ \\
29305  &          $54_{-11}^{+16}$ &           $61_{-19}^{+8}$ &          $349_{-4}^{+44}$ &         $564_{-58}^{+61}$ &       $0.6_{-0.1}^{+0.1}$ \\
38823  &           $80_{-53}^{+7}$ &          $71_{-63}^{+11}$ &    $7590_{-460}^{+39130}$ &         $114_{-13}^{+20}$ &         $67_{-11}^{+324}$ \\
40312  &          $63_{-13}^{+22}$ &          $59_{-45}^{+12}$ &      $1291_{-94}^{+2800}$ &         $724_{-63}^{+68}$ &       $1.8_{-0.2}^{+4.1}$ \\
49976  &          $69_{-29}^{+18}$ &          $63_{-57}^{+20}$ &   $7530_{-1690}^{+48620}$ &         $382_{-43}^{+46}$ &    $19.7_{-4.6}^{+134.5}$ \\
54118  &          $58_{-15}^{+27}$ &    $88.2_{-15.2}^{+-0.1}$ &     $5810_{-960}^{+1380}$ &         $432_{-43}^{+50}$ &      $13.4_{-2.1}^{+3.1}$ \\
56022  &          $50_{-17}^{+26}$ &          $56_{-34}^{+15}$ &        $712_{-91}^{+476}$ &        $1252_{-85}^{+88}$ &       $0.6_{-0.1}^{+0.4}$ \\
62140  &          $70_{-19}^{+18}$ &    $89.5_{-11.8}^{+-0.2}$ &     $5110_{-330}^{+1050}$ &         $349_{-58}^{+68}$ &      $14.6_{-1.6}^{+2.0}$ \\
65339  &          $55_{-11}^{+18}$ &      $89.1_{-3.9}^{+0.7}$ &   $18120_{-2700}^{+3540}$ &                $186\pm22$ &          $97_{-11}^{+15}$ \\
72968  &          $51_{-11}^{+18}$ &                   $8\pm5$ &     $1620_{-290}^{+1250}$ &         $241_{-33}^{+36}$ &       $6.7_{-1.8}^{+6.5}$ \\
74067  &          $58_{-20}^{+28}$ &          $57_{-48}^{+15}$ &     $3439_{-70}^{+12198}$ &         $440_{-49}^{+55}$ &      $7.8_{-0.8}^{+28.8}$ \\
83368  &          $69_{-10}^{+17}$ &     $87.7_{-31.3}^{+0.0}$ &      $2400_{-470}^{+540}$ &         $453_{-36}^{+42}$ &       $5.3_{-1.1}^{+1.1}$ \\
96616  &          $74_{-20}^{+14}$ &          $44_{-39}^{+24}$ &     $1260_{-210}^{+7780}$ &         $693_{-80}^{+89}$ &      $1.8_{-0.5}^{+11.8}$ \\
103192 &           $60_{-9}^{+14}$ &             $6_{-5}^{+6}$ &     $1380_{-320}^{+1080}$ &        $856_{-92}^{+104}$ &       $1.6_{-0.5}^{+1.6}$ \\
108662 &           $80_{-32}^{+8}$ &          $17_{-16}^{+30}$ &    $3110_{-770}^{+54640}$ &         $238_{-28}^{+31}$ &    $12.5_{-3.2}^{+225.1}$ \\
108945 &           $80_{-26}^{+9}$ &           $80_{-73}^{+7}$ &      $870_{-150}^{+4980}$ &         $782_{-56}^{+58}$ &       $1.1_{-0.2}^{+6.3}$ \\
109026 &            $15_{-6}^{+8}$ &          $65_{-12}^{+10}$ &     $2480_{-660}^{+1590}$ &       $490_{-130}^{+190}$ &       $5.0_{-1.2}^{+2.6}$ \\
112185 &          $56_{-11}^{+16}$ &          $71_{-21}^{+12}$ &         $327_{-62}^{+79}$ &         $460_{-42}^{+43}$ &       $0.7_{-0.1}^{+0.2}$ \\
112413 &          $48_{-21}^{+35}$ &           $82_{-42}^{+4}$ &     $3460_{-690}^{+2290}$ &         $245_{-36}^{+39}$ &      $14.1_{-3.2}^{+8.8}$ \\
118022 &            $27_{-5}^{+6}$ &            $58_{-7}^{+6}$ &      $3650_{-370}^{+610}$ &                $308\pm22$ &      $11.9_{-1.3}^{+2.0}$ \\
119213 &          $60_{-26}^{+27}$ &          $25_{-23}^{+24}$ &    $2620_{-660}^{+28240}$ &         $497_{-66}^{+77}$ &      $5.3_{-1.8}^{+58.8}$ \\
120198 &           $48_{-8}^{+11}$ &          $63_{-16}^{+13}$ &      $1600_{-360}^{+410}$ &       $890_{-120}^{+140}$ &       $1.8_{-0.4}^{+0.5}$ \\
124224 &           $46_{-8}^{+10}$ &            $82_{-5}^{+4}$ &      $4460_{-630}^{+780}$ &      $2020_{-200}^{+240}$ &       $2.2_{-0.3}^{+0.3}$ \\
128898 &            $42_{-6}^{+7}$ &          $23_{-16}^{+15}$ &      $1430_{-270}^{+310}$ &         $266_{-20}^{+22}$ &       $5.4_{-1.1}^{+1.5}$ \\
130559 &            $18_{-5}^{+6}$ &            $67_{-7}^{+6}$ &      $2360_{-440}^{+770}$ &        $716_{-89}^{+100}$ &       $3.3_{-0.6}^{+1.0}$ \\
137909 &           $84_{-27}^{+5}$ &          $75_{-70}^{+11}$ &    $2380_{-230}^{+25570}$ &         $111_{-16}^{+17}$ &    $19.7_{-2.3}^{+220.0}$ \\
137949 &                         - &                         - &                         - &             $0.27\pm0.03$ &                         - \\
140160 &          $60_{-11}^{+18}$ &    $88.4_{-20.7}^{+-0.1}$ &      $1180_{-260}^{+290}$ &         $811_{-62}^{+69}$ &       $1.5_{-0.3}^{+0.3}$ \\
140728 &          $46_{-10}^{+13}$ &            $87_{-3}^{+1}$ &      $2300_{-360}^{+460}$ &      $1080_{-140}^{+170}$ &       $2.1_{-0.3}^{+0.3}$ \\
148112 &          $58_{-16}^{+27}$ &                   $6\pm6$ &     $1090_{-340}^{+5680}$ &         $650_{-79}^{+83}$ &       $1.7_{-0.7}^{+9.3}$ \\
148898 &                  $30\pm5$ &            $71_{-5}^{+4}$ &      $2580_{-330}^{+440}$ &         $802_{-62}^{+67}$ &       $3.2_{-0.4}^{+0.5}$ \\
151199 &          $61_{-13}^{+23}$ &          $53_{-42}^{+14}$ &      $880_{-140}^{+2010}$ &         $684_{-80}^{+87}$ &       $1.3_{-0.3}^{+3.1}$ \\
152107 &          $50_{-12}^{+17}$ &                  $17\pm8$ &     $4930_{-690}^{+3040}$ &                $359\pm22$ &      $13.7_{-2.3}^{+8.8}$ \\
170000 &            $48_{-4}^{+5}$ &            $70_{-5}^{+4}$ &      $1750_{-160}^{+140}$ &        $1142_{-56}^{+49}$ &       $1.5_{-0.1}^{+0.1}$ \\
187474 &           $86_{-32}^{+3}$ &          $72_{-42}^{+16}$ &     $7210_{-710}^{+6310}$ &    $0.62_{-0.08}^{+0.10}$ &   $11590_{-1810}^{+9550}$ \\
188041 &                         - &                         - &                         - &    $6.36_{-0.65}^{+0.71}$ &                         - \\
201601 &                         - &                         - &                         - &           $0.038\pm0.003$ &                         - \\
203006 &          $51_{-11}^{+18}$ &      $89.3_{-1.5}^{+0.6}$ &      $4640_{-750}^{+890}$ &       $653_{-100}^{+118}$ &       $7.1_{-0.8}^{+1.0}$ \\
220825 &           $40_{-9}^{+10}$ &            $80_{-5}^{+3}$ &      $1700_{-260}^{+400}$ &         $722_{-66}^{+79}$ &       $2.4_{-0.3}^{+0.5}$ \\
221760 &            $47_{-7}^{+9}$ &            $84_{-5}^{+4}$ &         $343_{-43}^{+47}$ &         $373_{-33}^{+34}$ &               $0.9\pm0.1$ \\
223640 &           $84_{-28}^{+5}$ &          $12_{-12}^{+27}$ &  $3740_{-1030}^{+128220}$ &         $315_{-35}^{+37}$ &    $11.3_{-3.3}^{+395.1}$ \\
		\noalign{\vskip0.5mm}
		\hline \\
	\end{tabular*}
	\end{center}
\end{table*}
\renewcommand{\arraystretch}{1.0}

%% file: tbl_det_obs_full.tex
\tablecaption{Observations of confirmed mCP stars -- those stars for which at least one definite 
	detection was obtained based on the criterion proposed by \citet{Donati1997}. Columns 1 to 5 contain 
	the HD number, instrument used to obtain the observation (ESP~=~ESPaDOnS, MUS~=~MuSiCoS, and 
	NAR~=~NARVAL), HJD, rotational phase, and the derived $\langle B_{\rm z}\rangle$ value and its 
	associated uncertainty.}\label{tbl:det_obs_tbl_full}
\tablefirsthead{
	\hline
	\hline
	\noalign{\vskip0.5mm}
		HD & Inst. & HJD & Phase & $\langle B_z\rangle$ & HD & Inst. & HJD & Phase & $\langle B_z\rangle$ \\
		   &       &     &       & (kG)                 &    &       &     &       & (kG)                 \\
	\noalign{\vskip0.5mm}
	\hline
	\noalign{\vskip0.5mm}
}
\tablehead{
	\multicolumn{6}{l}{continued from previous page}\\
	\hline
	\hline
	\noalign{\vskip0.5mm}
		HD & Inst. & HJD & Phase & $\langle B_z\rangle$ & HD & Inst. & HJD & Phase & $\langle B_z\rangle$ \\
		   &       &     &       & (kG)                 &    &       &     &       & (kG)                 \\
	\noalign{\vskip0.5mm}
	\hline	
	\noalign{\vskip0.5mm}
}
\tabletail{
	\noalign{\vskip0.5mm}\hline
	\multicolumn{6}{l}{continued on next page}\\
}
\tablelasttail{
	\noalign{\vskip0.5mm}\hline
}
\centering
\begin{mpsupertabular*}{1.0\columnwidth}{@{\extracolsep{\fill}}l c c c c @{\hskip 1.1cm} c c c c r@{\extracolsep{\fill}}}
 15089 &      MUS &    3040.343 &      0.259 &           $223\pm93$ &        &      MUS &    3883.398 &      0.308 &           $151\pm68$ \\
       &      MUS &    3586.543 &      0.077 &           $450\pm23$ &        &      MUS &    3891.405 &      0.489 &           $526\pm93$ \\
       &      MUS &    3589.652 &      0.864 &           $506\pm18$ &        &      MUS &    3894.399 &      0.799 &          $-188\pm73$ \\
       &      MUS &    3590.561 &      0.386 &          $-258\pm20$ &        &      ESP &    7264.753 &      0.204 &          $-177\pm24$ \\
       &      MUS &    3591.597 &      0.981 &           $509\pm19$ &        &      ESP &    7284.802 &      0.679 &           $196\pm21$ \\
       &      MUS &    3594.551 &      0.678 &          $-166\pm24$ &        &      ESP &    7285.774 &      0.430 &           $463\pm40$ \\
       &      MUS &    3607.557 &      0.150 &           $441\pm32$ &        &      ESP &    7287.736 &      0.944 &          $-500\pm23$ \\
       &      MUS &    3616.513 &      0.296 &            $11\pm24$ &        &      ESP &    7289.704 &      0.463 &           $464\pm24$ \\
 15144 &      MUS &    2253.385 &      0.716 &          $-568\pm13$ & 148112 &      MUS &    2856.386 &      0.347 &          $-168\pm20$ \\
       &      MUS &    2254.408 &      0.057 &          $-619\pm12$ &        &      MUS &    3586.413 &      0.159 &          $-236\pm25$ \\
       &      MUS &    3410.324 &      0.620 &          $-567\pm18$ &        &      MUS &    3588.399 &      0.812 &          $-167\pm14$ \\
       &      MUS &    3613.503 &      0.392 &          $-551\pm15$ &        &      MUS &    3589.394 &      0.139 &          $-233\pm33$ \\
       &      MUS &    3615.559 &      0.078 &          $-612\pm10$ &        &      MUS &    3598.384 &      0.092 &          $-225\pm14$ \\
       &      MUS &    3617.570 &      0.749 &          $-586\pm11$ &        &      MUS &    3618.364 &      0.655 &          $-170\pm18$ \\
 18296 &      ESP &    7556.127 &      0.190 &            $91\pm19$ &        &      ESP &    7523.032 &      0.330 &          $-144\pm10$ \\
       &      ESP &    7561.124 &      0.923 &           $169\pm19$ &        &      ESP &    7560.973 &      0.794 &          $-218\pm15$ \\
       &      ESP &    7610.142 &      0.918 &        $195.7\pm9.7$ &        &      ESP &    7563.737 &      0.702 &          $-170\pm10$ \\
 24712 &      MUS &     857.333 &      0.180 &           $765\pm13$ & 148898 &      ESP &    7264.745 &      0.370 &           $-30\pm18$ \\
       &      MUS &    1924.360 &      0.830 &           $763\pm12$ &        &      ESP &    7285.720 &      0.409 &           $-94\pm20$ \\
       &      MUS &    3247.675 &      0.052 &          $1033\pm17$ &        &      ESP &    7287.729 &      0.275 &           $176\pm13$ \\
 56022 &      ESP &    7325.149 &      0.210 &           $139\pm32$ &        &      ESP &    7297.700 &      0.571 &       $-117.7\pm9.8$ \\
       &      ESP &    7438.845 &      0.941 &           $195\pm24$ & 151199 &      MUS &    3864.579 &      0.701 &           $-26\pm56$ \\
       &      ESP &    7443.892 &      0.434 &           $-81\pm36$ &        &      MUS &    3866.496 &      0.746 &            $18\pm57$ \\
       &      ESP &    7447.848 &      0.739 &            $86\pm26$ &        &      MUS &    3874.543 &      0.136 &          $-179\pm47$ \\
 72968 &      MUS &    3748.583 &      0.774 &        $346.4\pm8.4$ &        &      MUS &    3880.530 &      0.402 &            $49\pm48$ \\
       &      MUS &    3749.549 &      0.945 &        $343.0\pm7.3$ &        &      MUS &    3883.512 &      0.029 &          $-246\pm50$ \\
       &      MUS &    3755.429 &      0.985 &           $307\pm16$ &        &      MUS &    3885.510 &      0.119 &          $-225\pm49$ \\
       &      MUS &    3756.527 &      0.179 &        $323.5\pm7.0$ &        &      MUS &    3890.483 &      0.831 &          $-244\pm50$ \\
       &      ESP &    7416.994 &      0.763 &        $334.1\pm5.2$ &        &      MUS &    3893.479 &      0.466 &            $80\pm74$ \\
       &      ESP &    7498.720 &      0.221 &        $335.2\pm3.8$ &        &      ESP &    7264.762 &      0.511 &           $120\pm27$ \\
       &      ESP &    7500.781 &      0.586 &        $266.8\pm2.5$ & 152107 &      MUS &    3601.389 &      0.107 &          $1278\pm16$ \\
 74067 &      ESP &    7330.146 &      0.046 &          $1024\pm11$ &        &      MUS &    3617.372 &      0.251 &           $912\pm12$ \\
       &      ESP &    7331.099 &      0.352 &          $-147\pm10$ &        &      MUS &    3747.719 &      0.041 &          $1349\pm19$ \\
       &      ESP &    7348.156 &      0.828 &           $748\pm38$ &        &      MUS &    3758.748 &      0.900 &          $1194\pm17$ \\
       &      ESP &    7415.993 &      0.605 &          $-303\pm12$ &        &      MUS &    3759.711 &      0.150 &          $1164\pm16$ \\
       &      ESP &    7440.875 &      0.592 &          $-370\pm34$ &        &      MUS &    3760.763 &      0.423 &           $636\pm19$ \\
       &      ESP &    7445.857 &      0.192 &           $562\pm26$ &        &      MUS &    3761.760 &      0.681 &           $850\pm13$ \\
       &      ESP &    7446.838 &      0.506 &          $-480\pm28$ &        &      MUS &    3774.753 &      0.049 &          $1365\pm16$ \\
 96616 &      ESP &    7358.169 &      0.668 &           $-58\pm15$ &        &      MUS &    3866.586 &      0.856 &          $1114\pm15$ \\
       &      ESP &    7441.990 &      0.172 &           $213\pm16$ &        &      MUS &    3872.548 &      0.401 &           $683\pm14$ \\
       &      ESP &    7444.970 &      0.399 &          $-153\pm22$ & 170000 &      MUS &    3601.424 &      0.435 &          $-186\pm56$ \\
       &      ESP &    7447.937 &      0.620 &          $-101\pm13$ &        &      MUS &    3607.362 &      0.894 &          $560\pm130$ \\
       &      ESP &    7448.964 &      0.043 &           $325\pm14$ &        &      MUS &    3608.371 &      0.482 &          $-206\pm86$ \\
108662 &      MUS &    1612.541 &      0.893 &          $-647\pm11$ &        &      MUS &    3615.369 &      0.559 &          $-179\pm59$ \\
       &      MUS &    3747.645 &      0.408 &            $44\pm13$ &        &      MUS &    3872.511 &      0.366 &         $-250\pm110$ \\
       &      MUS &    3750.726 &      0.015 &          $-625\pm11$ &        &      MUS &    3874.576 &      0.569 &           $-98\pm82$ \\
       &      MUS &    3756.707 &      0.193 &          $-645\pm12$ &        &      MUS &    3891.554 &      0.460 &          $-176\pm68$ \\
       &      MUS &    3757.581 &      0.365 &          $-200\pm14$ &        &      NAR &    7621.404 &      0.412 &          $-236\pm23$ \\
       &      MUS &    3758.588 &      0.564 &           $262\pm17$ &        &      NAR &    7622.371 &      0.975 &           $403\pm22$ \\
       &      MUS &    3759.749 &      0.792 &          $-633\pm13$ &        &      NAR &    7623.420 &      0.587 &          $-195\pm30$ \\
       &      MUS &    3864.378 &      0.399 &            $-4\pm15$ &        &      NAR &    7624.424 &      0.171 &           $436\pm22$ \\
       &      MUS &    3891.370 &      0.715 &          $-398\pm17$ &        &      NAR &    7625.417 &      0.750 &           $163\pm25$ \\
108945 &      MUS &    1612.576 &      0.328 &           $111\pm42$ &        &      NAR &    7630.451 &      0.682 &           $-10\pm29$ \\
       &      MUS &    3405.675 &      0.217 &            $-7\pm49$ &        &      NAR &    7631.520 &      0.305 &           $-42\pm21$ \\
       &      MUS &    3410.674 &      0.653 &            $18\pm51$ &        &      NAR &    7634.390 &      0.978 &           $431\pm23$ \\
       &      MUS &    3752.664 &      0.327 &           $187\pm51$ &        &      NAR &    7635.370 &      0.548 &          $-148\pm27$ \\
       &      MUS &    3755.575 &      0.745 &           $-22\pm94$ &        &      NAR &    7638.400 &      0.314 &           $-80\pm22$ \\
       &      MUS &    3756.589 &      0.239 &          $-120\pm39$ &        &      NAR &    7639.406 &      0.900 &           $369\pm24$ \\
112185 &      MUS &    3748.683 &      0.160 &            $84\pm10$ &        &      NAR &    7640.393 &      0.475 &          $-288\pm21$ \\
       &      MUS &    3749.648 &      0.349 &        $-47.9\pm8.6$ &        &      NAR &    7644.313 &      0.758 &           $147\pm27$ \\
       &      MUS &    3752.737 &      0.956 &        $101.1\pm5.9$ &        &      NAR &    7734.235 &      0.145 &           $497\pm31$ \\
112413 &      MUS &    3202.378 &      0.438 &           $734\pm11$ &        &      NAR &    7735.233 &      0.727 &            $79\pm29$ \\
119213 &      MUS &    3406.516 &      0.105 &           $641\pm50$ &        &      NAR &    7736.233 &      0.309 &           $-68\pm21$ \\
       &      MUS &    3409.673 &      0.394 &            $68\pm53$ &        &      NAR &    7737.234 &      0.892 &           $393\pm24$ \\
       &      MUS &    3411.615 &      0.186 &           $517\pm44$ &        &      NAR &    7801.720 &      0.461 &          $-299\pm24$ \\
       &      MUS &    3412.613 &      0.594 &           $169\pm55$ &        &      NAR &    7802.707 &      0.036 &           $459\pm22$ \\
       &      MUS &    3746.637 &      0.935 &           $647\pm40$ &        &      NAR &    7803.706 &      0.618 &           $-77\pm26$ \\
       &      MUS &    3749.679 &      0.176 &           $577\pm31$ &        &      NAR &    7804.696 &      0.195 &           $371\pm25$ \\
       &      MUS &    3752.704 &      0.411 &            $79\pm36$ & 176232 &      MUS &    2827.591 &            &        $396.4\pm8.9$ \\
       &      MUS &    3757.619 &      0.417 &            $15\pm41$ &        &      MUS &    2835.485 &            &           $383\pm11$ \\
       &      MUS &    3758.625 &      0.828 &           $453\pm55$ &        &      MUS &    2854.413 &            &           $401\pm52$ \\
       &      MUS &    3760.664 &      0.660 &           $299\pm51$ &        &      MUS &    3217.404 &            &           $363\pm42$ \\
       &      MUS &    3761.640 &      0.059 &           $373\pm78$ &        &      MUS &    3585.441 &            &        $371.7\pm8.6$ \\
       &      MUS &    3762.668 &      0.478 &            $13\pm91$ &        &      MUS &    3589.436 &            &        $369.0\pm8.6$ \\
       &      MUS &    3767.676 &      0.522 &           $149\pm40$ &        &      MUS &    3595.388 &            &        $365.3\pm6.8$ \\
       &      MUS &    3768.625 &      0.910 &          $580\pm310$ &        &      MUS &    3881.556 &            &           $344\pm20$ \\
       &      MUS &    3769.647 &      0.327 &           $173\pm31$ &        &      MUS &    3887.492 &            &        $349.2\pm8.3$ \\
       &      MUS &    3773.574 &      0.930 &           $643\pm38$ &        &      MUS &    3892.524 &            &        $346.6\pm7.4$ \\
       &      ESP &    7522.776 &      0.270 &           $430\pm81$ &        &      MUS &    3893.521 &            &        $339.8\pm8.7$ \\
120198 &      MUS &    1601.676 &      0.529 &          $-287\pm62$ &        &      ESP &    7239.990 &            &        $244.9\pm4.7$ \\
       &      ESP &    7522.837 &      0.391 &           $-93\pm15$ & 187474 &      ESP &    7554.103 &      0.126 &         $-1539\pm16$ \\
124224 &      MUS &    3746.675 &      0.213 &          $580\pm220$ & 188041 &      MUS &    2838.532 &      0.103 &          $1294\pm24$ \\
       &      MUS &    3756.751 &      0.565 &         $-710\pm260$ &        &      MUS &    2848.513 &      0.147 &          $1266\pm22$ \\
       &      MUS &    3757.657 &      0.305 &         $-360\pm220$ &        &      MUS &    2857.421 &      0.187 &          $1231\pm23$ \\
       &      MUS &    3758.662 &      0.234 &          $280\pm250$ &        &      MUS &    3210.444 &      0.763 &          $1130\pm25$ \\
       &      MUS &    3759.635 &      0.104 &         $1170\pm300$ &        &      MUS &    3213.481 &      0.776 &          $1145\pm24$ \\
130559 &      MUS &    3406.743 &      0.544 &           $-64\pm30$ &        &      ESP &    7239.981 &      0.752 &          $1173\pm21$ \\
       &      MUS &    3411.649 &      0.115 &          $-432\pm20$ &        &      ESP &    7263.818 &      0.858 &          $1318\pm22$ \\
       &      MUS &    3746.712 &      0.727 &          $-276\pm16$ &        &      ESP &    7564.127 &      0.199 &          $1226\pm18$ \\
       &      MUS &    3755.725 &      0.451 &           $-81\pm21$ & 201601 &      ESP &    7239.998 &      0.135 &       $-948.0\pm9.0$ \\
       &      MUS &    3757.693 &      0.482 &           $-64\pm16$ &        &      ESP &    7564.132 &      0.144 &       $-907.1\pm8.7$ \\
       &      MUS &    3759.671 &      0.519 &           $-32\pm18$ & 203006 &      ESP &    7236.944 &      0.205 &          $-345\pm25$ \\
       &      MUS &    3760.708 &      0.062 &          $-416\pm20$ &        &      ESP &    7239.989 &      0.641 &           $723\pm15$ \\
       &      MUS &    3768.712 &      0.257 &          $-295\pm16$ &        &      ESP &    7261.783 &      0.917 &         $-1019\pm16$ \\
       &      MUS &    3770.716 &      0.308 &          $-213\pm20$ &        &      ESP &    7264.821 &      0.350 &           $720\pm21$ \\
       &      MUS &    3774.688 &      0.390 &          $-139\pm14$ &        &      ESP &    7284.829 &      0.784 &          $-241\pm21$ \\
       &      MUS &    3880.431 &      0.811 &          $-374\pm18$ &        &      ESP &    7296.836 &      0.446 &           $987\pm21$ \\
       &      MUS &    3893.383 &      0.599 &           $-99\pm19$ &        &      ESP &    7327.765 &      0.030 &         $-1119\pm14$ \\
140160 &      MUS &    1606.632 &      0.760 &         $-200\pm160$ & 217522 &      ESP &    7287.834 &            &       $-686.7\pm5.6$ \\
       &      MUS &    3864.501 &      0.581 &          $620\pm200$ & 220825 &      MUS &    3585.630 &      0.396 &          $-196\pm20$ \\
       &      MUS &    3889.556 &      0.280 &            $15\pm88$ &        &      MUS &    3588.558 &      0.458 &          $-265\pm19$ \\
       &      ESP &    7239.821 &      0.615 &           $278\pm38$ &        &      MUS &    3589.625 &      0.209 &           $175\pm18$ \\
       &      ESP &    7261.774 &      0.371 &           $246\pm35$ &        &      MUS &    3590.531 &      0.847 &           $272\pm17$ \\
       &      ESP &    7265.754 &      0.865 &          $-216\pm45$ &        &      MUS &    3598.585 &      0.517 &          $-286\pm21$ \\
       &      ESP &    7284.726 &      0.753 &           $-26\pm40$ &        &      MUS &    3607.515 &      0.805 &           $232\pm33$ \\
       &      ESP &    7326.684 &      0.045 &          $-312\pm35$ & 221760 &      ESP &    7262.006 &      0.694 &           $-17\pm10$ \\
       &      ESP &    7407.173 &      0.481 &           $230\pm46$ &        &      ESP &    7326.820 &      0.532 &        $-71.8\pm8.7$ \\
140728 &      MUS &    3864.537 &      0.750 &           $-54\pm75$ &        &      ESP &    7328.763 &      0.857 &         $57.3\pm9.5$ \\
       &      MUS &    3866.424 &      0.207 &          $-165\pm71$ &        &      ESP &    7330.817 &      0.201 &         $32.9\pm8.8$ \\
       &      MUS &    3872.445 &      0.855 &         $-460\pm110$ & 223640 &      ESP &    7611.144 &      0.509 &          $-148\pm11$ \\
\end{mpsupertabular*}